\RequirePackage{lineno} 
\documentclass[aps,prc,superscriptaddress,nofootinbib,showpacs,floatfix,singlecolumn]{revtex4}
\usepackage{graphicx} 
\usepackage{float} 
\usepackage{amssymb}
\usepackage{url}
\usepackage[colorlinks,linkcolor=blue,citecolor=blue,filecolor=black,urlcolor=blue]{hyperref}
\usepackage{cancel}
\usepackage{MnSymbol}
\usepackage{multirow}
\usepackage{bm}

\newcommand{\lr}[1]{\left\langle #1\right\rangle}
\newcommand{\llrr}[1]{\left\llangle #1\right\rrangle}
\newcommand{\pT} {\ensuremath{p_{\mathrm{T}}}}

\newcommand{\bQ} {\bm Q}
\newcommand{\bq} {\bm q}
\newcommand{\pp}{\mbox{$pp$}}
\newcommand{\pPb}{\mbox{$p$+Pb}}

\newcommand{\pA}{\mbox{$p$+A}}

\newcommand{\nch}{\mbox{$N_{\mathrm{ch}}$}}
\newcommand{\nchb}{\mbox{$N_{\textrm{ch}}^{\textrm{sel}}$}}

\newcommand{\ETfcal}{\mbox{$E_{\mathrm{T}}^{{\scriptscriptstyle \mathrm{Pb}}}$}}

\newcommand{\Dphi}{\mbox{$\Delta \phi$}}
\newcommand{\Deta}{\mbox{$\Delta \eta$}}

\usepackage{color}
\usepackage{harpoon}
\definecolor{my}{rgb}{1, 0, 0}
\begin{document} 
\title{Revealing long-range multi-particle collectivity in small collision systems via subevent cumulants}
\newcommand{\sunysb}{Department of Chemistry, Stony Brook University, Stony Brook, NY 11794, USA}
\newcommand{\bnl}{Physics Department, Brookhaven National Laboratory, Upton, NY 11796, USA}{
\newcommand{\polish}{Institute of Nuclear Physics Polish Academy of Sciences, Krakow, Poland}
\author{Jiangyong Jia}\email[Correspond to\ ]{jjia@bnl.gov}
\affiliation{\sunysb}\affiliation{\bnl}
\author{Mingliang Zhou}\email[Correspond to\ ]{mingliang.zhou@stonybrook.edu}
\affiliation{\sunysb}
\author{Adam Trzupek}\affiliation{\polish}

\begin{abstract}
Multi-particle azimuthal cumulants, often used to study collective flow in high-energy heavy-ion collisions, have recently been applied in small collision systems such as $\pp$ and $p$+A to extract the second-order azimuthal harmonic flow $v_2$. Recent observation of four-, six- and eight-particle cumulants with ``correct sign'' $c_2\{4\}<0, c_2\{6\}>0,  c_2\{8\}<0$ and approximate equality of the inferred single-particle harmonic flow, $v_2\{4\}\approx v_2\{6\}\approx v_2\{8\}$, have been used as strong evidence for a collective emission of all soft particles produced in the collisions. We show that these relations in principle could be violated due to the non-Gaussianity in the event-by-event fluctuation of flow and/or non-flow. Furthermore, we show, using $\pp$ events generated with the PYTHIA model, that $c_2\{2k\}$ obtained with standard cumulant method are dominated by non-flow from dijets. An alternative cumulant method based on two or more $\eta$-separated subevents is proposed to suppress the dijet contribution. The new method is shown to be able to recover a flow signal as low as 4\% imposed on the PYTHIA events, independently of how the event activity class is defined. Therefore the subevent cumulant method offers a more robust way of studying collectivity based on the existence of long-range azimuthal correlations between multiple distinct $\eta$ ranges. The prospect of using the subevent cumulants to study collective flow in A+A collisions, in particular its longitudinal dynamics, is discussed. 
\end{abstract}
\pacs{25.75.Dw} \maketitle
\section{Introduction}
\label{sec:1}
High energy heavy-ion collisions at the Relativistic Heavy Ion Collider (RHIC) and the Large Hadron Collider (LHC) create a strongly-interacting nuclear matter that exhibits many interesting characteristics. One such characteristic is the collimated emission of particle pairs with small azimuthal-angle separation, $\Dphi$, that extends over a large range of pseudorapidity differences, $\Deta$. This so called ``ridge'' correlation was first observed in A+A collisions~\cite{Adare:2008ae,Abelev:2009af,Alver:2009id,ALICE:2011ab,Aad:2012bu,Chatrchyan:2013kba}, and later was also observed in $\pp$~\cite{Khachatryan:2010gv,Aad:2015gqa,Aaboud:2016yar} and $\pA$ collisions~\cite{CMS:2012qk,Abelev:2012ola,Aad:2012gla,Adare:2013piz,Aad:2014lta,Khachatryan:2015waa}. In A+A collisions, the ridge is believed to be the consequence of collective emission of particles in the azimuthal direction, and the collectivity is generated in the final state after local thermalization, described by relativistic viscous hydrodynamic models~\cite{Heinz:2013th,Gale:2013da,Luzum:2013yya,Jia:2014jca}. For small systems such as $\pp$ and $\pA$ collisions, the origin of the ridge is less clear~\cite{Dusling:2015gta}. Current efforts are focused on understanding whether the ridge in small systems reflects global collectivity of the event~\cite{Loizides:2016tew}, and if so whether it is of hydrodynamic origin similar to A+A collisions~\cite{Bozek:2013uha} or it is created in the initial state from gluon saturation~\cite{Dusling:2013qoz}.

The ridge signal from two-particle correlation (2PC) is characterized by a Fourier decomposition $\sim 1+2v_n^2\cos (n \Dphi)$, where the $v_n$ denotes the single-particle anisotropy harmonics. The second-order coefficient $v_2$ is by far the largest, followed by $v_3$. In small collision systems, the extraction of the ridge signal requires a careful removal of a large contribution from dijets, which is estimated from 2PC in very low multiplicity events and then subtracted from higher multiplicity events~\cite{Abelev:2012ola,Aad:2012gla,Aad:2014lta,Aad:2015gqa,Aaboud:2016yar,Khachatryan:2016txc}. On the other hand, since collectivity is intrinsically a multi-particle phenomenon, it can be probed more directly using the multi-particle correlations (or cumulants) technique~\cite{Borghini:2000sa}. One of the perceived hallmark features of collectivity is the observation of positively defined signal from $2k$-particle correlation $v_2\{2k\}, k\geq2$, which has been measured for $v_2\{4\}$, $v_2\{6\}$ and $v_2\{8\}$ in high-multiplicity $\pp$ and $\pPb$ collisions~\cite{Aad:2013fja,Khachatryan:2015waa,Khachatryan:2016txc}. However, this perception could be wrong in small collision systems, where the non-flow correlations can be as large as or bigger than the genuine long-range ridge correlations. As we show in this paper, the sign of $v_n\{2k\}$ is sensitive to the event-by-event (ebye) fluctuations of flow and non-flow, and positive definiteness of $v_n\{2k\}$ in general is not required for flow correlations (also discussed in Ref.~\cite{Jia:2014pza}), and it could also be the result of non-flow correlations. 

Multi-particle cumulants suppress short-range correlations, but does not completely remove them. In fact $v_2\{4\}$ is observed to change sign at a smaller number of charged particle, $\nch$, in $pp$ and $\pPb$ collisions~\cite{Aad:2013fja,Khachatryan:2015waa,Khachatryan:2016txc}. Recently ATLAS observed~\cite{ATLAS:2016ics} that the $\nch$ value where sign-change happens and the magnitude of $v_2\{4\}$ depend on how the event classes are chosen for the calculation of cumulants. In this paper, we show that the choice of event class influences the probability distribution of non-flow, and consequently the non-flow contribution to the $v_2\{4\}$. An improved cumulant method is proposed to further suppress non-flow, and therefore reduce the sensitivity of $v_2\{4\}$ to non-flow fluctuations. In this method, cumulants are constructed from particles in several subevents separated in $\eta$. A subevent cumulant idea was proposed in Ref.~\cite{Adler:2002pu}, where the particles in the event are divided ``{\it randomly}'' into four subevents. However, the purpose there was not to suppress non-flow, but instead to circumvent a brute-force nested-loop calculation.

The structure of the paper is as follows. In sections~\ref{sec:2}-~\ref{sec:4}, we discuss the contributions of flow and non-flow to multi-particle cumulants and how such contributions depend on the probability distributions (or event-by-event fluctuations) of flow and non-flow. In sections~\ref{sec:5} and \ref{sec:6}, we introduce the subevent cumulant method, and demonstrate the robustness of the method using simulations based on PYTHIA8. The performance is compared to standard cumulant method focusing on four-particle correlations. In section~\ref{sec:7}, we discuss the implication of our findings for the interpretation of collectivity in small systems, as well as possible measurements enabled by our method, in particular for probing the longitudinal dynamics of collectivity. The formula for higher-order cumulants, and for symmetric and asymmetric cumulants are given in the Appendix.

\section{Event-by-event distribution of flow and non-flow}
\label{sec:2}
The azimuthal anisotropy of the particle production in an event can be characterized by Fourier expansion of the underlying probability distribution $\emph{P}(\phi)$ in azimuthal angle $\phi$:  
\begin{eqnarray}
\label{eq:b1}
\emph{P}(\phi) = \frac{1}{2\pi} \sum_{n=-\infty}^{\infty} {\bm v}_{n}e^{-in\phi},\;\;\; {\bm v}_n = v_n e^{in\Phi_n}\;,
\end{eqnarray}
where $v_n$ and $\Phi_n$ are magnitude and phase, respectively. In heavy ion collisions, flow harmonics ${\bm v}_n$ vary event to event, and can be described by a probability distribution $p({\bm v}_n)$~\cite{Luzum:2013yya,Jia:2014jca}. In A+A collisions, the flow fluctuations are close to Gaussian or equivalently the distribution of $v_n$ is Bessel-Gaussian after integrating out the $\phi$ angle~\cite{Voloshin:1994mz}:
\begin{eqnarray}
\label{eq:b2}
p({\bm v}_n) =\frac{1}{2\pi\delta^2_{n}}e^{-\left|{\bm v}_n-v_n^{\;_0}\right|^2 \big{/}\left(2\delta^2_{n}\right)},\; p(v_n) =\frac{v_n}{\delta_{n}^2}e^{-\frac{(v_n)^2+(v_n^{0})^2}{2\delta_{n}^2}} I_0\left(\frac{v_n^{0}v_n}{\delta_{n}^2}\right).
\end{eqnarray}
The parameter $v_n^{\;_0}$ reflects the component driven by the average geometry of the overlap region; it is expected to be sizable only for $n=2$.

Due to the finite number of particles $M$ produced in each event, harmonic flow can only be estimated from the observed flow vector ${\bm Q}_n$ or per-particle normalized flow vector ${\bm q}_n$:
\begin{eqnarray}
\label{eq:b4}
{\bm Q}_n \equiv \sum_i  e^{in\phi_i} = Q_n e^{in\Psi_n},\; {\bm q}_n \equiv \frac{\sum_i  e^{in\phi_i}}{M}= q_n e^{in\Psi_n}
\end{eqnarray}
where the sum runs over the particles in the event, $\phi_i$ are their azimuthal angles and $\Psi_n$ is the event plane. The magnitude and direction of ${\bm q}_n$ differ from those for the true flow, due to azimuthal fluctuations associated with finite particle multiplicity, denoted by ${\boldsymbol s}_{\mathrm{n}}^{\rm{stat}}$, as well as non-flow from various short-range correlations, denoted by ${\boldsymbol s}_{\mathrm{n}}$,
\begin{eqnarray}
\label{eq:b5}
{\bm q}_n = {\bm v}_{n}+{\boldsymbol s}_{\mathrm{n}}+{\boldsymbol s}_{\mathrm{n}}^{\rm{stat}}
\end{eqnarray}

In heavy ion collisions, all three components, flow, non-flow and finite number effects fluctuate event to event. Therefore the probability distributions for ${\bm q}_n$ and ${\bm v}_{n}$ can be related to each other by random smearing functions that reflect the non-flow and statistical fluctuation.
\begin{eqnarray}
\label{eq:b6}
p({\bm q}_n) = p({\bm v}_{n})\otimes p({\boldsymbol s}_{n})\otimes p({\boldsymbol s}_{\mathrm{n}}^{\rm{stat}})
\end{eqnarray}
The statistical fluctuation component usually cancels out after averaging over many events in two- or multi-particle correlation analyses (see the discussion in the next section). The non-flow fluctuations are more complicated, and they depend on the number of short-range sources, the particle multiplicity in each source and possible correlations between different sources (for example dijets)~\footnote{In this definition, the special case where the cluster sizes are fixed and only only the number of clusters fluctuates event by event, is considered as fluctuation of non-flow.}.

\section{Contribution of flow and non-flow to multi-particle cumulants}
\label{sec:3}

Let us first consider the case where there are only flow correlations for events with finite multiplicity. The moment of $p(v_n)$ distribution can be extracted from multi-particle correlations: A $2k$-particle azimuthal correlator is obtained by averaging over all unique combinations in one event then over all events~\cite{Borghini:2001vi,Bilandzic:2010jr}:
\begin{eqnarray}
\label{eq:c1}
\llrr{2k} = \llrr{e^{in\sum_{j=1}^{k}(\phi_{2j-1}-\phi_{2j})}}=\lr{v_n^{2k}}.
\end{eqnarray}
where $\left\langle x^{2k}\right\rangle\equiv\int x^{2k} p(x)dx$ is the $2k$-th moment of the probability distribution for $x$, and we have used the fact that statistical fluctuations (${\boldsymbol s}_{\mathrm{n}}^{\rm{stat}}$) drop out after averaging over many events. The $2k$-particle cumulant is then obtained by proper combination of correlations involving $\leq2k$ number of particles, whose expression can be obtained with the following generating function~\cite{Borghini:2001vi}:
\begin{eqnarray}
\label{eq:c2}
\ln \lr{e^{z({\bm v}_n+{\bm v}_n^*)}} = \ln\left(\sum_{k=1}^{\infty} \frac{z^{2k}}{k!^2}\lr{({\bm v}_n{\bm v}_n^*)^k}\right) =  \ln\left(\sum_{k=1}^{\infty} \frac{z^{2k}}{k!^2}\llrr{2k}\right) \equiv \sum_{k=1}^{\infty} \frac{z^{2k}}{k!^2}c_n\{2k\}\\
\end{eqnarray}
The formula for the first three are~\cite{Borghini:2001vi}:
\begin{eqnarray}\nonumber
c_n\{2\} &=& \left\langle\left\langle 2\right\rangle\right\rangle\\\nonumber
c_n\{4\} &=& \left\langle\left\langle 4\right\rangle\right\rangle-2\left\langle\left\langle 2\right\rangle\right\rangle^2\\\label{eq:c3}
c_n\{6\} &=& \left\langle\left\langle 6\right\rangle\right\rangle-9\left\langle\left\langle 4\right\rangle\right\rangle\left\langle\left\langle 2\right\rangle\right\rangle+12\left\langle\left\langle 2\right\rangle\right\rangle^3
\end{eqnarray}
which leads to the following cumulant-based ${\it definition}$ of harmonic flow $v_n$~\footnote{This definition of $v_n\{2k\}$ assumes the sign of $c_n\{2k\}$ is negative for even $k$, and positive for odd $k$, which is not true for an arbitrary $p(v_n)$ distribution~\cite{Jia:2014pza}.}:
\begin{eqnarray}\label{eq:c4}
v_n\{2\} = \sqrt{c_n\{2\}},\;v_n\{4\} =\sqrt[4]{-c_n\{4\}},\;v_n\{6\} = \sqrt[6]{c_n\{6\}/4}.
\end{eqnarray}
Eqs.~\ref{eq:c1}--\ref{eq:c4} play a crucial role for understanding how the flow and non-flow contribute to multi-particle correlations, so we shall discuss them in more detail below.

Let's consider the usual expressions for two- and four-particle correlators in terms of ${\bm q}_n$ and ${\bm Q}_n$ for one event with $M$ particles~\cite{Bilandzic:2010jr,Bilandzic:2012wva},
\begin{eqnarray}
\left\langle 2\right\rangle &=& \left\langle e^{in(\phi_1-\phi_2)}\right\rangle = \frac{Q_n^2-M}{M(M-1)} = \frac{q_n^2-\tau}{1-\tau}\\\nonumber
\left\langle 4\right\rangle &=& \left\langle e^{in(\phi_1+\phi_2-\phi_3-\phi_4)}\right\rangle 
= \frac{Q_n^4-2\mbox{Re}[{\bm Q}_{2n}{\bm Q}_n^{*2}]-4(M-2)Q_n^2+2M(M-3)+Q_{2n}^2}{M(M-1)(M-2)(M-3)} \\\label{eq:c5}
&&\hspace*{2.7cm}= \frac{q_n^4-2\tau(\mbox{Re}[{\bm q}_{2n}{\bm q}_n^{*2}] +2q_n^2)+\tau^2(2+8q_n^2+q_{2n}^2)-6\tau^3}{(1-\tau)(1-2\tau)(1-3\tau)}
\end{eqnarray}
The advantage of using per-particle normalized flow vector ${\bm q}_n$ is that all quantities in the equation are smaller than one, and the terms can be sorted in powers of $\tau=1/M\ll1$ and $q_{kn}\sim q_{n}^k$~\cite{Jia:2014pza}. The event-by-event weights are slightly modified from those in Ref.~\cite{Bilandzic:2010jr}: $W_{\lr{2}}=M(M-1)/2$, $W_{\lr{4}}=M(M-1)(M-2)(M-3)/4!$ etc~\footnote{Since the correct weight should consider only unique $2k$-particle combinations instead of all permutations, the weight for 2k-particle correlation need to be divided by $(2k)!$. However since all events are affected by the same factor, this does not matter much in practice except for discussing the statistical power between different cumulant methods.}. Each term has a simple interpretation. The two-particle correlator $\left\langle 2\right\rangle$ has two terms: the first term $Q_n^2$ contains $M^2$ pairs, and the second term corresponds to the contribution of $M$ duplicate pairs: $\phi_1=\phi_2$. Four-particle correlator $\left\langle 4\right\rangle$ has $P_{M,4}=M(M-1)(M-2)(M-3)$ quadruplets, expressed as $Q_n^4$ ($M^4$ quadruplets) minus contributions from quadruplets where the same particle appears more than once, e.g., combinations such as $\phi_1=\phi_2=\phi_3=\phi_4$, $\phi_1=\phi_2=\phi_3 \neq \phi_4$,  $\phi_1=\phi_2 \neq \phi_3 \neq \phi_4$ ... etc. In total, there are $M^4-P_{M,4}$ quadruplets containing duplicated particles. The removal of duplicate combinations insures that the statistical fluctuation associated with finite particle multiplicity drops out from ${\bm q}_n$ after averaging over many events.

When only flow correlations are present, cumulants are fully determined from moments of the $p(v_n)$:
\begin{eqnarray}\nonumber
c_n\{2\}&=& \left\langle v_n^2\right\rangle\\\nonumber
c_n\{4\}&=&\left\langle v_n^4\right\rangle-2\left\langle v_n^2\right\rangle^2\\
c_n\{6\}&=&\left\langle v_n^6\right\rangle-9\left\langle v_n^4\right\rangle\left\langle v_n^2\right\rangle+12\left\langle v_n^2\right\rangle^3\\\nonumber
&...&
\end{eqnarray}
For any $p(v_n)$ distribution, it is expected that $v_n\{2\}\geq v_n\{4\}$ since $v_n\{2\}^4-v_n\{4\}^4= \lr{(v_n^2-\lr{v_n^2})^2}\geq0$.

If the distribution of ${\bm v}_n$ is the Gaussian function defined in Eq.~\ref{eq:b2}, flow harmonics defined by cumulants have a simple expression~\cite{Voloshin:2007pc}:
\begin{eqnarray}
\label{eq:c6}
v_n\{2k\} &=& \left\{\begin{array}{ll} \sqrt{\left(v_n^{\mathrm{0}}\right)^2+2\delta^2_{n}} & k=1\\
v_n^{\mathrm{0}} & k>1\\   \end{array}\right.
\end{eqnarray}
while the $v_n\{2\}$ also includes contribution from flow fluctuations, the higher-order cumulants $v_n\{4,6,..\}$ measure the component associated with average geometry.

In small systems, due to finite number of sources in the initial density distribution, flow fluctuations are expected to deviate from Gaussian, and were suggested to follow a power function~\cite{Bzdak:2013rya,Yan:2013laa}. 
\begin{eqnarray}
\label{eq:b3}
p(v_n) =2\frac{\alpha}{\kappa} \frac{v_n}{\kappa} (1-(\frac{v_n}{\kappa})^2)^{\alpha-1}
\end{eqnarray}
where $\alpha$ is related to the number of sources for particle production in the initial state, and $\kappa$ is related to hydrodynamic response to the initial eccentricity $\epsilon_n$, $\kappa = v_n/\epsilon_n$. The higher-order cumulants $v_n\{4,6,..\}$ are expected to quickly converge to a non-zero value reflecting the influence of the non-Gaussian tail (controlled by $\alpha$) of the power distribution.

When both flow and non-flow are present and are un-correlated with each other, the generating function Eq.~\ref{eq:c2} becomes:
\begin{eqnarray}
\label{eq:c7}
\ln \lr{e^{z({\bm q}_n+{\bm q}_n^*)}} = \ln \lr{e^{z({\bm v}_n+{\bm v}_n^*)}}+\ln \lr{e^{z({\bm s}_n+{\bm s}_n^*)}}= \sum_{k=1}^{\infty} \frac{z^{2k}}{k!^2}\left(c_n\{2k,v\}+c_n\{2k,s\}\right),
\end{eqnarray}
and the measured cumulant is the sum of the separate contributions from cumulant of flow and cumulant of non-flow:
\begin{eqnarray}
\label{eq:c8}
c_n\{2k\}=c_n\{2k,v\}+c_n\{2k,s\}.
\end{eqnarray}

Equation~\ref{eq:c8} provides a simple way to understand the influence of non-flow to $c_n\{2k\}$. In large collision systems, the number of non-flow sources is large and proportional to $M$. Since the orientations of these sources are weakly correlated, non-flow fluctuation $p(s_n)$ is expected to approach Gaussian with a width that scales as $1/M$ (confirmed in HIJING simulation~\cite{Jia:2013tja}). In this case, the non-flow contributions to four-particle or higher-order cumulants are naturally suppressed, e.g., $c_n\{2k,s\}\approx 0$ for $k>1$. However, in small collision systems, the $p(s_n)$ distribution is expected to be highly non-Gaussian since the number of sources is small but strongly fluctuates event to event, and the particle multiplicity in a single source can be large compared to $M$. In this case, the contribution from non-flow could be large, and the sign and magnitude of $c_n\{2k\}$ depend on the nature of $p(v_n)$ and $p(s_n)$. 

\section{Fluctuations of flow and non-flow and the sign of cumulants}
\label{sec:4}
The $c_n\{2k\}$ are more fundamental than the $v_n\{2k\}$; the latter is well motivated only when flow distribution is close to Gaussian. As we shall discuss below (also see~\cite{Jia:2014pza}), it is easy to find $p(v_n)$ distributions for which $\left\langle v_n^4\right\rangle>2\left\langle v_n^2\right\rangle^2$ such that $c_n\{4\}>0$ and $v_n\{4\}$ become undefined~\footnote{This is also true for higher-order cumulants and Lee-Yang Zero method~\cite{Bhalerao:2003xf}, whose existence is only proven for narrow fluctuations ($v_n^{\rm{max}}/v_n^{\rm{min}}\lesssim2.3$) or nearly Gaussian fluctuations~\cite{Bhalerao:2003xf}.}. Furthermore due to the large power relating the two quantities, i.e., $c_n\{2k\} \propto (v_n\{2k\})^{2k}$, a large change in $c_n\{2k\}$ usually leads to a small change in $v_n\{2k\}$ for large $k$. A factor of two change in $c_n\{8\}$, for example, only leads to 9\% change in $v_n\{8\}$. Given the large uncertainty (more than $10\%$) in current measurements of $v_2\{4\}$, $v_2\{6\}$ and $v_2\{8\}$ in small systems, they cannot yet place strong constraints on $p(v_n)$~\footnote{To distinguish between Gaussian and power distributions, for example, a precision of few percents is required for $v_2\{4\}$, $v_2\{6\}$, and $v_2\{8\}$~\cite{Yan:2013laa}.}.

We consider a simple case where the data consist of two types of events: The first type has finite correlation signal $v_0$ with a probability of $a$, and the second type has zero signal with a probability of $1-a$.
\begin{eqnarray}
\label{eq:c9}
p(x;a) &=& a\delta(x-v_0)+(1-a) \delta(x),\; 0<a<1.
\end{eqnarray}
It can be shown that for certain range of $a$, the $c_{n}\{2k\}$ for $k>1$ always have the ``wrong'' sign such that $v_{n}\{2k\}$ are undefined. For example for $a=1/3$, $c_n\{4\}= \frac{1}{9}v_0^4$, $c_n\{6\}= -\frac{2}{9}v_0^6$, $c_n\{8\}= \frac{71}{9}v_0^8$. Note that we make no explicit distinction between flow and non-flow, but rather just require the events to have the same underlying {\it p.d.f} $\emph{P}(\phi)$.  This can be achieved in a simulation where events are generated with fixed multiplicity $M$ and $a=1/3$. In the case of flow, one could generate $N$ events with the same $v_0$, but with a random phase event-to-event; these events are then combined with $2N$ events that are generated with zero signal. In the case of non-flow, one could generate $N$ events combined with the same dijet event but rotated randomly in azimuth. The multiplicity of the dijet is chosen such that its effective harmonic signal in the merged event is also $v_0$; these events are then combined with $2N$ events generated with zero signal. It is clear that both cases should give the same $c_{n}\{2k\}$ value.

Another important property of cumulants is that they are not additive (see Ref.~\cite{Jia:2014jca}) due to the presence of non-linear term such as $\lr{v_n^2}^2$ in$ c_n\{4\}$, i.e. the cumulants calculated for ensemble A+B $c_n\{2k,$A+B$\}$ is not equal to the average of the cumulants calculated separately for ensemble A $c_n\{2k,$A$\}$ and ensemble B $c_n\{2k,$B$\}$. To see this, consider a pure fluctuation driven scenario of Eq.~\ref{eq:b2} by setting $v_n^0=0$: $p(v_n)\propto v_n e^{-v_n^2/(2\delta_n^2)}$. The events for $p(v_n)$ are divided into two equal halves, A and B, according to the magnitude of $v_n$ (this is possible in reality via event-shape engineering technique). Then we have $c_n\{2k\}=0$ for $k>1$, but both $c_n\{2k,$A$\}$ and $c_n\{2k,$B$\}$ are negative. As we will show later, this is one of the reason why multi-particle cumulants are sensitive to the event class definition. 

These examples demonstrate that the values of $c_n\{2k\}$ depend only on the probability distribution of flow and non-flow via Eq.~\ref{eq:c8}. Having the ``correct sign'' for $c_n\{2k\}$, i.e. negative for even-$k$ and positive for odd-$k$, in principle, is neither a necessary nor sufficient condition for the correlations being dominated by a global collectivity. On the other hand, it is entirely possible that Gaussian and power distributions could be a reasonable guess of the initial geometry in $\pp$ collisions. In such cases, the ``correct signs'' of $c_n\{2k\}$ are still respected, but it is not yet proven they are the only models. ATLAS has shown that $c_2\{4\}$ calculated with the standard cumulant method in $pp$ collisions is sensitive to the event class definition~\cite{ATLAS:2016ics}. This is because, as shown below, $c_2\{4\}$ has a significant non-flow contribution, and the sensitivity of $c_2\{4\}$ to event class definition is simply due to the change of the underlying non-flow distribution when the event class is changed and the fact that multi-particle cumulants are not additive (discussed above). The main source of non-flow in $pp$ collisions is dijets, which are responsible for the large non-flow peak at $\Dphi\sim\pi$ observed in two-particle correlations~\cite{Aad:2015gqa}. In the following we introduce a subevent cumulant method that further suppresses the non-flow correlations and as well as isolates genuine long-range collectivity that correlates particles in several distinct rapidity ranges. 

\section{Subevent cumulants}
\label{sec:5}

In this paper we focus mainly on four-particle cumulants, but the generalization to higher-order is straightforward (formulae for some six- and eight-particle cumulants are provided in Appendixes B and C). The basic idea for subevents cumulants is shown in Figure~\ref{fig:0}a. The event is divided into three non-overlapping rapidity ranges, labeled as $a$,$b$ and $c$. Four-particle correlators are constructed by choosing two particles from subevent $a$ and one particle each from subevents $b$ and $c$. Dijets contributions, the main source for non-flow, are suppressed, since they can only produce particles in two subevents, except for the small chance of one jet falling on the boundary between two subevents. To increase the statistical power of the events, we include quadruplets obtained by permutations of $a$, $b$ and $c$. For comparison, we also consider cumulants based on two subevents as shown in Figure~\ref{fig:0}b. In this case, two particles each are chosen from $b$ and $c$, which effectively suppresses contributions from intra-jet correlations, but not inter-jet correlations when the two jets from a dijet land in different subevents. 
\begin{figure}[!h]
\begin{center}
\includegraphics[width=0.6\linewidth]{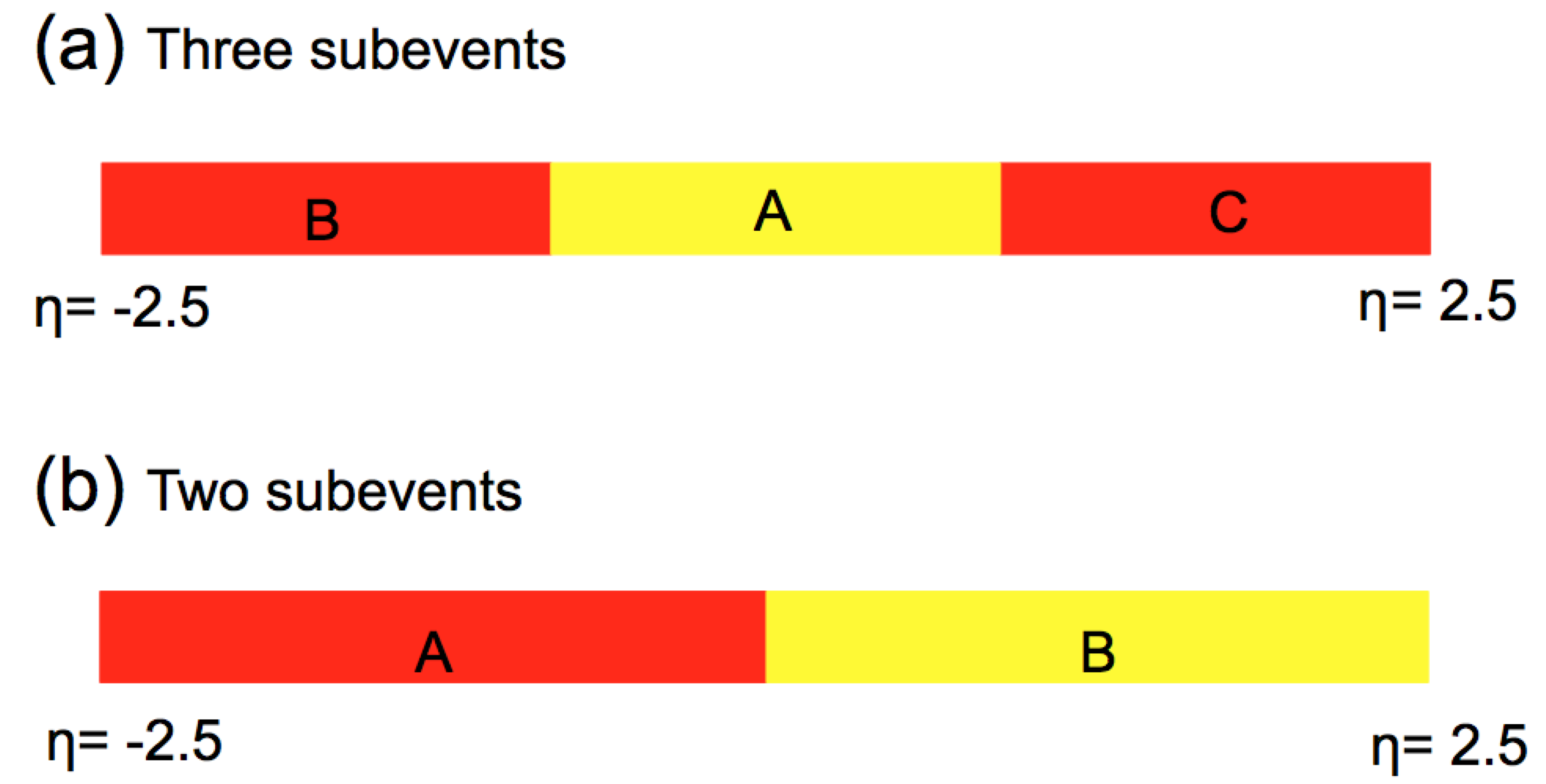}
\end{center}
\caption{\label{fig:0} The $\eta$ ranges for the subevents in three-subevent method (a) and two-subevent method (b).} 
\end{figure}

\subsection{Two-subevent cumulants}
\label{sec:5.1}
Two different types of four-particle correlators can be constructed from two subevents, expressed as $e^{in(\phi_{1}^a+\phi_{2}^a -\phi_{3}^b-\phi_{4}^b)}$ or $e^{in(\phi_{1}^a+\phi_{2}^b -\phi_{3}^a-\phi_{4}^b)}$. In the first type,  particles in subevent $a$ are conjugated with particles in subevent $b$, and are denoted as ``$(aa, b^*b^*)$'', with the comma separating the subevents. We will use a short-handed notation ``$a,a|b,b$'' or simply ``$2a|2b$''. In the second type, the two particles in each subevent are conjugated with each other, and are denoted as ``$(aa^*, bb^*)$'' or simply ``$a,b|a,b$''.

Multi-particle correlators based on two subevents, each having $M_a$ and $M_b$ particles, can be derived by keeping track of the duplicates terms~\cite{Bilandzic:2010jr,Bilandzic:2013kga} similar to those in Eq.~\ref{eq:c5}. The four-particle correlator of the first type and the related two-particle correlator, as well as their event-by-event weights have the following expressions in terms of flow vectors:
\begin{eqnarray}
\label{eq:d0}
\lr{2}_{a|b}&\equiv&\lr{e^{in(\phi_{1}^a-\phi_{2}^b)}} = \frac{{\bm Q}_{n,a}{\bm Q}^*_{n,b}}{M_aM_b} = {\bm q}_{n,a}{\bm q}^*_{n,b}\\\nonumber
\lr{4}_{a,a|b,b}&\equiv&\lr{e^{in(\phi_{1}^a+\phi_{2}^a -\phi_{3}^b-\phi_{4}^b)}} = \frac{({\bm Q}_{n,a}^2-{\bm Q}_{2n,a})({\bm Q}_{n,b}^{2}-{\bm Q}_{2n,b})^*}{M_a(M_a-1)\;M_b(M_b-1)}=\frac{({\bm q}_{n,a}^2-\tau_a{\bm q}_{2n,a})({\bm q}_{n,b}^{2}-\tau_b{\bm q}_{2n,b})^*}{(1-\tau_a)(1-\tau_b)}\\\label{eq:d1}
&&\hspace*{2.7cm}=\frac{{\bm q}_{n,a}^2{\bm q}_{n,b}^{*2}-\tau_a{\bm q}_{2n,a}{\bm q}_{n,b}^{*2}-\tau_b{\bm q}_{n,a}^2{\bm q}_{2n,b}^*+\tau_a\tau_b{\bm q}_{2n,a}{\bm q}_{2n,b}^*}{(1-\tau_a)(1-\tau_b)}\\
W_{\lr{2}_{a|b}}&=&M_aM_b\;,\;\;\; W_{\lr{4}_{a,a|b,b}}=M_a(M_a-1)\;M_b(M_b-1)/4
\end{eqnarray}
For simplicity we have dropped the label ``Re'', but it should be understood that only the real component of the $\lr{2k}$ is kept. The superscripts $o=a,b$ denote the two subevents, and ${\bm Q}_{n,o}$ and ${\bm q}_{n,o}$ represent the flow vector and normalized flow vectors, respectively:
\begin{eqnarray}
\label{eq:d2}
{\bm Q}_{n,o}\equiv \sum_i  e^{in\phi_i^{o}},\;{\bm q}_{n,o}\equiv \frac{\sum_i  e^{in\phi_i^o}}{M_o},\; \tau_o=\frac{1}{M_o},\;\;\; o=a\;\; \mbox{or}\;\; b
\end{eqnarray}
The last three terms in the numerator of Eq.~\ref{eq:d1} account for three types of quadruplets containing duplicated particles, $\phi_{1}=\phi_{2}$, $\phi_{3}=\phi_{4}$, and $\phi_{1}=\phi_{2}$ \& $\phi_{3}=\phi_{4}$, respectively. There are a total of $M_a^2M_b^2-M_a(M_a-1)\;M_b(M_b-1)$ such quadruplets.

Following the diagrammatic approach and the notation used in Ref.~\cite{Borghini:2000sa}, multi-particle correlators, after averaging over many events, can be expanded into contributions from flow and non-flow, assuming they are uncorrelated:
\begin{eqnarray}
\llrr{2}_{a|b} &\equiv&\llrr{e^{in(\phi_{1}^a-\phi_{2}^b)}} = \lr{e^{in(\phi_1^a-\phi_2^b)}}_c + \lr{{\bm v}_{n,a}{\bm v}_{n,b}^*}\\\label{eq:d3}
\llrr{4}_{a,a|b,b} &\equiv& \llrr{e^{in(\phi_1^a+\phi_2^a-\phi_3^b-\phi_4^b)}} =\lr{e^{in(\phi_1^a+\phi_2^a-\phi_3^b-\phi_4^b)}}_c + \lr{{\bm v}^2_{n,a}{\bm v}^{*2}_{n,b}}+ \underline{2 \lr{e^{in(\phi_1^a-\phi_2^b)}}^2_c+4\lr{e^{in(\phi_1^a-\phi_2^b)}}_c\lr{{\bm v}_{n,a}{\bm v}_{n,b}^*}}\;,
\end{eqnarray}
where the ``connected'' terms denoted by $\lr{ }_c$ represent genuine multi-particle correlations from non-flow, and we have kept the complex notation for flow vectors, e.g. $\lr{{\bm v}_{n,a}{\bm v}_{n,b}^*}=\lr{v_{n,a}v_{n,b}\cos n(\Phi_n^a-\Phi_n^b)}$, to keep track of the twist of event-plane angles and fluctuations of flow magnitudes. The last two terms in the four-particle correlator with underline are non-flow terms involving only two particles, which can be removed if the four-particle cumulant is defined as:
\begin{eqnarray}
\label{eq:d4}
c_n^{a,a|b,b}\{4\} \equiv \llrr{4}_{a,a|b,b}- 2\llrr{2}_{a|b}^2 = \lr{e^{in(\phi_1^a+\phi_2^a-\phi_3^b-\phi_4^b)}}_c + \lr{{\bm v}^2_{n,a}{\bm v}^{*2}_{n,b}}-2\lr{{\bm v}_{n,a}{\bm v}_{n,b}^*}^2\;.
\end{eqnarray}
The remaining non-flow contribution, denoted by the first term, has to connect two particles in subevent $a$ with the other two particles in subevent $b$.

The second type of four-particle correlation has the following expressions:
\begin{eqnarray}
&&\lr{2}_{a|a}\equiv\lr{e^{in(\phi_{1}^a-\phi_{2}^a)}} = \frac{Q_{n,a}^2-M_a}{M_a(M_a-1)} = \frac{q_{n,a}^2-\tau_a}{1-\tau_a}\;, \;\;\;\;\lr{2}_{b|b}\equiv\lr{e^{in(\phi_{1}^b-\phi_{2}^b)}} = \frac{Q_{n,b}^2-M_b}{M_b(M_b-1)} = \frac{q_{n,b}^2-\tau_b}{1-\tau_b}\\\label{eq:d5}
&&\lr{4}_{a,b|a,b}\equiv\lr{e^{in(\phi_{1}^a+\phi_{2}^b -\phi_{3}^a-\phi_{4}^b)}} = \frac{(Q_{n,a}^2-M_a)(Q_{n,b}^{2}-M_b)}{M_a(M_a-1)\;M_b(M_b-1)} = \frac{(q_{n,a}^2-\tau_a)(q_{n,b}^{2}-\tau_b)}{(1-\tau_a)(1-\tau_b)}
\end{eqnarray}
In this case, one can show that the four particle cumulants should be defined as:
\begin{eqnarray}
\label{eq:d6}
c_n^{a,b|a,b}\{4\} \equiv \llrr{4}_{a,b|a,b}- \llrr{2}_{a|a}\llrr{2}_{b|b}-\llrr{2}_{a|b}^2 = \lr{e^{in(\phi_1^a+\phi_2^b-\phi_3^a-\phi_4^b)}}_c + \lr{{\bm v}^2_{n,a}{\bm v}^{*2}_{n,b}}-\lr{v_{n,a}^2}\lr{v_{n,b}^2}-\lr{{\bm v}_{n,a}{\bm v}_{n,b}^*}^2\;.
\end{eqnarray}
The second type of four-particle cumulant has a larger contribution from non-flow, since a single jet could correlate two particles in one subevent, as indicated by $\llrr{2}_{a|a}$ and $\llrr{2}_{b|b}$ that need to be subtracted from $\llrr{4}_{a,b|a,b}$.

\subsection{Three-subevent cumulants}
\label{sec:5.2}
There are also two types of four-particle correlators constructed from three subevents, which can be expressed as either $e^{in(\phi_{1}^a+\phi_{2}^a -\phi_{3}^b-\phi_{4}^c)}$ or $e^{in(\phi_{1}^a+\phi_{2}^b -\phi_{3}^a-\phi_{4}^c)}$. In the first type, particles in $a$ are conjugated with particles in $b$ and $c$, and are denoted as ``$(b^*, aa, c^*)$'' or ``$a,a|b,c$''. In the second correlator, the two particles from subevent $a$ are conjugated with each other, and is denoted as ``$(b, aa^*, c^*)$'' or ``$a,b|a,c$''.

The first type of four-particle correlation can be written as:
\begin{eqnarray}\nonumber
\lr{4}_{a,a|b,c}&\equiv&\lr{e^{in(\phi_{1}^a+\phi_{2}^a -\phi_{3}^b-\phi_{4}^c)}} = \frac{({\bm Q}_{n,a}^2-{\bm Q}_{2n,a}){\bm Q}_{n,b}^{*}{\bm Q}_{n,c}^{*}}{M_a(M_a-1)M_bM_c}=\frac{({\bm q}_{n,a}^2-\tau_a{\bm q}_{2n,a}){\bm q}_{n,b}^*{\bm q}_{n,c}^*}{1-\tau_a},\;W_{\lr{4}_{a,a|b,c}}=M_a(M_a-1)M_bM_c/2\;.\\\label{eq:d7}
\end{eqnarray}
It can be expanded into contributions from flow and non-flow after averaging over many events:
\begin{eqnarray}
\label{eq:d8}
\llrr{4}_{a,a|b,c}&\equiv&\llrr{e^{in(\phi_{1}^a+\phi_{2}^a -\phi_{3}^b-\phi_{4}^c)}} = \lr{e^{in(\phi_1^a+\phi_2^a-\phi_3^b-\phi_4^c)}}_c + \lr{{\bm v}^2_{n,a}{\bm v}^*_{n,b}{\bm v}^*_{n,c}}\\
&&+\underline{2 \lr{e^{in(\phi_1^a-\phi_2^b)}}_c\lr{e^{in(\phi_1^a-\phi_2^c)}}_c+2\lr{e^{in(\phi_1^a-\phi_2^b)}}_c\lr{{\bm v}_{n,a}{\bm v}_{n,c}^*}+2\lr{e^{in(\phi_1^a-\phi_2^c)}}_c\lr{{\bm v}_{n,a}{\bm v}_{n,b}^*}}
\end{eqnarray}
The second line contains non-flow involving only two particles, which can be removed by the following definition of the four-particle cumulant:
\begin{eqnarray}
\label{eq:d9}c_n^{a,a|b,c}\{4\} \equiv \llrr{4}_{a,a|b,c}- 2\llrr{2}_{a|b}\llrr{2}_{a|c} = \lr{e^{in(\phi_1^a+\phi_2^a-\phi_3^b-\phi_4^b)}}_c + \lr{{\bm v}^2_{n,a}{\bm v}^*_{n,b}{\bm v}^*_{n,c}}-2\lr{{\bm v}_{n,a}{\bm v}_{n,b}^*}\lr{{\bm v}_{n,a}{\bm v}_{n,c}^*}
\end{eqnarray}

The second type of four-particle correlation has the following expression:
\begin{eqnarray}
\label{eq:d10}\lr{4}_{a,b|a,c}&\equiv&\lr{e^{in(\phi_{1}^a+\phi_{2}^b -\phi_{3}^a-\phi_{4}^c)}} = \frac{(Q_{n,a}^2-M_a){\bm Q}_{n,b}{\bm Q}_{n,c}^*}{M_a(M_a-1)\;M_bM_c} = \frac{(q_{n,a}^2-\tau_a){\bm q}_{n,b}{\bm q}_{n,c}^*}{1-\tau_a}
\end{eqnarray}
The corresponding four-particle cumulant should be defined as:
\begin{eqnarray}
\nonumber
c_n^{a,b|a,c}\{4\} &\equiv& \llrr{4}_{a,b|a,c}- \llrr{2}_{a|a}\llrr{2}_{b|c}-\llrr{2}_{a|b}\llrr{2}_{a|c} \\\label{eq:d11}
&=& \lr{e^{in(\phi_1^a+\phi_2^b-\phi_3^a-\phi_4^c)}}_c + \lr{{\bm v}^2_{n,a}{\bm v}^*_{n,b}{\bm v}^*_{n,c}}-\lr{v_{n,a}^2}\lr{{\bm v}_{n,b}{\bm v}_{n,c}^*}-\lr{{\bm v}_{n,a}{\bm v}_{n,b}^*}\lr{{\bm v}_{n,a}{\bm v}_{n,c}^*}
\end{eqnarray}
This second type of four-particle cumulant has more contribution from non-flow, since a single jet correlates two particles in the same subevent $a$, as indicated by $\llrr{2}_{a|a}$ that need to be subtracted from $\llrr{4}_{a,b|a,c}$.

\subsection{Statistical power of different methods}
The statistical power of the cumulant method is controlled by the number of unique combinations in each event. When the event multiplicity $M$ is large, the number of unique quadruplets for $\lr{4}$ is expected to be $W \approx M^4/4! = M^4/24$ for standard cumulant, $W \approx (M/2)^2/2!(M/2)^2/2!  = M^4/64$ for the two-subevent method and $W \approx 3\times (M/3)^2/2!(M/3)(M/3)  = M^4/54$ for the three-subevent method, where the factor 3 takes into account the three permutations of  $a$, $b$ and $c$. Therefore in comparison to the standard cumulant method, the two-subevent and three-subevent methods loose a factor of 2.7 and 2.3 statistics, respectively. It is also interesting to note that the three-subevent method has more statistical power than the two-subevent method. On the other hand, the number of unique pairs for $\lr{2}$ is $M^2/2$, $(M/2)^2 = M^2/4$ and $(M/3)^2=M^2/9$ for standard, two-subevent (see Eq.~\ref{eq:d4}) and three-subevent (see Eq.~\ref{eq:d9}) methods, respectively. In the end, the total statistical uncertainty on $c_n\{4\}$ is expected to be smallest for the standard method, but it is non-trivial to decide whether the two-subevent method or the three-subevent method has better statistical precision for $c_n\{4\}$.

\section{Simulation setup}
\label{sec:6}
Particle production in $pp$ collisions is often described by QCD-inspired models implemented in Monte Carlo (MC) event generators such as PYTHIA~\cite{Sjostrand:2007gs}. The PYTHIA model contains significant non-flow correlations from jets, dijets and resonance decays but has no genuine long-range ridge correlations. In this paper, 200 million $pp$ collisions at $\sqrt{s}=13$ TeV are generated with PYTHIA8. Multi-particle cumulants based on standard method as well as subevent methods are calculated to quantify how they are biased by non-flow correlations as a function of charged particle multiplicity. Furthermore, flow signal is added to the generated event using a flow afterburner~\cite{Masera}, and the performance for recovering the input flow signal is studied.

In a typical azimuthal correlation analysis, multi-particle correlators $\lr{2k}$ are calculated for particles passing selection criteria $X_1$, and are then averaged over many events with the same number of charged particles $\nch$ passing another selection criteria $X_2$ to obtain $\llrr{2k}$ and then $c_n\{2k\}$~\cite{ATLAS:2016ics}. In order to present the cumulants along a common $x$-axis, the $\nch$ based on criteria $X_2$ is often mapped to a common event activity measure, typically $\nch$ passing yet another criteria $X_3$. If non-flow is sufficiently suppressed, $c_n\{2k\}$ should measure the collectivity of the entire event and therefore should not depend on the intermediate criteria $X_2$~\footnote{This was observed for $v_n$ obtained 2PC analysis in $p$+Pb collisions with peripheral subtraction~\cite{Aad:2014lta}: the $v_n$ obtained in bins of forward total energy $\ETfcal$, but then mapped to $\nch$ based on correlation between $\ETfcal$ and $\nch$, is found to agree well with $v_n$ obtained by binning events directly in $\nch$.}. However, the ATLAS Collaboration observed that $c_2\{4\}$ calculated in both PYTHIA and with $pp$ data depends sensitively on criteria $X_2$~\cite{ATLAS:2016ics}, and sensitivity was thought to arise from relative multiplicity fluctuation between $X_2$ and $X_3$. The real reason, as we shall argue below, is due to change in event-by-event non-flow distribution $p(s_n)$~\footnote{As we discussed before, the statistical component ${\boldsymbol s}_{\mathrm{n}}^{\rm{stat}}$ associated with multiplicity fluctuation does not influence the measured $v_n$.}.  We show that the three-subevent method nearly completely removes non-flow contributions, such that the resulting $c_2\{4\}$ is independent of $X_2$. 
\begin{table}[!h]
\centering
\caption{The setup used in this analysis. The first column shows the three cumulant methods and corresponding  $\eta$ ranges, the other three columns shows the $\pT$ selection used for calculating azimuthal correlation $\lr{2k}$, for selecting events used for $c_n\{2k\}$, and for defining event activity used to represent the $x$-axis, respectively. See the text for more detail.}
\label{tab:1}
\begin{tabular}{|c||c|c|c|}\hline 
 \multirow{2}{*}{Cumulant types}& $X_1$: criteria for $\lr{2k}$        & $X_2$: criterial for $\nchb$  & $X_3$: criteria for $\nch$\\
                             &                                       & used in event averaging       &  to represent the $x$-axis \tabularnewline\hline
 standard method          &\multirow{6}{1.3in}{\;\;\;\;\mbox{$0.3<\pT<3$ GeV} \mbox{\;\;\;\;$0.5<\pT<5$ GeV}}& \multirow{6}{1.3in}{\;\;\;\;same as for $X_1$ \mbox{\;\;\;\;$\pT>0.2$ GeV} \mbox{\;\;\;\;$\pT>0.4$ GeV}  \mbox{\;\;\;\;$\pT>0.6$ GeV}}    & \\
  $|\eta|<2.5$               &                                                    &                  & \\\cline{1-1}
 two-subevent method         &                                                    &                  &$\pT>0.4$ GeV\\
$\eta\in[-2.5,0],[0,2.5]$    &                                                    &                  &\\\cline{1-1}
 three-subevent method       &                                                    &                  &\\
$\eta\in[-2.5,-0.83],[-0.83,0.83],[0.83,2.5]$&                                    &                  &\\\hline             
\end{tabular}
\end{table}

Table~\ref{tab:1} summarizes the types of cumulants and charged particle selection criteria studied in PYTHIA8. For each method, cumulants are calculated in two $\pT$ ranges ($X_1$) for event classes defined by the number of charged particles in four $\pT$ ranges ($X_2$), denoted by $\nchb$.  For each combination, the $c_n\{2k\}$ are calculated for events with a fixed $\nchb$ multiplicity, the results for one-particle-width bins in $\nchb$ are then combined to broader $\nchb$ intervals. When presenting the final results, the $\lr{\nchb}$ values for given $\nchb$ interval are mapped to the average number of charged particles with $\pT>0.4$ GeV ($X_3$), denoted by $\lr{\nch}$.

\section{Results}
\label{sec:7}
\begin{figure}[h!]
\begin{center}
\includegraphics[width=1\linewidth]{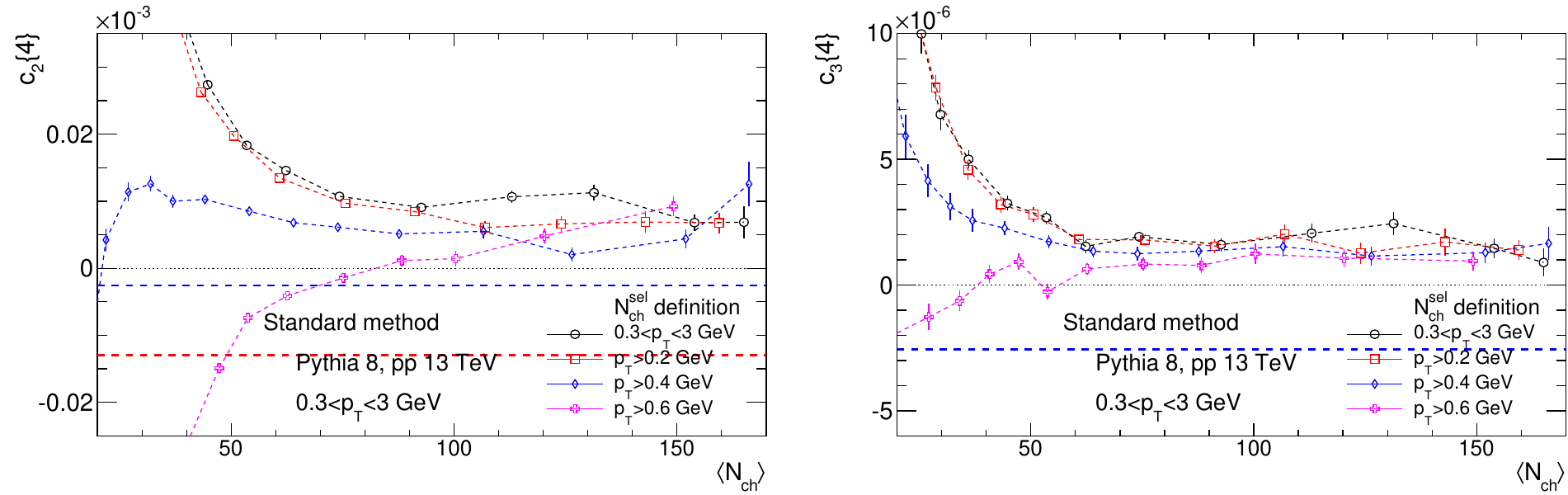}
\end{center}
\caption{\label{fig:1} The $c_2\{4\}$ (left panel) and $c_3\{4\}$ (right panel) calculated for particles in $0.3<\pT<3$ GeV with the standard cumulant method. The event averaging is performed for $\nchb$ calculated for various $\pT$ selections as indicated in the figure, which is then mapped to $\lr{\nch}$, the average number of charged particles with $\pT>0.4$ GeV.}
\end{figure}
Figure~\ref{fig:1} shows the four-particle cumulants obtained with the standard method using charged particles in \mbox{$0.3<\pT<3$~GeV}. The $\nchb$ for calculating $\llrr{2k}$ is defined in $0.3<\pT<3$ GeV, $\pT>0.2$ GeV, $\pT>0.4$ GeV, or $\pT>0.6$ GeV, which are then mapped to $\lr{\nch}$. The $c_2\{4\}$ values are found to differ significantly between the four choices of $\nchb$, therefore confirming the observation made by the ATLAS Collaboration~\cite{ATLAS:2016ics}. The red and blue dashed lines indicate expected $c_n\{4\}$ values corresponding to 4\% and 6\% flow signal. Clearly non-flow contributions in the standard method are too large for a meaningful measurement of $v_2$. In fact, when $\nchb$ is defined in the $\pT$ range of $\pT>0.6$ GeV, the sign of $c_2\{4\}$ is negative at small $\lr{\nch}$ values. The non-flow contribution to $v_3$ is significantly smaller as dijets are not expected to contribute significantly to $v_3$; a $v_3\{4\}$ signal larger than 4\% is measurable in the standard method.

\begin{figure}[h!]
\begin{center}
\includegraphics[width=1\linewidth]{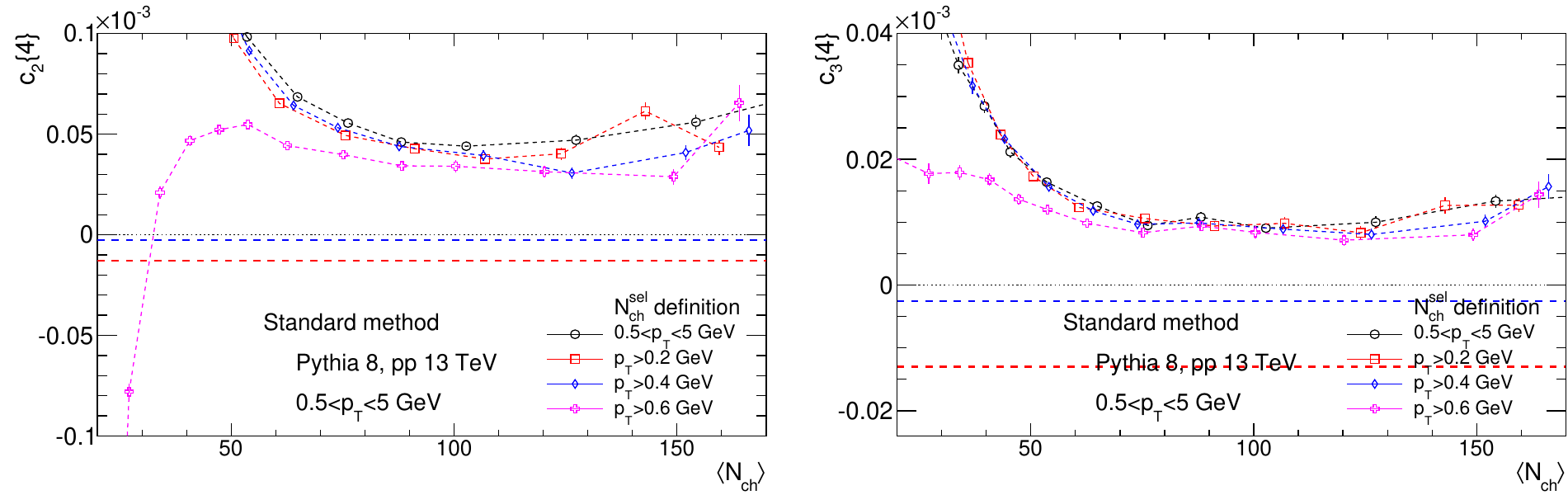}
\end{center}
\caption{\label{fig:2} The $c_2\{4\}$ (left panel) and $c_3\{4\}$ (right panel) calculated for particles in $0.5<\pT<5$ GeV with the standard cumulant method. The event averaging is performed for $\nchb$ calculated for various $\pT$ selections as indicated in the figure, which is then mapped to $\lr{\nch}$, the average number of charged particles with $\pT>0.4$ GeV.}
\end{figure}

Figure~\ref{fig:2} shows the $c_n\{4\}$ obtained with the standard method using charged particles in a higher transverse momentum range $0.5<\pT<5$ GeV. The $c_2\{4\}$ is much larger in all cases, which is consistent with a larger non-flow contribution expected from dijets. Figures~\ref{fig:3} and \ref{fig:4} show $c_n\{4\}$ obtained from the two-subevent and three-subevent methods (the first type), respectively. The two-subevent method greatly reduces the values of $c_n\{4\}$ compared to the standard method, and the dependence on the choice of $\nchb$ is also much smaller. This behavior is expected since the contributions from single jets or resonance decays are suppressed. Nevertheless, the magnitude of the residual non-flow signal is still comparable to a 4\%--6\% flow signal. In contrast, the three-subevent method almost completely suppresses the non-flow. For both $c_2\{4\}$ and $c_3\{4\}$, the magnitudes of the residual non-flow are less than 4\% of flow signal. This means that the three-subevent method should be sensitive to genuine long-range flow signal as small as 4\%.  
\begin{figure}[h!]
\begin{center}
\includegraphics[width=0.9\linewidth]{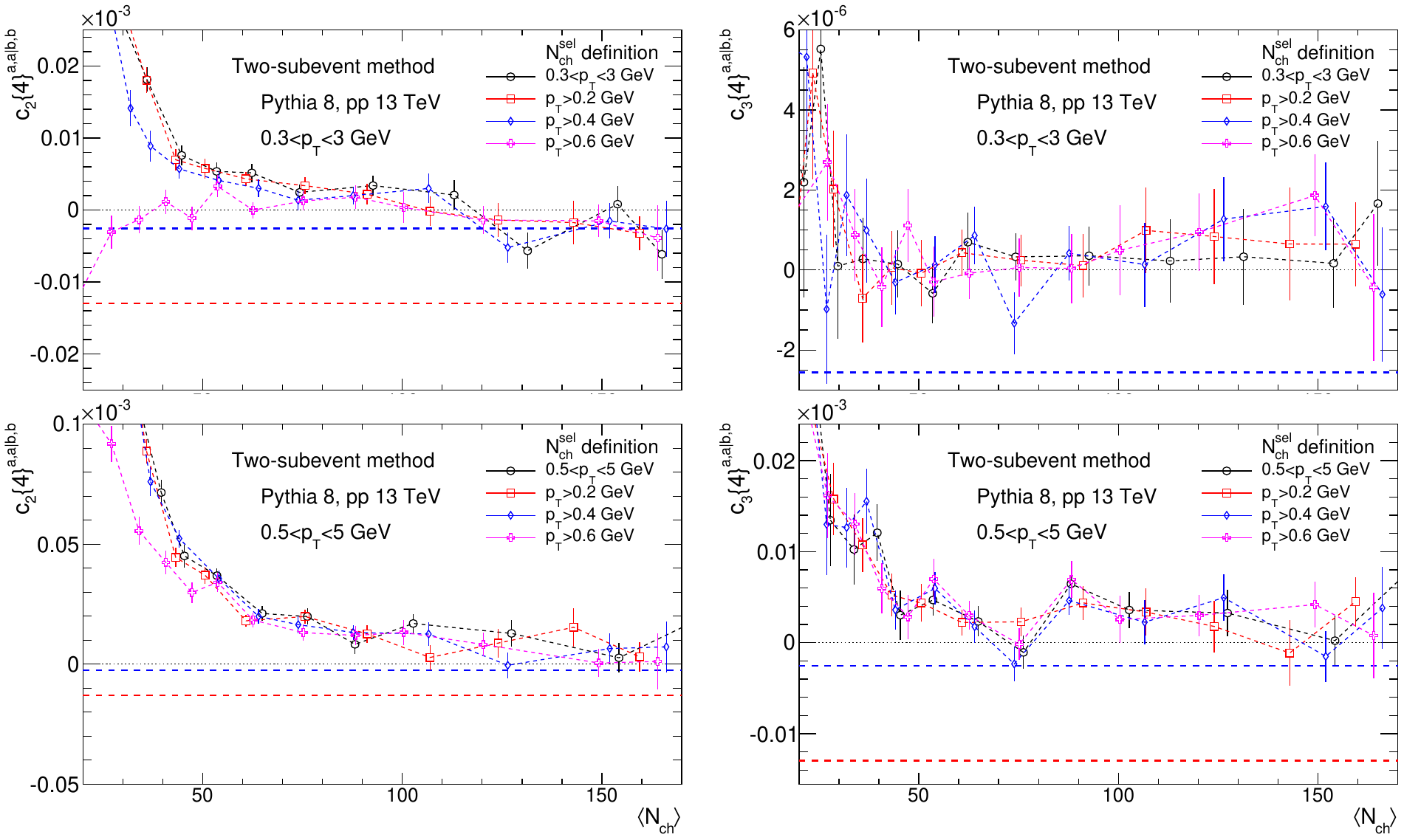}
\end{center}
\caption{\label{fig:3} The $c_2\{4\}$ (left panels) and $c_3\{4\}$ (right panels) calculated for particles in $0.3<\pT<3$ GeV (top panels) or $0.5<\pT<5$ GeV (bottom panels) with the two-subevent cumulant method. The event averaging is performed for $\nchb$ calculated for various $\pT$ selections as indicated in the figure, which is then mapped to $\lr{\nch}$, the average number of charged particles with $\pT>0.4$ GeV.}
\end{figure}

\begin{figure}[h!]
\begin{center}
\includegraphics[width=0.9\linewidth]{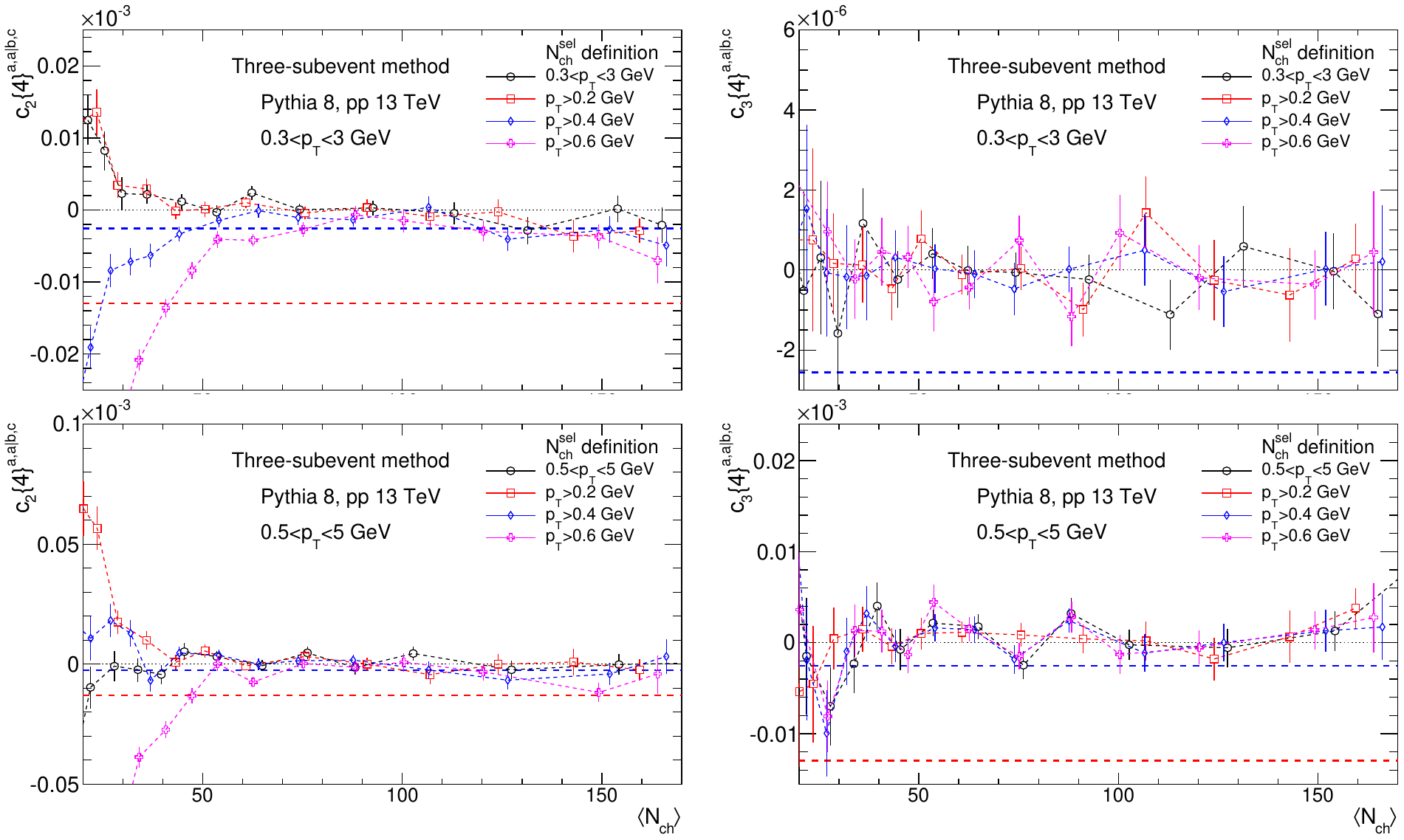}
\end{center}
\caption{\label{fig:4} The $c_2\{4\}$ (left panels) and $c_3\{4\}$ (right panels) calculated for particles in $0.3<\pT<3$ GeV (top panels) or $0.5<\pT<5$ GeV (bottom panels) with the three-subevent cumulant method. The event averaging is performed for $\nchb$ calculated for various $\pT$ selections as indicated in the figure, which is then mapped to $\lr{\nch}$, the average number of charged particles with $\pT>0.4$ GeV.}
\end{figure}

Figure~\ref{fig:5} shows a direct comparison between the standard method and the two- and three-subevent methods for $v_2\{4\}$ in two $\pT$ ranges. The three-subevent method has the best performance in suppressing the non-flow effects.

To quantify the performance of the three methods for recovering the underlying flow signal, a flow afterburner~\cite{Masera} is used to add a constant $v_2$ or $v_3$ signal to the generated PYTHIA events. Figure~\ref{fig:6} shows the calculated $c_2\{4\}$ with 4\% or 6\% $v_2$ imposed on the generated events. In the case 4\% input flow, only the three-subevent method can recover the input. In the case of 6\% input flow, the two-subevent method can also recover the input flow for $\lr{\nch}>80$.
\begin{figure}[!h]
\begin{center}
\includegraphics[width=1\linewidth]{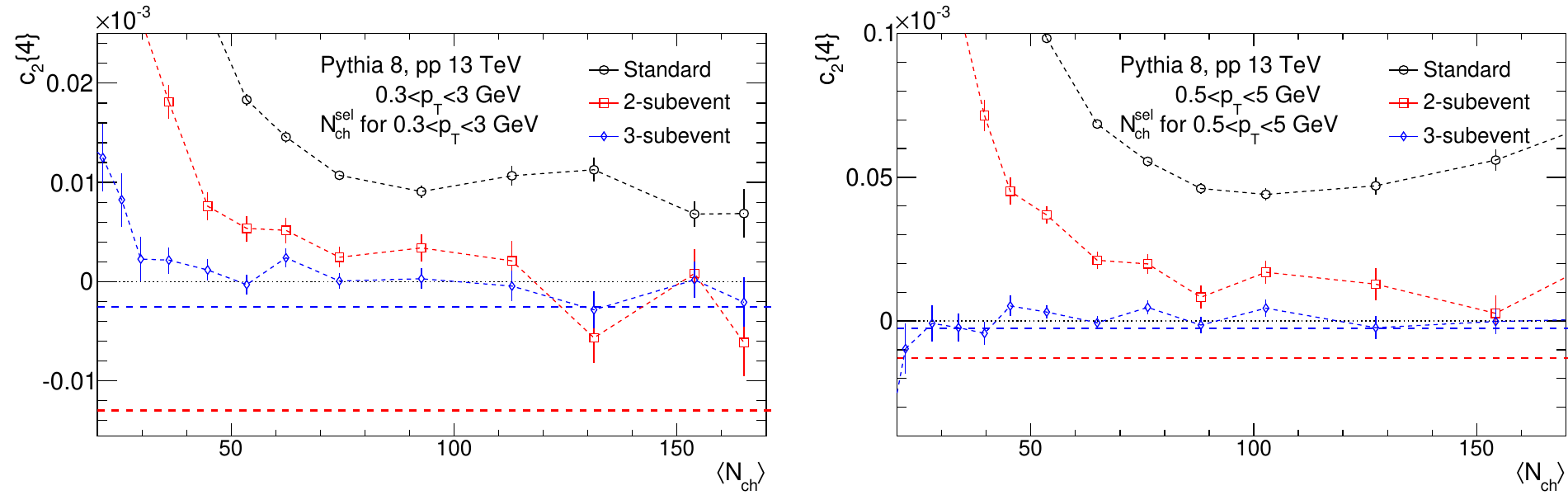}
\end{center}
\caption{\label{fig:5} The $c_2\{4\}$ calculated for particles in $0.3<\pT<3$ GeV (left panel) or $0.5<\pT<5$ GeV (right panel) compared between the three cumulant methods. The event averaging is performed for $\nchb$ calculated for same $\pT$ range, which is then mapped to $\lr{\nch}$, the average number of charged particles with $\pT>0.4$ GeV.}
\end{figure}
\begin{figure}[!h]
\begin{center}
\includegraphics[width=1\linewidth]{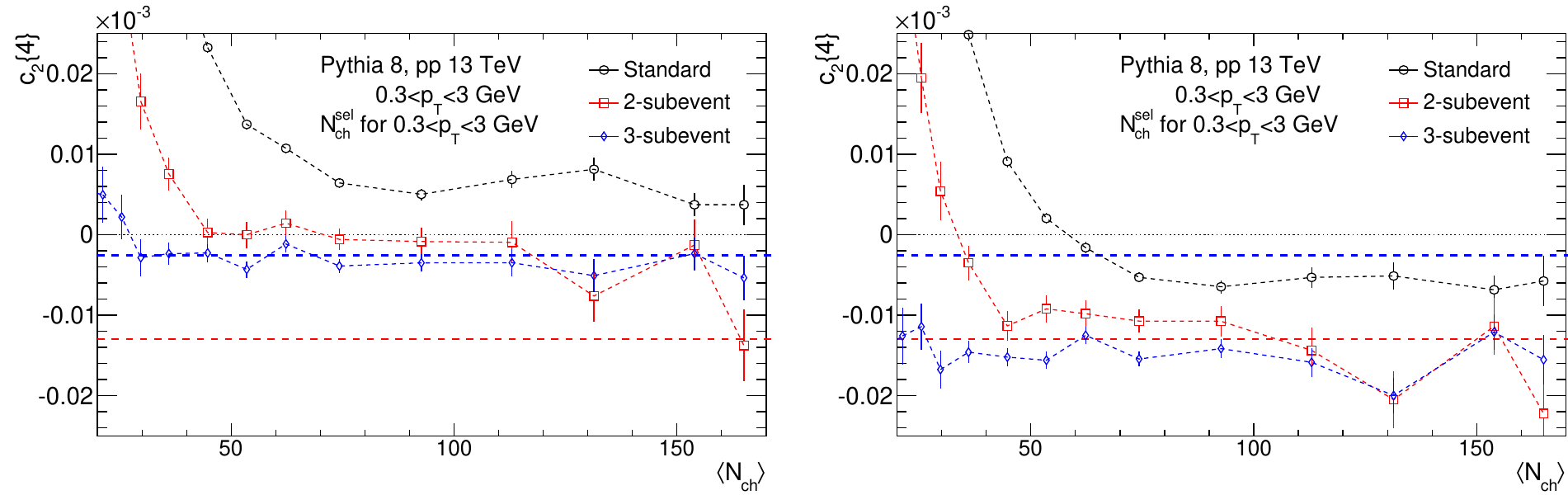}
\end{center}
\caption{\label{fig:6} The $c_2\{4\}$ calculated for particles in $0.3<\pT<3$ GeV compared among the three cumulant methods with 4\% (left panel) or 6\% (right panel) $v_2$ imposed. The event averaging is based on $\nchb$ calculated for the same $\pT$ range, which is then mapped to $\lr{\nch}$, the average number of charged particles with $\pT>0.4$ GeV.}
\end{figure}

In the cumulant analysis, it has been argued that $c_n\{2k\}$ should be calculated for each fixed $\nchb$ bin to minimize the event-by-event variation of particle multiplicity, then should be averaged over broader $\nchb$ intervals~\cite{Bilandzic:2010jr,Chatrchyan:2013nka}. Figure~\ref{fig:6b} shows $c_2\{4\}$ for charged particles in $0.3<\pT<3$ GeV for events with $40\leq\nch<80$, where $\nch$ is the number of charged particles also in $0.3<\pT<3$ GeV. The $c_2\{4\}$ values are obtained with $\nch$ bin widths of 1, 5, 10, 20 and 40, respectively. The $c_2\{4\}$ values for each case are then averaged over the $40\leq\nch<80$ interval to give a single $c_2\{4\}$ value. The difference of $c_2\{4\}$ for a given bin-width from $c_2\{4\}$ for unit-bin-width is then plotted. Clearly, the standard method is much more sensitive to the bin-width than the two-subevent and three-subevent methods. This sensitivity is due to the fact that the standard method still has significant non-flow, whose event-by-event fluctuation also has significant non-Gaussianity. Therefore the residual non-flow has significant dependence on the bin-width in the standard method~\footnote{Even if $c_2\{4\}$ is calculated in unit $\nch$ bin, there is still significant residual non-flow (as shown by Figs~\ref{fig:2} and \ref{fig:3}).}. On the other hand, since non-flow contributions are significantly further suppressed in the subevent methods, the nature of the non-flow fluctuations no longer matter much for the subevent methods.

\begin{figure}[!h]
\begin{center}
\includegraphics[width=0.6\linewidth]{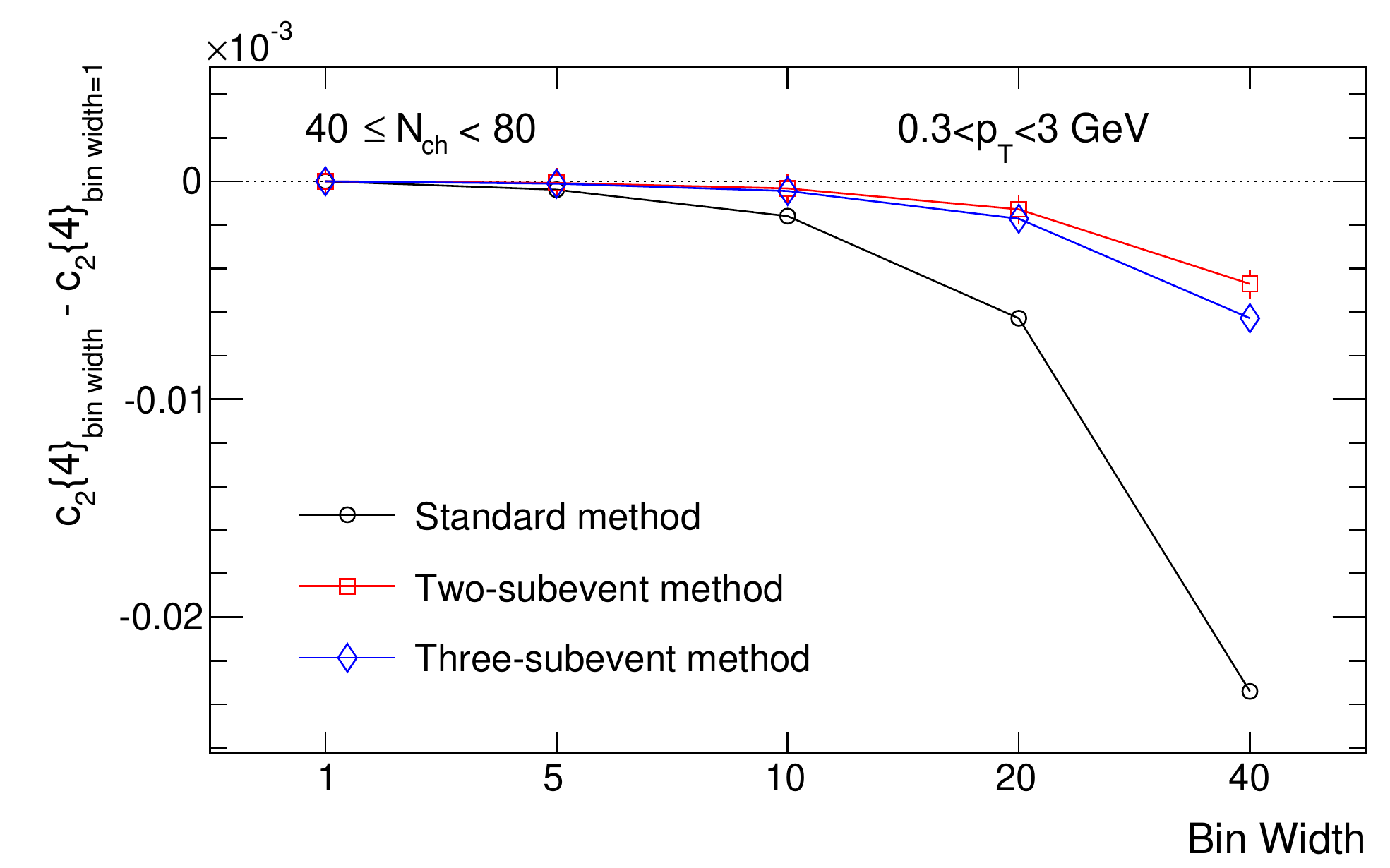}
\end{center}
\caption{\label{fig:6b} The $c_2\{4\}$ calculated in different bin-widths of $\nch(0.3<\pT<3~\mbox{GeV})$ and then averaged over the $40\leq\nch(0.3<\pT<3~\mbox{GeV})<80$ interval. The difference from the unit-bin-width case is plotted as a function of bin-width.}
\end{figure}

Finally, we also studied the $c_n\{4\}$ using a four subevent method, where $|\eta|<2.5$ range is divided into four equal $\eta$ ranges $a,b,c,d$ and the cumulant is calculated as $c_n^{a,b|c,d}\{4\} \equiv \llrr{4}_{a,b|c,d}- 2\llrr{2}_{a|c}\llrr{2}_{b|d}$. Despite the significantly larger statistical uncertainties, the values of $c_n\{4\}$ converge to that from the three-subevent method~\footnote{Another test is also performed, where we increase the $\eta$ range to $|\eta|<5$ to include more particles, in that case, the four-subevent method agrees with the three-subevent method perfectly.}.

\section{Discussion and outlook}
\label{sec:8}
Ever since its discovery in small collision systems, the ridge phenomenon has been an area of insensitive study in the high-energy and heavy-ion physics community. The current debate is centered on the underlying multi-particle dynamics and collective nature of the long-range correlations, in particular whether the ridge is a global property of the event, and what is its connection to the collective flow in large collision systems~\cite{Dusling:2015gta,Schlichting:2016sqo}. The main challenge is to disentangle correlations involving all particles (flow or collectivity) from those involving a subset of the particles in restricted $\eta$ and $\phi$ space (non-flow). The two-particle correlation method based on peripheral subtraction or the template fit method~\cite{Aad:2014lta,Aad:2015gqa,Aaboud:2016yar} can successfully remove the non-flow, but it does not address the multi-particle nature of the ridge.  Results from multi-particle azimuthal cumulants, namely negative $c_n\{4\}$ and $c_n\{8\}$, positive $c_n\{6\}$, and the relation $v_n\{4\}\approx v_n\{6\}\approx v_n\{8\}$, have been used as a ``{\it definition}'' of collectivity~\cite{Khachatryan:2015waa,Khachatryan:2016txc}. However this is true only if the flow fluctuations are relatively narrow or close to Gaussian and non-flow contributions to $c_2\{2k\}$ are small. A recent study from ATLAS~\cite{ATLAS:2016ics} shows that in $pp$ collisions, the standard cumulant method is contaminated by non-flow correlations over the entire measured $\nch$ range, which further complicates the relationship between different $v_n\{2k\}$s.

In this context, our paper addresses the issue of collectivity in small collision systems from two fronts. First, we clarify the statistical nature of multi-particle cumulants, and show that azimuthal cumulants are affected not only by the event-averaged flow and non-flow, but more importantly by the event-by-event fluctuations of flow and/or non-flow. We then propose alternative cumulant methods based on two or three subevents in non-overlapping $\eta$ ranges, and demonstrate that these methods can further suppress non-flow correlations. 

When cumulant methods were first proposed in 2000~\cite{Borghini:2000sa}, the importance of non-gaussian flow fluctuations was not recognized. If flow fluctuations or flow distribution $p(v_n)$ are close to Gaussian, all higher-order cumulants should give the same answer $v_n\{4\}=v_n\{6\}...$, and the difference between $v_n\{2\}$ and $v_n\{4\}$ reflects the width of $p(v_n)$~\cite{Voloshin:2007pc}. However when $p(v_n)$ strongly deviates from Gaussian, $v_n\{2k\}$ (and also the Lee-Yang zero method~\cite{Bhalerao:2003xf}) may be mathematically undefined and one should instead rely on the original cumulant quantity $c_n\{2k\}$~\cite{Jia:2014pza}, which is always well defined. For example consider
\begin{eqnarray}\\\nonumber
v_n\{4\} =\sqrt[4]{-c_n\{4\}} = \sqrt[4]{2\lr{v_n^2}^2-\lr{v_n^4}} 
\end{eqnarray}
which is mathematically undefined only if $2\lr{v_n^2}^2<\lr{v_n^4}$. If this relation is violated due to the underlying $p(v_n)$, the perceived hallmark relation for collectivity, $v_n\{4\}\approx v_n\{6\}\approx v_n\{8\}$, is no longer meaningful.

The contribution of non-flow to cumulants can be discussed following a similar probabilistic approach. The contributions of flow and non-flow to cumulants are additive:
\begin{eqnarray}\\\nonumber
c_n\{2k\}=c_n\{2k,\mbox{flow}\}+c_n\{2k,\mbox{non-flow}\}.
\end{eqnarray}
The sign of $c_n\{2k,\mbox{non-flow}\}$ can be either positive or negative depending on the shape of event-by-event fluctuations of non-flow $p(s_n)$. The results from ATLAS and our follow-up study from PYTHIA8 suggest that the value and sign of $c_n\{4\}$ in $pp$ collisions are driven mainly by the non-flow component in the standard cumulant method. The values of $c_n\{2k,\mbox{non-flow}\}$ are found to be very sensitive to the particle selection criteria $\pT> \pT^{\rm{min}}$ used to define event for averaging: $c_n\{4\}$ is positive when $\pT^{\rm{min}}=0.2$ and 0.4 GeV, but is negative for $\pT^{\rm{min}}=0.6$ GeV in low multiplicity events. This suggests that the $p(s_n)$ in PYTHIA8 is very sensitive to $\pT$ of the particles, as expected from jets and dijets.  It is also interesting to point out that by choosing the $\pT$ range used to define $\nchb$, one may be able to ``dial'' the $c_n\{2k,\mbox{non-flow}\}$ to be very close to zero, and thereby expose the genuine flow signal. Unfortunately one can not do this in a model independent way, therefore the observed negative $c_2\{4\}$ by CMS Collaboration~\cite{Khachatryan:2016txc} can be interpreted as a negative $c_n\{2k,\mbox{flow}\}$ from collectivity, if and only if the non-flow contribution $c_n\{2k,\mbox{non-flow}\}$ can be shown to be small~\footnote{Indeed, the $c_n\{2k,\mbox{non-flow}\}$ from PYTHIA8 model is large and can be negative depending on the choice of $\pT$ range. On the other hand, no model that include only non-flow can describe the observation in the $pp$ data.}. 

Motivated by the discussions above, we have developed a cumulant method based on two- or three $\eta$-separated subevents. The two-subevent method suppresses the non-flow from single jet fragmentation, while the three-subevent method also suppresses correlations between the two jets in a dijet.  The performance of the two methods is quantified using PYTHIA8 simulation with only non-flow, as well as with physical flow added with an afterburner. The three-subevent method is shown to give consistent results between different criteria used to define event classes for averaging, while the two-subevent method still has some sensitivity. The three-subevent method is able to recover flow signal as small as 4\% for $v_2$ and $v_3$ for events with $\lr{\nch}$ as low as 30. The subevent method also offers a more robust definition of collectivity in $pp$ or $p$+A collisions, i.e. the existence of long-range ridge signal that correlates between three or more distinct $\eta$ ranges. The method also provides a strong test of the gluon saturation models used to describe $c_2\{4\}$ obtained with the standard cumulant method, where it was argued that the sign change of $c_2\{4\}$ could be due to non-Gaussian correlations of the domains of strong QCD fields in the initial state~\cite{Dumitru:2014yza,Dumitru:2014yza,Lappi:2015vta}. It would be interesting to see wether or not such color domains can simultaneously correlate three or more pseudorapidity ranges and contribute to subevent cumulants. 

Looking into the future, there are several improvements or applications of the subevent cumulant method in both small as well as large collision systems:
\begin{itemize}
\item Rapidity gap between subevents: this can further suppress short-range sources falling on the boundary between two subevents. Furthermore, by requiring the same rapidity gap as in 2PC analysis, $v_n\{4\}$ can be compared directly with $v_n\{2\}$ obtained via a peripheral subtraction procedure.
\item Four or higher number of subevents: this can further reduce non-flow and prove that the azimuthal collectivity indeed exists between particles in four or more distinct $\eta$ ranges. But this also reduce significantly the statistics.
\item Subevents selected in more exclusive way: for example one could reject events containing jets or dijets. By doing so, the large tail in the non-flow distribution $p(s_n)$ should be suppressed and $c_n\{2k,\mbox{non-flow}\}$ is reduced. One could also have one subevent containing all the reconstructed jets (several $\eta$ slices each one containing one or more jets), and the rest of the $\eta$ range is defined as the second subevent. This way the non-flow contributions from jets and dijets are contained in one subevent.
\item Differential flow: the procedure for calculating $v_n\{4\}(p_{T},\eta)$ is straightforward in the three-subevent method. One just needs to restrict the particle in one subevent to a certain $\pT$ or $\eta$ range.
\item Flow longitudinal dynamics: for example, ignoring the non-flow terms, subevent cumulants such as Eqs.~\ref{eq:d4}, \ref{eq:d6}, \ref{eq:d8} or \ref{eq:d9}, are sensitive to decorrelation effects between different $\eta$ ranges: 
\begin{eqnarray}
\label{eq:e1}
\lr{{\bm v}_{n,a}{\bm v}_{n,b}^*} &\approx& \llrr{2}_{a|b}+\mbox{peripheral~subtraction\; or\; template\; fit}\\\label{eq:e2}
\lr{v^2_{n,a}}\lr{v^2_{n,b}} &\approx&  c_n^{a,a|b,b}\{4\}-c_n^{a,b|a,b}\{4\}+\lr{{\bm v}_{n,a}{\bm v}_{n,b}^*}^2\\\label{eq:e3}
\lr{{\bm v}^2_{n,a}{\bm v}^{*2}_{n,b}} &\approx&  2c_n^{a,b|a,b}\{4\}-c_n^{a,a|b,b}\{4\}+2\lr{v_{n,a}^2}\lr{v_{n,b}^2}
\end{eqnarray}
First, $\lr{{\bm v}_{n,a}{\bm v}_{n,b}^*}=\lr{v_{n,a}v_{n,b}\cos n(\Phi_n^a-\Phi_n^b)}$ can be calculated from two-particle correlations (see Eq.~\ref{eq:d0}), supplemented with peripheral subtraction and template fit~\cite{Aad:2014lta,Aad:2015gqa,Aaboud:2016yar} to reject the non-flow contributions. $\lr{v^2_{n,a}}\lr{v^2_{n,b}}$ and $\lr{{\bm v}^2_{n,a}{\bm v}^{*2}_{n,b}}$ are then obtained from Eqs.~\ref{eq:e2} and \ref{eq:e3}. The study of flow fluctuations based on moments were discussed before in Ref.~\cite{Bhalerao:2014xra}, but subevent cumulants allow us to directly measure these moments including longitudinal flow decorrelation effects, without relying on a reference detector as was usually done in experiments~\cite{Khachatryan:2015oea}.

\item Generalization of subevent cumulants to symmetric cumulants or event-plane correlations~\cite{Bilandzic:2013kga}: Some examples are discussed in Appendixes E and F; they provide access to the longitudinal dynamics of the initial state (e.g. correlation between $v_2$ and $v_3$ ) and final state mode-mixing (e.g. correlation between $v_2$ and $v_4$). 
\end{itemize}
We thank A. Bilandzic for clarification on the framework of direct cumulants. We acknowledge Sooraj Radhakrishnan for help in the early preparation of this paper. We appreciate valuable comments and fruitful discussions with R. Lacey. This research is supported by the NSF Grant No. PHY-1613294.

\section*{Appendix}
\label{sec:a1}
\subsection{Performance of 2nd type of two- and three-subevent cumulants}
We also calculate cumulants based on the second type of two-subevent and three-subevent methods mentioned in Section~\ref{sec:5}. The results, as shown in Figs.~\ref{fig:a1} and \ref{fig:a2}, are found to be very sensitive to the particle $\pT$ selection criteria used to define events for averaging, suggesting that these correlators have more contributions from non-flow. The larger non-flow in these correlators can be traced to the fact that the two particles from the same subevent are conjugated with each other as shown in Eqs.~\ref{eq:d6} and \ref{eq:d11} (terms containing $\llrr{2}_{a|a}$ or $\llrr{2}_{b|b}$).

\begin{figure}[h!]
\begin{center}
\includegraphics[width=1\linewidth]{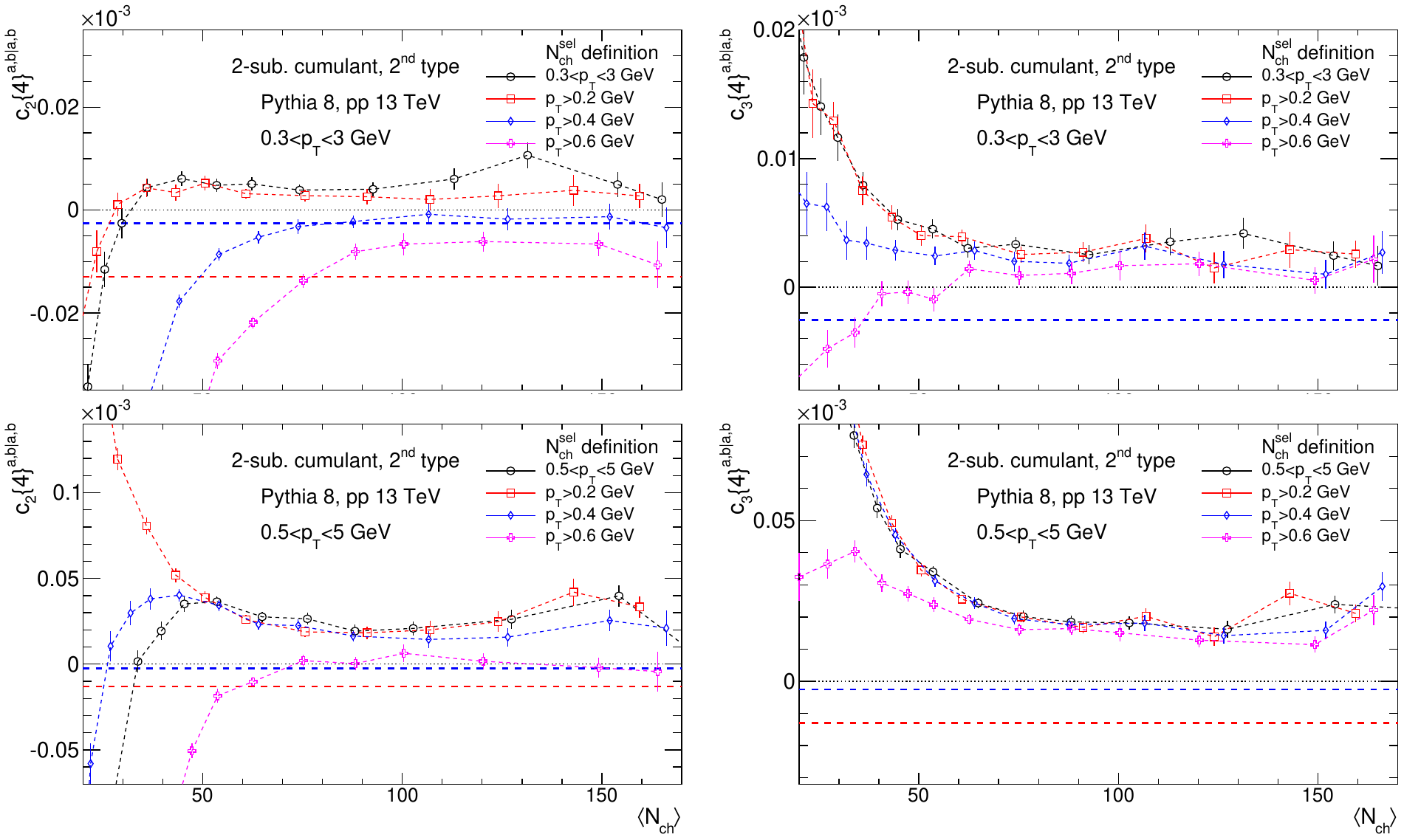}
\end{center}
\caption{\label{fig:a1} The $c_2\{4\}$ (left panels) and $c_3\{4\}$ (right panels) calculated for particles in $0.3<\pT<3$ GeV (top panels) or $0.5<\pT<5$ GeV (bottom panels) with the second type of two-subevent cumulant method (Eq.~\ref{eq:d6}). The event averaging is performed for $\nchb$ calculated for various $\pT$ selections as indicated in the figure, which is then mapped to $\lr{\nch}$, the average number of charged particles with $\pT>0.4$ GeV.}
\end{figure}
\begin{figure}[h!]
\begin{center}
\includegraphics[width=1\linewidth]{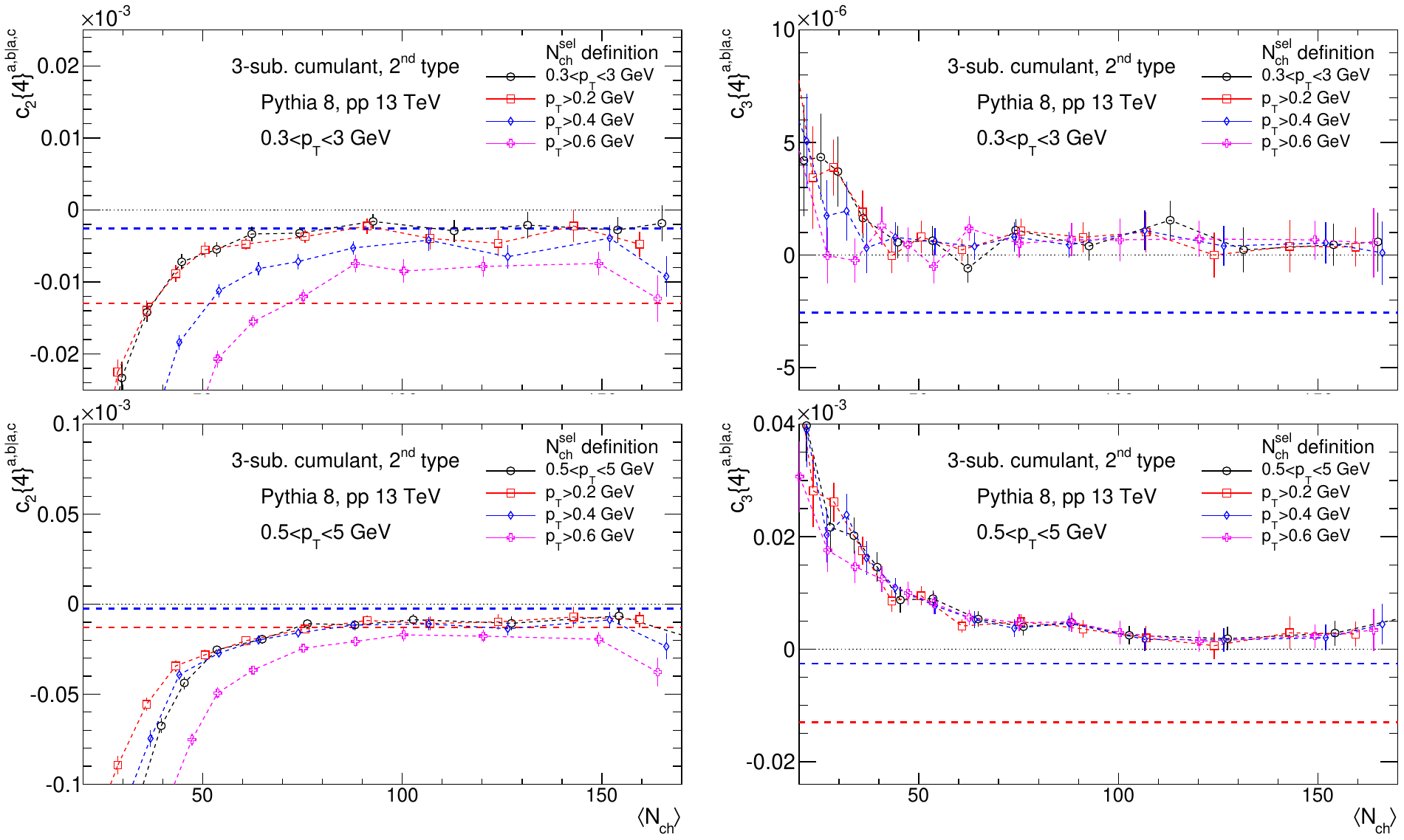}
\end{center}
\caption{\label{fig:a2} The $c_2\{4\}$ (left panels) and $c_3\{4\}$ (right panels) calculated for particles in $0.3<\pT<3$ GeV (top panels) or $0.5<\pT<5$ GeV (bottom panels) with the second type of three-subevent cumulant method (Eq.~\ref{eq:d11}). The event averaging is performed for $\nchb$ calculated for various $\pT$ selections as indicated in the figure, which is then mapped to $\lr{\nch}$, the average number of charged particles with $\pT>0.4$ GeV.}
\end{figure}

\subsection{Six- and eight-particle cumulants based on two subevents}
The cumulants based on two subevents can be generated for more than four particles:
\begin{eqnarray}
\nonumber
\lr{6}_{3a|3b} &\equiv& \lr{e^{in\sum_{i=1}^{3}(\phi_{2i-1}^a-\phi_{2i}^b)}} = \frac{({\bm q}_{n,a}^3-3\tau_a{\bm q}_{2n,a}{\bm q}_{n,a}+2\tau_a^2{\bm q}_{3n,a})({\bm q}_{n,b}^3-3\tau_a{\bm q}_{2n,b}{\bm q}_{n,b}+2\tau_b^2{\bm q}_{3n,b})^*}{(1-\tau_a)(1-2\tau_a)(1-\tau_b)(1-2\tau_b)}\\
c_{n}^{3a|3b}\{6\} &\equiv & \llrr{6}_{3a|3b}-9\llrr{4}_{2a|2b}\llrr{2}_{a|b}+12\llrr{2}_{a|b}^3
\end{eqnarray}
The non-flow contribution to $\llrr{6}_{3a|3b}$ is straightforward to derive, but the expression is quite lengthy, so we will not show it here. But one can show that only flow and intrinsic six-particle non-flow contribute to this definition of cumulants:
\begin{eqnarray}
c_{n}^{3a|3b}\{6\} = \lr{e^{in\sum_{i=1}^{3}(\phi_{2i-1}^a-\phi_{2i}^b)}}_c+ \lr{{\bm v}^3_{n,a}{\bm v}^{*3}_{n,b}}-9\lr{{\bm v}^2_{n,a}{\bm v}^{*2}_{n,b}}\lr{{\bm v}_{n,a}{\bm v}^*_{n,b}}+12\lr{{\bm v}_{n,a}{\bm v}^*_{n,b}}^3
\end{eqnarray}

Similarly the eight-particle cumulant can be derived as,
\small{\begin{eqnarray}
\nonumber
\lr{8}_{4a|4b} &\equiv& \frac{({\bm q}_{n,a}^4-6\tau_a{\bm q}_{2n,a}{\bm q}_{n,a}^2+3\tau_a^2{\bm q}_{2n,a}^2+8\tau_a^2{\bm q}_{3n,a}{\bm q}_{n,a}-6\tau_a^3{\bm q}_{4n,a})({\bm q}_{n,b}^4-6\tau_b{\bm q}_{2n,b}{\bm q}_{n,b}^2+3\tau_b^2{\bm q}_{2n,b}^2+8\tau_b^2{\bm q}_{3n,b}{\bm q}_{n,b}-6\tau_b^3{\bm q}_{4n,b})^*}{(1-\tau_a)(1-2\tau_a)(1-3\tau_a)(1-\tau_b)(1-2\tau_b)(1-3\tau_b)}\\\label{eq:ap1}
c_{n}^{4a|4b}\{8\} &\equiv& \llrr{8}_{4a|4b}-16\llrr{6}_{3a|3b}\llrr{2}_{a|b}-18\llrr{4}_{2a|2b}^2+144\llrr{4}_{2a|2b}\llrr{2}_{a|b}^2 -144\llrr{2}_{a|b}^4
\end{eqnarray}}\normalsize
The procedures for deriving $\lr{6}_{3a|3b}$ and $\lr{8}_{4a|4b}$ are provided in Appendix H. The correctness of these formulas has been cross-checked by comparing the results with exact calculation using nested-loops.

By combining information from subevent cumulants of different orders (Eqs.\ref{eq:e1}--\ref{eq:e3}), one can systematically calculate higher-order flow moments including the longitudinal dynamics, such as $\lr{{\bm v}^3_{n,a}{\bm v}^{*3}_{n,b}}$ and $\lr{{\bm v}^4_{n,a}{\bm v}^{*4}_{n,b}}$,  with minimum contribution from non-flow effects.

\subsection{Six-particle cumulants based on three subevents}
We shall consider only six-particle correlations where two particles are chosen from each subevent, which leads to two types of correlators ``$(bb^*,aa^*,cc^*)$'' and ``$(bb,aa^*,c^*c^*)$''. In the first type, the two particles in each subevent are conjugated with each other. In the second type, particles in subevent $b$ are conjugated with particles in subevent $c$, while the two particles in subevent $a$ are conjugated with each other. These two types can also be denoted as ``$a,b,c|a,b,c$'' and ``$a,b,b|a,c,c$'', respectively.

The expression for cumulants can be derived using the diagrammatic approach in Ref.~\cite{Borghini:2000sa}, except that we also need to keep track of the subevent indexes explicitly. The results for six-particle cumulants are:
\begin{eqnarray}
\nonumber
c_{n}^{a,b,c|a,b,c}\{6\} &=& \llrr{6}_{a,b,c|a,b,c}-\left(\llrr{4}_{a,b|a,b}\llrr{2}_{c|c}+\llrr{4}_{b,c|b,c}\llrr{2}_{a|a}+\llrr{4}_{a,b|a,b}\llrr{2}_{b|b}\right.\\\nonumber
&&\left.+2\llrr{4}_{a,b|b,c}\llrr{2}_{c|a}+2\llrr{4}_{a,b|a,c}\llrr{2}_{c|b}+2\llrr{4}_{a,c|b,c}\llrr{2}_{b|a}\right)+2\left(\llrr{2}_{a|a}\llrr{2}_{b|b}\llrr{2}_{c|c}\right.\\\nonumber
&&\left.+\llrr{2}_{a|b}\llrr{2}_{b|c}\llrr{2}_{c|a}+\llrr{2}_{a|c}\llrr{2}_{b|a}\llrr{2}_{c|b}+\llrr{2}_{a|a}\llrr{2}_{b|c}^2+\llrr{2}_{b|b}\llrr{2}_{a|c}^2+\llrr{2}_{c|c}\llrr{2}_{a|b}^2\right)
\end{eqnarray}
\begin{eqnarray}
\nonumber
c_{n}^{a,b,b|a,c,c}\{6\} &=& \llrr{6}_{a,b,b|a,c,c}-\left(\llrr{4}_{b,b|c,c}\llrr{2}_{a|a}+4\llrr{4}_{a,b|a,c}\llrr{2}_{b|c}+2\llrr{4}_{a,b|c,c}\llrr{2}_{b|a}+2\llrr{4}_{b,b|a,c}\llrr{2}_{a|c}\right)\\
&&+4\llrr{2}_{a|a}\llrr{2}_{b|c}^2+8\llrr{2}_{a|c}\llrr{2}_{b|a}\llrr{2}_{b|c}\;.
\end{eqnarray}
The $c_{n}^{a,b,b|a,c,c}\{6\}$ has a smaller contribution from non-flow.

\subsection{Cumulants based on four subevents}
Four-particle correlations, where one particle is taken from each subevent can be written as~\cite{Adler:2002pu}:
\begin{eqnarray}
\nonumber
\lr{4}_{a,b|c,d}&\equiv&\lr{e^{in(\phi_{1}^a+\phi_{2}^b -\phi_{3}^c-\phi_{4}^d)}} = \frac{{\bm Q}_{n,a}{\bm Q}_{n,b}{\bm Q}_{n,c}^*{\bm Q}_{n,d}^*}{M_aM_bM_cM_d} = {\bm q}_{n,a}{\bm q}_{n,b}{\bm q}_{n,c}^*{\bm q}_{n,d}^*\\
c_n^{a,b|c,d}\{4\} &\equiv& \llrr{4}_{a,b|c,d}- \llrr{2}_{a|c}\llrr{2}_{b|d}-\llrr{2}_{a|d}\llrr{2}_{b|c}\;.
\end{eqnarray}

One interesting six-particle correlation based on four subevents is $(cc,a,b*,d^*d^*)$ or ``$a,c,c|b,d,d$'':
\begin{eqnarray}
\nonumber
\lr{6}_{a,2c|b,2d}&\equiv&\lr{e^{in(\phi_{1}^a+\phi_{2}^c+\phi_{3}^c -\phi_{4}^b-\phi_{5}^d-\phi_{6}^d)}} = \frac{{\bm q}_{n,a}{\bm q}_{n,b}^*({\bm q}_{n,c}^2-\tau_c{\bm q}_{2n,c})({\bm q}_{n,d}^{2}-\tau_d{\bm q}_{2n,d})^*}{(1-\tau_c)(1-\tau_d)}\\\nonumber
c_n^{a,2c|c,2d}\{6\} &\equiv& \llrr{6}_{a,2c|b,2d}- \left(4\llrr{4}_{ac|bd}\llrr{2}_{c|d}+2\llrr{4}_{ac|dd}\llrr{2}_{c|b}+2\llrr{4}_{cc|bd}\llrr{2}_{a|d}+\llrr{4}_{cc|dd}\llrr{2}_{a|b}\right)\\
&&+4\llrr{2}_{a|b}\llrr{2}_{c|d}^2+8\llrr{2}_{a|d}\llrr{2}_{c|b}\llrr{2}_{c|d}
\end{eqnarray}
To minimize the residual non-flow, one can put subevents $a$ and $b$ in between subevents $c$ and $d$, and also require a small rapidity gap between $a$ and $b$.

\subsection{Symmetric cumulants}

The subevent cumulants can be generalized to symmetric cumulants first proposed in Ref.~\cite{Bilandzic:2013kga}. There are three variants of two-subevent symmetric cumulants:
\begin{eqnarray}
\nonumber
\lr{4}_{n_a,m_a|n_b,m_b}&\equiv&\lr{e^{i(n\phi_{1}^a+m\phi_{2}^a -n\phi_{3}^b-m\phi_{4}^b)}} = \frac{({\bm q}_{n,a}{\bm q}_{m,a}-\tau_a{\bm q}_{n+m,a})({\bm q}_{n,b}{\bm q}_{m,b}-\tau_b{\bm q}_{n+m,b})^*}{(1-\tau_a)(1-\tau_b)}\\\nonumber
\lr{4}_{n_a,m_b|m_a,n_b}&\equiv&\lr{e^{i(n\phi_{1}^a+m\phi_{2}^b -m\phi_{3}^a-n\phi_{4}^b)}} =  \frac{({\bm q}_{n,a}{\bm q}_{m,a}^*-\tau_a{\bm q}_{n-m,a} )({\bm q}_{m,b}{\bm q}_{n,b}^*-\tau_b{\bm q}^*_{n-m,b})}{(1-\tau_a)(1-\tau_b)}\\\nonumber
\lr{4}_{n_a,m_b|n_a,m_b}&\equiv&\lr{e^{i(n\phi_{1}^a+m\phi_{2}^b -n\phi_{3}^a-m\phi_{4}^b)}} =  \frac{({\bm q}_{n,a}{\bm q}_{n,a}^*-\tau_a)({\bm q}_{m,b}{\bm q}_{m,b}^*-\tau_b)}{(1-\tau_a)(1-\tau_b)}\\\label{eq:ab1}
\rm{SC}_{n_a,m_a|n_b,m_b}\{4\} &\equiv& \llrr{4}_{n_a,m_a|n_b,m_b}- \llrr{2}_{n_a|n_b}\llrr{2}_{m_a|m_b} \approx \lr{{\bm v}_{n,a}{\bm v}^*_{n,b}{\bm v}_{m,a}{\bm v}^*_{m,b}}-\lr{{\bm v}_{n,a}{\bm v}^*_{n,b}}\lr{{\bm v}_{m,a}{\bm v}^*_{m,b}}\\\label{eq:ab2}
\rm{SC}_{n_a,m_b|m_a,n_b}\{4\} &\equiv& \llrr{4}_{n_a,m_b|m_a,n_b}- \llrr{2}_{n_a|n_b}\llrr{2}_{m_a|m_b} \approx \lr{{\bm v}_{n,a}{\bm v}^*_{n,b}{\bm v}^*_{m,a}{\bm v}_{m,b}}-\lr{{\bm v}_{n,a}{\bm v}^*_{n,b}}\lr{{\bm v}_{m,a}{\bm v}^*_{m,b}}\\\label{eq:ab3}
\rm{SC}_{n_a,m_b|n_a,m_b}\{4\} &\equiv& \llrr{4}_{n_a,m_b|n_a,m_b}- \llrr{2}_{n_a|n_a}\llrr{2}_{m_b|m_b} \approx \lr{v_{n,a}^2v_{m,b}^2}-\lr{v_{n,a}^2}\lr{v_{m,b}^2}
\end{eqnarray}
where $n> m$ and the notation combines harmonic number and subevent index. These three types can also be schematically labeled as $(n_am_a,n_b^*m_b^*)$, $(n_am_a*, m_bn_b^*)$, $(n_an_a^*,m_bm_b^*)$, respectively. It is interesting to note that contributions of flow are expected to be nearly identical for Eq.~\ref{eq:ab1} and Eq.~\ref{eq:ab2}, and they are sensitive to both flow magnitude fluctuations and event-plane twist effects. On the other hand, the correlator in Eq.~\ref{eq:ab3} is only sensitive to the fluctuation of the flow magnitudes.

Since terms such as $\lr{{\bm v}_{n,a}{\bm v}^*_{n,b}}$, $\lr{{\bm v}_{m,a}{\bm v}^*_{m,b}}$, $\lr{v_{n,a}^2}$, $\lr{v_{m,b}^2}$ can be calculated from subevent cumulants involving harmonic flow of the same order via Eqs.~\ref{eq:e1}--\ref{eq:e3}, one can also calculate directly $\lr{v_{n,a}^2v_{m,b}^2}$ and $\lr{{\bm v}_{n,a}{\bm v}^*_{n,b}{\bm v}_{m,a}{\bm v}^*_{m,b}}$ by combining cumulants and symmetric cumulants, and therefore gain knowledge about the decorrelations between harmonics of different order. 

Similarly, one can write down the formulas for symmetric cumulants from three subevents. There are four possibilities, $(n_b^*,n_am_a,m_c^*)$, $(n_b^*,n_am_a^*,m_c)$, $(m_b,n_an_a^*,m_c^*)$ and $(n_b,m_am_a^*,n_c^*)$, with the following formula:
\begin{eqnarray}
\nonumber
\lr{4}_{n_a,m_a|n_b,m_c}&\equiv&\lr{e^{i(n\phi_{1}^a+m\phi_{2}^a -n\phi_{3}^b-m\phi_{4}^c)}} = \frac{({\bm q}_{n,a}{\bm q}_{m,a}-\tau_a{\bm q}_{n+m,a}){\bm q}_{n,b}^*{\bm q}_{m,c}^*}{(1-\tau_a)}\\\nonumber
\lr{4}_{n_a,m_c|m_a,n_b}&\equiv&\lr{e^{i(n\phi_{1}^a+m\phi_{2}^c -m\phi_{3}^a-n\phi_{4}^b)}} =  \frac{({\bm q}_{n,a}{\bm q}_{m,a}^*-\tau_a{\bm q}_{n-m,a} ){\bm q}_{m,c}{\bm q}_{n,b}^*}{(1-\tau_a)}\\\nonumber
\lr{4}_{n_a,m_b|n_a,m_c}&\equiv&\lr{e^{i(n\phi_{1}^a+m\phi_{2}^b -n\phi_{3}^a-m\phi_{4}^c)}} =  \frac{({\bm q}_{n,a}{\bm q}_{n,a}^*-\tau_a){\bm q}_{m,b}{\bm q}_{m,c}^*}{(1-\tau_a)}\\\nonumber
\lr{4}_{m_a,n_b|m_a,n_c}&\equiv&\lr{e^{i(m\phi_{1}^a+n\phi_{2}^b -m\phi_{3}^a-n\phi_{4}^c)}} =  \frac{({\bm q}_{m,a}{\bm q}_{m,a}^*-\tau_a){\bm q}_{n,b}{\bm q}_{n,c}^*}{(1-\tau_a)}\\\nonumber
\rm{SC}_{n_a,m_a|n_b,m_c}\{4\} &\equiv& \llrr{4}_{n_a,m_a|n_b,m_c}- \llrr{2}_{n_a|n_b}\llrr{2}_{m_a|m_c} \approx \lr{{\bm v}_{n,a}{\bm v}^*_{n,b}{\bm v}_{m,a}{\bm v}^*_{m,c}}-\lr{{\bm v}_{n,a}{\bm v}^*_{n,b}}\lr{{\bm v}_{m,a}{\bm v}^*_{m,c}}\\\nonumber
\rm{SC}_{n_a,m_c|m_a,n_b}\{4\} &\equiv& \llrr{4}_{n_a,m_c|m_a,n_b}- \llrr{2}_{n_a|n_b}\llrr{2}_{m_c|m_a} \approx \lr{{\bm v}_{n,a}{\bm v}^*_{n,b}{\bm v}^*_{m,a}{\bm v}_{m,c}}-\lr{{\bm v}_{n,a}{\bm v}^*_{n,b}}\lr{{\bm v}_{m,a}{\bm v}^*_{m,c}}\\\nonumber
\rm{SC}_{n_a,m_b|n_a,m_c}\{4\} &\equiv& \llrr{4}_{n_a,m_b|n_a,m_c}- \llrr{2}_{n_a|n_a}\llrr{2}_{m_b|m_c} \approx \lr{v_{n,a}^2{\bm v}_{m,b}{\bm v}^*_{m,c}}-\lr{v_{n,a}^2}\lr{{\bm v}_{m,b}{\bm v}^*_{m,c}}\\
\rm{SC}_{m_a,n_b|m_a,n_c}\{4\} &\equiv& \llrr{4}_{m_a,n_b|m_a,n_c}- \llrr{2}_{m_a|m_a}\llrr{2}_{n_b|n_c} \approx \lr{v_{m,a}^2{\bm v}_{n,b}{\bm v}^*_{n,c}}-\lr{v_{m,a}^2}\lr{{\bm v}_{n,b}{\bm v}^*_{n,c}}
\end{eqnarray}

\subsection{Asymmetric cumulants (event-plane correlators)}
We consider two interesting cases, the correlation between $v_2$ and $v_4$
\begin{eqnarray}
\label{eq:af1}
\lr{3}_{2_a,2_a|4_b}&\equiv&\lr{e^{i(2\phi_{1}^a+2\phi_{2}^a -4\phi_{3}^b)}} = \frac{({\bm q}_{2,a}^2-\tau_a{\bm q}_{4,a}){\bm q}_{4,b}^*}{(1-\tau_a)} \\\label{eq:af2}
\lr{3}_{2_a,2_b|4_a}&\equiv&\lr{e^{i(2\phi_{1}^a+2\phi_{2}^b -4\phi_{3}^a)}} = \frac{({\bm q}_{2,a}{\bm q}^*_{4,a}-\tau_a{\bm q}^*_{2,a}){\bm q}_{2,b}}{(1-\tau_a)},
\end{eqnarray}
and correlation between $v_2$, $v_3$ and $v_5$
\begin{eqnarray}
\label{eq:af3}
\lr{3}_{2_a,3_a|5_b}&\equiv&\lr{e^{i(2\phi_{1}^a+3\phi_{2}^a -5\phi_{3}^b)}} = \frac{({\bm q}_{2,a}{\bm q}_{3,a}-\tau_a{\bm q}_{5,a}){\bm q}_{5,b}^*}{(1-\tau_a)}\\\label{eq:af4}
\lr{3}_{2_a,3_b|5_a}&\equiv&\lr{e^{i(2\phi_{1}^a+3\phi_{2}^b -5\phi_{3}^a)}} = \frac{({\bm q}_{2,a}{\bm q}^*_{5,a}-\tau_a{\bm q}^*_{3,a}){\bm q}_{3,b}}{(1-\tau_a)}\\\label{eq:af5}
\lr{3}_{2_b,3_a|5_a}&\equiv&\lr{e^{i(2\phi_{1}^b+3\phi_{2}^a -5\phi_{3}^a)}} = \frac{({\bm q}_{3,a}{\bm q}^*_{5,a}-\tau_a{\bm q}^*_{2,a}){\bm q}_{2,b}}{(1-\tau_a)}
\end{eqnarray}
Depending on how the particles are arranged, removal of the duplicate terms (or autocorrelation) in subevent $a$ may play an important role. If the all ${\bm q}$ are placed in subevent $a$ and all ${\bm q}^*$ are placed in subevent $b$ (Eqs.~\ref{eq:af1} and \ref{eq:af3}), the autocorrelation effect is on the order of $\tau_a v_{n+m}/(v_nv_m)\sim 1/M_a$ as first derived by Bhalerao, Ollitrault and Pal.~\cite{Bhalerao:2013ina}, which is small when multiplicity in the subevent is large (e.g. in A+A collisions). On the other hand, when ${\bm q}$ and  ${\bm q}^*$ appear in the same subevent, the autocorrelation contribution is on the order of $\tau_a v_{|n-m|}/(v_nv_m)$, which could be large since $v_{|n-m|}\gg v_nv_m$. Therefore explicit removal of duplicate terms is important if the multiplicity in the subevent is small, for example if the $\eta$ range is narrow or in a small collision system, or if ${\bm q}$ and  ${\bm q}^*$ appear in the same subevent.

Various combinations of subevents for the same correlators also contain different information on the non-linear mode-mixing effects and longitudinal decorrelation effects. It is well-established that the $v_4$ and $v_5$ in A+A collisions contain a linear contribution associated with initial conditions and a mode-mixing contribution due to non-linear hydrodynamic response~\cite{Gardim:2011xv,Teaney:2012ke,Aad:2014fla,Aad:2015lwa,Qiu:2012uy}:
\begin{eqnarray}
\label{eq:af6}
{\bm v}_4 = {\bm v}_{4\rm{L}} +\beta_{2,2} {\bm v}_2^2\;,\;\;{\bm v}_5 = {\bm v}_{5\rm{L}}+\beta_{2,3} {\bm v}_2{\bm v}_3\;,
\end{eqnarray}

If the non-flow contribution can be ignored, we have
\begin{eqnarray}
\nonumber
\lr{3}_{2_a,2_a|4_b}&=& \lr{{\bm v}_{2,a}^2{\bm v}^*_{4,b}}\hspace*{0.5cm} = \hspace*{0.6cm} \lr{{\bm v}_{2,a}^2{\bm v}^*_{\rm{4L},b}}+\beta_{2,2} \lr{{\bm v}_{2,a}^2{\bm v}^{2*}_{2,b}}\hspace*{0.5cm}\approx \lr{{\bm v}_{2}^2{\bm v}^*_{\rm{4L}}}+\beta_{2,2} \lr{{\bm v}_{2,a}^2{\bm v}^{2*}_{2,b}} \\\label{eq:af8}
\lr{3}_{2_a,2_b|4_a}&=& \lr{{\bm v}_{2,a}{\bm v}^*_{4,a}{\bm v}_{2,b}}= \lr{{\bm v}_{2,a}{\bm v}_{2,b}{\bm v}^*_{\rm{4L},a}}+ \beta_{2,2} \lr{v^2_{2,a}{\bm v}_{2,a}{\bm v}^*_{2,b}}\approx \lr{{\bm v}_{2}^2{\bm v}^*_{\rm{4L}}}+ \beta_{2,2} \lr{v^2_{2}{\bm v}_{2,a}{\bm v}^*_{2,b}}\;,
\end{eqnarray}
These two correlators are sensitive to the second moments and first moment of the decorrelations between the two subevents, respectively. In central A+A collisions, where $v_4$ is dominated by the linear component $v_{4\rm{L}}$, the two correlators both measure $\lr{{\bm v}_{2}^2{\bm v}^*_{\rm{4L}}}$, which is sensitive to the correlations in the initial eccentricity, i.e.,
\begin{eqnarray}
\lr{{\bm v}_{2}^2{\bm v}^*_{\rm{4L}}}\propto \lr{{\bm \epsilon}_{2}^2{\bm \epsilon}^*_{\rm{4}}}
\end{eqnarray}
In non-central collisions, where the mode-mixing component is important, one expects that the two correlators deviate from each other, with $\lr{3}_{2_a,2_a|4_b}<\lr{3}_{2_a,2_b|4_a}$. The deviation is expected to increase with the rapidity separation between subevents $a$ and $b$.

Note that the relations in Eqs.~\ref{eq:af8} can be independently probed by directly measuring two four-particle correlators $\llrr{4}_{2a|2b}$, $\llrr{4}_{a,a|a,b}$ and their associated cumulants $c_2\{4\}^{2a|2b}$, $c_2\{4\}^{a,a|a,b}$ :
\begin{eqnarray}
c_2^{a,a|a,b}\{4\} =  \lr{v^2_{2,a}{\bm v}_{2,a}{\bm v}^*_{2,b}} - 2\lr{v^2_{2,a}} \lr{{\bm v}_{2,a}{\bm v}^*_{2,b}}.
\end{eqnarray}
This relation, together with Eqs.~\ref{eq:e1}--\ref{eq:e3}, can be used to calculate $\lr{{\bm v}_{2,a}^2{\bm v}^{2*}_{2,b}}$ and $\lr{v^2_{2}{\bm v}_{2,a}{\bm v}^*_{2,b}}$, and eventually the value of $\beta_{2,2}$.

Similarly, the three correlators involving $v_5$, $v_2$, and $v_3$ can be written as:
\begin{eqnarray}
\nonumber
\lr{3}_{2_a,3_a|5_b}&=& \lr{{\bm v}_{2,a}{\bm v}_{3,a}{\bm v}^*_{5,b}}= \lr{{\bm v}_{2,a}{\bm v}_{3,a}{\bm v}^*_{5\rm{L},b}}+ \beta_{2,3}\lr{{\bm v}_{2,a}{\bm v}^*_{2,b}{\bm v}_{3,a}{\bm v}^*_{3,b}} \approx \lr{{\bm v}_{2}{\bm v}_{3}{\bm v}^*_{5\rm{L}}}+ \beta_{2,3}\lr{{\bm v}_{2,a}{\bm v}^*_{2,b}{\bm v}_{3,a}{\bm v}^*_{3,b}}\\\nonumber
\lr{3}_{2_a,3_b|5_a}&=& \lr{{\bm v}_{2,a}{\bm v}^*_{5,a}{\bm v}_{3,b}}= \lr{{\bm v}_{2,a}{\bm v}_{3,b}{\bm v}^*_{5\rm{L},a}}+ \beta_{2,3}\lr{v_{2,a}^2{\bm v}_{3,a}{\bm v}^*_{3,b}}           \hspace*{0.6cm}        \approx  \lr{{\bm v}_{2}{\bm v}_{3}{\bm v}^*_{5\rm{L}}}+\beta_{2,3}\lr{v_{2}^2{\bm v}_{3,a}{\bm v}^*_{3,b}}\\\label{eq:af11}
\lr{3}_{2_b,3_a|5_a}&=& \lr{{\bm v}_{3,a}{\bm v}^*_{5,a}{\bm v}_{2,b}}= \lr{{\bm v}_{2,b}{\bm v}_{3,a}{\bm v}^*_{5\rm{L},a}}+ \beta_{2,3}\lr{v_{3,a}^2{\bm v}_{2,a}{\bm v}^*_{2,b}}           \hspace*{0.6cm}        \approx  \lr{{\bm v}_{2}{\bm v}_{3}{\bm v}^*_{5\rm{L}}}+\beta_{2,3}\lr{v_{3}^2{\bm v}_{2,a}{\bm v}^*_{2,b}}\;.
\end{eqnarray}
The first correlator is sensitive to the decorrelation of both $v_2$ and $v_3$, while the second and the third correlators are sensitive to the decorrelation of $v_3$ only and $v_2$ only, respectively. It would be interesting to measure the centrality and $\eta$ dependence of these three correlators to extract the linear component which is directly sensitive to the correlation in the initial eccentricity:
\begin{eqnarray}
\lr{{\bm v}_{2}{\bm v}_{3}{\bm v}^*_{5\rm{L}}} \propto \lr{{\bm \epsilon}_{2}{\bm \epsilon}_{3}{\bm \epsilon}^*_{5}}
\end{eqnarray}

\subsection{Particle weights}
Often in data analysis, the weight $w_i$ needs to be applied for each particle. The general framework for deriving the expression for weighted cumulants is given in Refs.~\cite{Bilandzic:2010jr,Bilandzic:2012wva,Bilandzic:2013kga,DiFrancesco:2016srj} or via a slightly simplified approach provided in Appendix H.  All correlators can be conveniently expressed in terms of weighted per-particle flow vector:
\begin{eqnarray}
{\bm q}_{n;k} \equiv \frac{\sum_i  w_i^ke^{in\phi_i}}{\sum_iw_i^k}, \tau_k \equiv \frac{\sum_iw_i^{k+1}}{(\sum_iw_i)^{k+1}}\;,\;{\bm q}_{kn}\equiv{\bm q}_{kn;k}
\end{eqnarray}
With this definition, the formulas for subevent correlators can be obtained by simply replacing $\tau^{k-1}{\bm q}_{kn}$ by $\tau_{k-1}{\bm q}_{kn;k}$ ($k\geq2$). Here are some examples:
\begin{eqnarray}
\nonumber
\lr{4}_{2a|2b}&=&\frac{({\bm q}_{n}^2-\tau_{1}{\bm q}_{2n})_a({\bm q}_{n}^{2}-\tau_{1}{\bm q}_{2n})_b^*}{(1-\tau_{1})_a(1-\tau_{1})_b}\\\nonumber
\lr{6}_{3a|3b} &=&\frac{({\bm q}_{n}^3-3\tau_{1}{\bm q}_{2n}{\bm q}_{n}+2\tau_{2}{\bm q}_{3n})_a({\bm q}_{n}^3-3\tau_{1}{\bm q}_{2n}{\bm q}_{n}+2\tau_{2}{\bm q}_{3n})_b^*}{(1-3\tau_{1}+2\tau_{2})_a(1-3\tau_{1}+2\tau_{2})_b}\\\nonumber
\lr{8}_{4a|4b} &=& \frac{({\bm q}_{n}^4-6\tau_{1}{\bm q}_{2n}{\bm q}_{n}^2+3\tau_{1}^2{\bm q}_{2n}^2+8\tau_{2}{\bm q}_{3n}{\bm q}_{n}-6\tau_{3}{\bm q}_{4n})_a({\bm q}_{n}^4-6\tau_{1}{\bm q}_{2n}{\bm q}_{n}^2+3\tau_{1}^2{\bm q}_{2n}^2+8\tau_{2}{\bm q}_{3n}{\bm q}_{n}-6\tau_{3}{\bm q}_{4n})_b^*}{(1-6\tau_{1}+3\tau_{1}^2+8\tau_{2}-6\tau_{3})_a(1-6\tau_{1}+3\tau_{1}^2+8\tau_{2}-6\tau_{3})_b}\\\label{eq:af12}
\end{eqnarray}
where we have moved the subevent labels $a$ and $b$ outside the parentheses to compactify the formulas, i.e., $(1-\tau_{1})_a\equiv 1-\tau_{1,a}$. The advantage of using per-particle flow vector is that all quantities are smaller than 1, and they can be ordered in terms of $\tau_k\sim 1/M^k$ or $q_{kn;k}\sim q_{n}^k$.

Formulas for two-particle correlators and four-particle correlators based on three-subevents are
\begin{eqnarray}
\nonumber
&&\lr{2}_{a|b} = {\bm q}_{n,a}{\bm q}^*_{n,b}\;,\lr{2}_{a|a} = \frac{(q_{n}^2-\tau_1)_a}{(1-\tau_1)_a}\;,\; \lr{4}_{a,b|a,b}= \frac{(q_{n}^2-\tau_1)_a(q_{n}^{2}-\tau_{1})_b}{(1-\tau_1)_a(1-\tau_1)_b}\;,\\
&&\lr{4}_{a,a|b,c}=\frac{({\bm q}_{n}^2-\tau_1{\bm q}_{2n})_a{\bm q}_{n,b}^*{\bm q}_{n,c}^*}{1-\tau_{1,a}}\;,
\lr{4}_{a,b|a,c}= \frac{(q_{n}^2-\tau_1)_a{\bm q}_{n,b}{\bm q}_{n,c}^*}{1-\tau_{1,a}}
\end{eqnarray}

For completeness, we also document the expression for standard cumulant, which again can be readily converted from the formulas given in Refs.~\cite{Bilandzic:2010jr,Bilandzic:2012wva,Bilandzic:2013kga} or by using the procedure given in Appendix H (Eqs.~\ref{eq:j9} and \ref{eq:j10}). The situation is slightly more complicated since there are two or more particles that are complex-conjugated with each other.
\begin{eqnarray}
\label{eq:af20}\lr{4} &=& \frac{q_n^4-2\tau_1({\bm q}_{2n}{\bm q}_n^{*2} +2q_{n}^2)+8\tau_2{\bm q}_{n;3}{\bm q}_n^*+\tau_1^2(2+q_{2n}^2)-6\tau_3 }{1-6\tau_1+8\tau_2+3\tau_1^2-6\tau_3}\\\nonumber
\lr{6} &=& \left(q_n^6-6\tau_1\bq_{2n}\bq_{n}^{*2}q_n^2+9\tau_1^2q^2_{2n}q^2_n+4\tau_2\bq_{3n}\bq_{n}^{*3}+18\tau_1^2\bq_{2n}\bq_{n}^{*2}-36\tau_3\bq_{2n;4}\bq_{n}^{*2}-36\tau_2\tau_1 \bq_{n;3}\bq_n\bq_{2n}^*+18\tau_1^2q_n^2\right.\\\nonumber
&&\hspace*{-0.2cm}-\;54\tau_3 q^2_n-72\tau_2\tau_1\bq_{n;3}\bq_{n}^*+36\tau_2^2q_{n;3}^2+144\tau_4\bq_{n;5}\bq_{n}^*-9\tau_1q_n^4+36\tau_2\bq_{n;3}\bq_{n}^*q_n^2-9\tau_1^3q_{2n}^2+36\tau_3\tau_1\bq_{2n;4}\bq_{2n}^*\\\nonumber
&&\hspace*{-0.2cm}\left.+\;54\tau_3\tau_1-6\tau_1^3-120\tau_5-12\tau_2\tau_1\bq_{3n}\bq_{2n}^*\bq_{n}^*+4\tau_2^2q_{3n}^2\right)/\\\label{eq:af21}
&&\left(1-15\tau_1+40\tau_2+45\tau_1^2 -90\tau_3-120\tau_2\tau_1-15\tau_1^3+144\tau_4+90\tau_3\tau_1+40\tau_2^2-120\tau_5\right)
\end{eqnarray}

The event-by-event weight for any correlator is also straightforward to write out. It is simply the denominator times the $\sum_iw_i$ in each subevent raised to the power of number of particles chosen from that subevent. Here are some examples:
\begin{eqnarray}
\nonumber
W_{\lr{4}_{2a|2b}} &=& (1-\tau_{1})_a(1-\tau_{1})_b (\sum_iw_i)_a^2(\sum_iw_i)_b^2/4\\\nonumber
W_{\lr{4}_{a,a|b,c}} &=& (1-\tau_{1})_a(\sum_iw_i)_a^2(\sum_iw_i)_b(\sum_iw_i)_c/2\\\label{eq:af22}
W_{\lr{4}} &=& (1-6\tau_1+8\tau_2+3\tau_1^2-6\tau_3)(\sum_iw_i)^4/24
\end{eqnarray}

If all particles have the same weights, $\lr{4}$ and $\lr{6}$ are simplified to:
\begin{eqnarray}
\nonumber
\lr{4} &=& \frac{q_n^4-2\tau{\bm q}_{2n}{\bm q}_n^{*2} -4\tau(1-2\tau)q_{n}^2+2\tau^2(1-3\tau)+\tau^2q_{2n}^2 }{(1-\tau)(1-2\tau)(1-3\tau)}\\\nonumber
\lr{6} &=& \left[q_n^6-6\tau{\bm q}_{2n}{\bm q}_{n}{\bm q}^{*3}_{n}+9\tau^2q_{2n}^2q_n^2+4\tau^2{\bm q}_{3n}{\bm q}^{*3}_{n}-12\tau^3{\bm q}_{3n}{\bm q}^{*}_{2n}{\bm q}^{*}_{n}+18\tau^2(1-4\tau){\bm q}_{2n}{\bm q}^{*2}_{n}-9\tau(1-4\tau)(q_n^4+\tau^2q_{2n}^2)\right.\\\nonumber
&&\left.+18\tau^2(1-2\tau)(1-5\tau)q_n^2-6\tau^3(1-4\tau)(1-5\tau)+4\tau^4q_{3n}^2\right]/[(1-\tau)(1-2\tau)(1-3\tau)(1-4\tau)(1-5\tau)]\\\label{eq:af23}
\end{eqnarray}

\subsection{Iterative procedure for deriving the equations for multi-particle correlation}
In the original paper on direct cumulants by A. Bilandzic, R. Sneilings and S. Voloshin ~\cite{Bilandzic:2010jr}, the expression for multi-particle correlator $\lr{2k}$ is obtained by expanding all lower-order correlators and solving many coupled linear equations to remove combinations with duplicated particle indexes.  This was later improved into an iterative procedure in Ref.~\cite{Bilandzic:2013kga} for general multi-particle correlations including mixed harmonics and the algorithm for calculating these correlators has been provided~\cite{code}. Here we recast that general iterative procedure into a simplified form to handle two specific classes of correlators used in this paper: the correlation involving the whole event used in the standard method and two-subevent correlations.
\subsubsection{unweighted case}
We introduce the following notation for a multi-particle correlator in an event with $M$ particles:
\begin{eqnarray}
\nonumber
\overline{k}_{k*n} &=&  \sum_{j_1\neq j_2\neq..\neq j_k=1}^M e^{in(\sum_{i=1}^k\phi_{j_i})}\\\label{eq:h2}
\overline{2k}_{k*n,-k*n} &=&  \sum_{j_1\neq j_2\neq..\neq j_{2k}=1}^M e^{in(\sum_{i=1}^k\phi_{j_{2i+1}}-\sum_{i=1}^k\phi_{j_{2i+2}})}
\end{eqnarray}
where, following convention in Ref.~\cite{Bilandzic:2010jr}, $k*n$ and $-k*n$ are short-hand notations for $n,n,..,n$ and $-n,-n,..,-n$, each with $k$ indexes. With this notation, the correlators in the standard method and two-subevent method (Eq.~\ref{eq:af12}) can be expressed as:
\begin{eqnarray}
\label{eq:h3}
\lr{2k}_{k*n,-k*n}\equiv \frac{\overline{2k}_{k*n,-k*n}}{M(M-1)...(M-2k+1)}\;, \lr{2k}_{ka|kb}\equiv \frac{(\overline{k}_{k*n})_a (\overline{k}_{k*n})_b^*}{M_a(M_a-1)...(M_a-k+1)M_b(M_b-1)...(M_b-k+1)}.
\end{eqnarray}
Note that the number of permutations can be written as:
\begin{eqnarray}
\label{eq:h3b}
M(M-1)...(M-2k+1) = \lim_{n\rightarrow0}\overline{2k}_{k*n,-k*n},\;\; M(M-1)...(M-k+1)=\lim_{n\rightarrow0}\overline{k}_{k*n}
\end{eqnarray}

To derive the expression for $\overline{k}_{k*n}$, we introduce an operator $\otimes$ and a set of expansion rules, which relate the number of permutations, $M(M-1)...(M-k+1)$, to the final expression in terms of flow vector ${\bm Q}_n$:
\begin{eqnarray}
\nonumber
&&\hspace*{3cm}\overline{k}_{k*n} = {\bm Q}_n\otimes({\bm Q}_n-1)\otimes...\otimes({\bm Q}_n-k+1)\\\label{eq:h4}
&&\left(\prod_{i=1}^{h}{\bm Q}_{p_in}\right)\otimes {\bm Q}_n = {\bm Q}_n\left(\prod_{i=1}^{h}{\bm Q}_{p_in}\right)\;, \left(\prod_{i=1}^{h}{\bm Q}_{p_in}\right)\otimes \left(\sum_{i=1}^{h} p_i\right) = \left(\sum_{i=1}^{h}  \left(p_i\frac{{\bm Q}_{(p_i+1)n}}{{\bm Q}_{p_in}}\right)\right)\prod_{i=1}^{h} {\bm Q}_{p_in}
\end{eqnarray}
The expansion rules explicitly keep track of the duplications as one goes from $k$-particle to $k+1$-particle correlation. The first few are:
\begin{eqnarray}
\label{eq:h5}
{\bm Q}_n\otimes 1={\bm Q}_{2n},\; {\bm Q}_n^2\otimes2 = 2{\bm Q}_{2n}{\bm Q}_n,\; {\bm Q}_{2n}\otimes2 = 2{\bm Q}_{3n},\; {\bm Q}_n^3\otimes3=3{\bm Q}_{2n}{\bm Q}_{n}^2,\; {\bm Q}_{2n}{\bm Q}_n\otimes3 = 2{\bm Q}_{3n}{\bm Q}_n + {\bm Q}_{2n}^2.
\end{eqnarray}
Therefore it is straightforward to write out the expression for $\overline{k}_{k*n}$,
\begin{eqnarray}
\nonumber
&&\overline{2}_{2*n}= {\bm Q}_n\otimes({\bm Q}_n-1) = {\bm Q}_n^2-{\bm Q}_n\otimes 1 = {\bm Q}_n^2 - {\bm Q}_{2n}\\\nonumber
&&\overline{3}_{3*n}= \overline{k}_{2*n}\otimes({\bm Q}_n-2) = {\bm Q}_n^3 - {\bm Q}_{2n}{\bm Q}_n-{\bm Q}_n^2\otimes2+{\bm Q}_{2n}\otimes2= {\bm Q}_n^3 - 3{\bm Q}_{2n}{\bm Q}_n+2{\bm Q}_{3n}\\\nonumber
&&\overline{4}_{4*n}= \overline{k}_{3*n}\otimes({\bm Q}_n-3) = {\bm Q}_n^4 - 3{\bm Q}_{2n}{\bm Q}_n^2+2{\bm Q}_{3n}{\bm Q}_n-{\bm Q}_n^3\otimes3 + 3{\bm Q}_{2n}{\bm Q}_n\otimes3-2{\bm Q}_{3n}\otimes3\\\label{eq:h6}
&&\hspace*{3.5cm} = {\bm Q}_n^4 - 6{\bm Q}_{2n}{\bm Q}_n^2 +3{\bm Q}_{2n}^2+8{\bm Q}_{3n}{\bm Q}_n-6{\bm Q}_{4n}
\end{eqnarray}

To derive the expression for $\overline{2k}_{k*n,-k*n}$ used in the standard method where all particle are chosen from the whole event, one just needs to extend the rules in Eq.~\ref{eq:h4} to include case where particle are conjugated with each other:
\begin{eqnarray}
\nonumber
&&\hspace*{0cm}\overline{2k}_{k*n,-k*n} = {\bm Q}_n\otimes({\bm Q}_{-n}-1^-)\otimes({\bm Q}_n-2^+)\otimes({\bm Q}_{-n}-3^-)\otimes...\otimes({\bm Q}_n-(2k-2)^+)\otimes({\bm Q}_{-n}-(2k-1)^-)\\\nonumber
&&\left(\prod_{i=1}^{h}{\bm Q}_{p_in,-m_in}\right)\otimes {\bm Q}_{\pm n} = {\bm Q}_{\pm n}\left(\prod_{i=1}^{h}{\bm Q}_{p_in,-m_in}\right), \; {\bm Q}_{pn,-mn} =  \sum_{i=1}^M e^{i(p-m)n\phi_{i}}\\\nonumber
&& \left(\prod_{i=1}^{h}{\bm Q}_{p_in,-m_in}\right)\otimes \left(\sum_{i=1}^{h} (p_i+m_i)\right)^+ = \left(\sum_{i=1}^{h}  \left((p_i+m_i)\frac{{\bm Q}_{(p_i+1)n,-m_in}}{{\bm Q}_{p_in,-m_in}}\right)\right)\prod_{i=1}^{h} {\bm Q}_{p_in,-m_in}\\\label{eq:h7}
&& \left(\prod_{i=1}^{h}{\bm Q}_{p_in,-m_in}\right)\otimes \left(\sum_{i=1}^{h} (p_i+m_i)\right)^- = \left(\sum_{i=1}^{h}  \left((p_i+m_i)\frac{{\bm Q}_{p_in,-(m_i+1)n}}{{\bm Q}_{p_in,-m_in}}\right)\right)\prod_{i=1}^{h} {\bm Q}_{p_in,-m_in}
\end{eqnarray}
where we have used superscripts $^+$ and $^-$ to indicate the sign of $\phi$ angle for the additional particle going from $k$-particle correlation to $k+1$-particle correlation. Some examples are:
\begin{eqnarray}
\nonumber
&&{\bm Q}_n\otimes 1^- = {\bm Q}_{n,-n},\;{\bm Q}_n{\bm Q}_{-n}\otimes2^+={\bm Q}_{2n}{\bm Q}_{-n}+{\bm Q}_n{\bm Q}_{n,-n},\;{\bm Q}_{n,-n}\otimes2^+=2{\bm Q}_{2n,-n},\;\\\nonumber
&&{\bm Q}_n^2{\bm Q}_{-n}\otimes3^-=2{\bm Q}_{n,-n}{\bm Q}_{n}{\bm Q}_{-n}+{\bm Q}_n^2{\bm Q}_{-2n},\;{\bm Q}_{2n}{\bm Q}_{-n}\otimes3^-=2{\bm Q}_{2n,-n}{\bm Q}_{-n}+{\bm Q}_{2n}{\bm Q}_{-2n},\\\label{eq:h8}
&&{\bm Q}_n{\bm Q}_{n,-n}\otimes3^-={\bm Q}_{n,-n}^2+2{\bm Q}_n{\bm Q}_{n,-2n},\;{\bm Q}_{2n,-n}\otimes3^-=3{\bm Q}_{2n,-2n},
\end{eqnarray}

\begin{eqnarray}
\nonumber
\overline{2}_{n,-n}   & =& {\bm Q}_n\otimes({\bm Q}_{-n}-1^-) = {\bm Q}_n{\bm Q}_{-n}-{\bm Q}_n\otimes 1^- = {\bm Q}_n{\bm Q}_{-n} - {\bm Q}_{n,-n}\\\nonumber
\overline{3}_{2*n,-n} & =& \overline{k}_{n,-n}\otimes({\bm Q}_n-2^+) =  {\bm Q}_n^2{\bm Q}_{-n}-{\bm Q}_{2n}{\bm Q}_{-n}-2{\bm Q}_{n,-n}{\bm Q}_n+2{\bm Q}_{2n,-n}\\\nonumber
\overline{4}_{2*n,-2*n}& =& \overline{k}_{2*n,-n}\otimes({\bm Q}_{-n}-3^-) =  {\bm Q}_n^2{\bm Q}_{-n}^2-({\bm Q}_{2n}{\bm Q}_{-n}^2+{\bm Q}_{-2n}{\bm Q}_n^2)-4{\bm Q}_{n,-n}{\bm Q}_n{\bm Q}_{-n}\\\label{eq:h9}
&&+{\bm Q}_{2n}{\bm Q}_{-2n}+2{\bm Q}_{n,-n}^2+4({\bm Q}_{2n,-n}{\bm Q}_{-n}+{\bm Q}_{n,-2n}{\bm Q}_n)-6{\bm Q}_{2n,-2n}
\end{eqnarray}

The expressions for higher-order correlation are rather lengthy, therefore it is helpful to define new operator $\otimes_{2k\rightarrow2k+2}$ that directly relates the $2k$- and $2k+2$-particle correlations: 
\begin{eqnarray}
\nonumber
&&\otimes({\bm Q}_{n}-(2k)^+)\otimes({\bm Q}_{-n}-(2k+1)^-) ={\bm Q}_{n}{\bm Q}_{-n}+\otimes_{2k\rightarrow2k+2}\\\label{eq:h10}
&&\otimes_{2k\rightarrow2k+2}\equiv-\otimes(2k)^+{\bm Q}_{-n}-{\bm Q}_{n}\otimes(2k+1)^-+\otimes(2k)^+\otimes(2k+1)^-
\end{eqnarray}
\begin{eqnarray}
\nonumber
\overline{6}_{3*n,-3*n}& =& \overline{4}_{2*n,-2*n}\otimes({\bm Q}_{n}-4^+)\otimes({\bm Q}_{-n}-5^-)=\overline{4}_{2*n,-2*n}{\bm Q}_{n}{\bm Q}_{-n}+{\bm Q}_n^2{\bm Q}_{-n}^2\otimes_{4\rightarrow6}-({\bm Q}_{2n}{\bm Q}_{-n}^2+cc)\otimes_{4\rightarrow6}\\\nonumber
&&\hspace*{-0.2cm}-\;4{\bm Q}_{n,-n}{\bm Q}_n{\bm Q}_{-n}\otimes_{4\rightarrow6}+{\bm Q}_{2n}{\bm Q}_{-2n}\otimes_{4\rightarrow6}+2{\bm Q}_{n,-n}^2\otimes_{4\rightarrow6}+4({\bm Q}_{2n,-n}{\bm Q}_{-n}+cc)\otimes_{4\rightarrow6}-6{\bm Q}_{2n,-2n}\otimes_{4\rightarrow6}\\\nonumber
&=& {\bm Q}_n^3{\bm Q}_{-n}^3-3(\bQ_{2n}\bQ_{-n}^2+cc)\bQ_n\bQ_{-n}+9\bQ_{2n}\bQ_{-2n}\bQ_n\bQ_{-n}+2(\bQ_{3n}\bQ_{-n}^3+cc)\\\nonumber
&&\hspace*{-0.2cm}+\;9(\bQ_{2n}\bQ_{-n}^2+cc)\bQ_{n,-n}-18(\bQ_{3n,-n}\bQ_{-n}^2+cc)-18(\bQ_{2n,-n}\bQ_{-2n}\bQ_n+cc)\\\nonumber
&&\hspace*{-0.2cm}+\;18\bQ_{n,-n}^2\bQ_n\bQ_{-n}-54{\bm Q}_{2n,-2n}{\bm Q}_n{\bm Q}_{-n}-36(\bQ_{2n,-n}\bQ_{-n}+cc)\bQ_{n,-n}+36\bQ_{2n,-n}\bQ_{n,-2n}+72(\bQ_{3n,-2n}\bQ_{-n}+cc)\\\nonumber
&&\hspace*{-0.2cm}-9\;\bQ_{n,-n}\bQ_n^2\bQ_{-n}^2+18(\bQ_{2n,-n}\bQ_{-n}+cc)\bQ_n\bQ_{-n}-9\bQ_{2n}\bQ_{-2n}\bQ_{n,-n}+18(\bQ_{3n,-n}\bQ_{-2n}+cc)\\\label{eq:h11}
&&\hspace*{-0.2cm}+\;54\bQ_{2n,-2n}\bQ_{n,-n}-6\bQ_{n,-n}^3-120\bQ_{3n,-3n}-6(\bQ_{3n}\bQ_{-2n}\bQ_{-n}+cc)+4\bQ_{3n}\bQ_{-3n}
\end{eqnarray}
where ``$cc$'' stands for complex conjugate.

\subsubsection{with particle weights}
All formula discussions above can be generalized to include particle weights, and we shall express all results using per-particle flow vectors:
\begin{eqnarray}
\nonumber
&& {\bm q}_{pn,-mn} =  \frac{\sum_{i=1}^M w_i^{p+m}e^{i(p-m)n\phi_{i}}}{\sum_{i=1}^M w_i^{p+m}},\;{\bm q}_{pn,-mn}\equiv {\bm q}_{(p-m)n;p+m}\\\nonumber
&& \tau_{k}=\frac{\sum_{i=1}^M w_i^{k+1}}{(\sum_{i=1}^Mw_i)^{k+1}},\; \tau_{0}=1\\\label{eq:j1}
&&{\bm Q}_{pn,-mn} = \tau_{p+m-1}{\bm q}_{pn,-mn}\;\left(\sum_iw_i\right)^{n+m}
\end{eqnarray}
After replacing ${\bm Q}$ with ${\bm q}$, the two-subevent correlation and corresponding rules can be rewritten as:
\begin{eqnarray}
\nonumber
&&\lr{2k}_{ka|kb}\equiv \left(\frac{\hat{k}_{k*n}}{\lim_{n\rightarrow0}\hat{k}_{k*n}}\right)_a\times\left(\frac{\hat{k}_{k*n}}{\lim_{n\rightarrow0}\hat{k}_{k*n}}\right)_b^*,\;\;\;\hat{k}_{k*n}={\bm q}_n\otimes({\bm q}_n-\tau)\otimes...\otimes({\bm q}_n-(k-1)\tau)\\\label{eq:j3}
&&\left(\prod_{i=1}^{h}{\bm q}_{p_in}\right)\otimes {\bm q}_n = \left(\prod_{i=1}^{h}{\bm q}_{p_in}\right){\bm q}_n,\;\left(\prod_{i=1}^{h}\tau_{p_i-1}{\bm q}_{p_in}\right)\otimes \left(\sum_{i=1}^{h} p_i\right)\tau = \left(\sum_{i=1}^{h}  \left(p_i\frac{\tau_{p_i}{\bm q}_{(p_i+1)n}}{\tau_{p_i-1}{\bm q}_{p_in}}\right)\right)\prod_{i=1}^{h} \tau_{p_i-1}{\bm q}_{p_in}\;.
\end{eqnarray}
Note that $\overline{k}_{k*n}=\hat{k}_{k*n}\left(\sum_iw_i\right)^k$, and therefore the factor $\left(\sum_iw_i\right)^k$ cancels out between the numerator and denominator of $\lr{2k}_{ka|kb}$. When particle weight is a constant, $\lim_{n\rightarrow0}\hat{k}_{k*n}=(1-\tau)(1-2\tau)...(1-(k-1)\tau), \tau=1/M$.

Similarly the expansion rules for $\lr{2k}_{k*n,-k*n}$ are given by:
\begin{eqnarray}
\nonumber
&&\lr{2k}_{k*n,-k*n}=\frac{\widehat{2k}_{k*n,-k*n}}{\lim_{n\rightarrow0}\widehat{2k}_{k*n,-k*n}}\\\nonumber
&&\widehat{2k}_{k*n,-k*n}\equiv{\bm q}_n\otimes({\bm q}_{-n}-\tau^-)\otimes({\bm q}_n-2\tau^+)\otimes({\bm q}_{-n}-3\tau^-)\otimes...\otimes({\bm q}_n-(2k-2)\tau^+)\otimes({\bm q}_{-n}-(2k-1)\tau^-)\\\nonumber
&&\left(\prod_{i=1}^{h}{\bm q}_{p_in,-m_in}\right)\otimes {\bm q}_{\pm n} = \left(\prod_{i=1}^{h}{\bm q}_{p_in,-m_in}\right){\bm q}_{\pm n}, \\\nonumber
&&\left(\prod_{i=1}^{h}\tau_{p_i+m_i-1}{\bm q}_{p_in,-m_in}\right)\otimes \left(\sum_{i=1}^{h} (p_i+m_i)\right)\tau^+ = \left(\sum_{i=1}^{h}  \left((p_i+m_i)\frac{\tau_{p_i+m_i}{\bm q}_{(p_i+1)n,-m_in}}{\tau_{p_i+m_i-1}{\bm q}_{p_in,-m_in}}\right)\right)\prod_{i=1}^{h} \tau_{p_i+m_i-1}{\bm q}_{p_in,-m_in}\\\nonumber
&& \left(\prod_{i=1}^{h}\tau_{p_i+m_i-1}{\bm q}_{p_in,-m_in}\right)\otimes \left(\sum_{i=1}^{h} (p_i+m_i)\right)\tau^- = \left(\sum_{i=1}^{h}  \left((p_i+m_i)\frac{\tau_{p_i+m_i}{\bm q}_{p_in,-(m_i+1)n}}{\tau_{p_i+m_i-1}{\bm q}_{p_in,-m_in}}\right)\right)\prod_{i=1}^{h} \tau_{p_i+m_i-1}{\bm q}_{p_in,-m_in}\\\label{eq:j7}
\end{eqnarray}

With these, we can rewrite Eqs.~\ref{eq:h8} and \ref{eq:h9} as:
\begin{eqnarray}
\nonumber
&&{\bm q}_n\otimes 1^- = \tau_1{\bm q}_{n,-n},\;{\bm q}_n{\bm q}_{-n}\otimes2^+=\tau_1({\bm q}_{2n}{\bm q}_{-n}+{\bm q}_n{\bm q}_{n,-n}),\;\tau_1{\bm q}_{n,-n}\otimes2^+=2\tau_2{\bm q}_{2n,-n},\;\\\nonumber
&&{\bm q}_n^2{\bm q}_{-n}\otimes3^-=2\tau_1({\bm q}_{n,-n}{\bm q}_{n}{\bm q}_{-n}+{\bm q}_n^2{\bm q}_{-2n}),\;\tau_1{\bm q}_{2n}{\bm q}_{-n}\otimes3^-=2\tau_2{\bm q}_{2n,-n}{\bm q}_{-n}+\tau_1^2{\bm q}_{2n}{\bm q}_{-2n},\\\label{eq:j8}
&&\tau_1{\bm q}_n{\bm q}_{n,-n}\otimes3^-=\tau_1^2{\bm q}_{n,-n}^2+2\tau_2{\bm q}_n{\bm q}_{n,-2n},\;\tau_2{\bm q}_{2n,-n}\otimes3^-=3\tau_3{\bm q}_{2n,-2n},\;
\end{eqnarray}

\begin{eqnarray}
\nonumber
\hat{2}_{n,-n}   & =& {\bm q}_n\otimes({\bm q}_{-n}-1^-) = {\bm q}_n{\bm q}_{-n}-{\bm q}_n\otimes 1^- = {\bm q}_n{\bm q}_{-n} - \tau_1{\bm q}_{n,-n}\\\nonumber
\hat{3}_{2*n,-n} & =& \hat{k}_{n,-n}\otimes({\bm q}_n-2^+) =  {\bm q}_n^2{\bm q}_{-n}-\tau_1{\bm q}_{2n}{\bm q}_{-n}-2\tau_1{\bm q}_n{\bm q}_{n,-n}+2\tau_2{\bm q}_{2n,-n}\\\nonumber
\hat{4}_{2*n,-2*n}& =& \hat{k}_{2*n,-n}\otimes({\bm q}_{-n}-3^-) =  {\bm q}_n^2{\bm q}_{-n}^2-\tau_1({\bm q}_{2n}{\bm q}_{-n}^2+{\bm q}_n^2{\bm q}_{-2n})-4\tau_1{\bm q}_n{\bm q}_{-n}{\bm q}_{n,-n}\\\nonumber
&&+\tau_1^2{\bm q}_{2n}{\bm q}_{-2n}+2\tau_1^2{\bm q}_{n,-n}^2+4\tau_2({\bm q}_{2n,-n}{\bm q}_{-n}+{\bm q}_n{\bm q}_{n,-2n})-6\tau_3{\bm q}_{2n,-2n}\\\nonumber
\hat{6}_{3*n,-3*n}&=& \bq_n^3\bq_{-n}^3-3\tau_1(\bq_{2n}\bq_{-n}^2+cc)\bq_n\bq_{-n}+9\tau_1^2\bq_{2n}\bq_{-2n}\bq_n\bq_{-n}+2\tau_2(\bq_{3n}\bq_{-n}^3+cc)\\\nonumber
&&\hspace*{-0.2cm}+\;9\tau_1^2(\bq_{2n}\bq_{-n}^2+cc)\bq_{n,-n}-18\tau_3(\bq_{3n,-n}\bq_{-n}^2+cc)-18\tau_2\tau_1(\bq_{2n,-n}\bq_{-2n}\bq_n+cc)+18\tau_1^2\bq_{n,-n}^2\bq_n\bq_{-n}\\\nonumber
&&\hspace*{-0.2cm}-\;54\tau_3{\bm q}_{2n,-2n}{\bm q}_n{\bm q}_{-n}-36\tau_2\tau_1(\bq_{2n,-n}\bq_{-n}+cc)\bq_{n,-n}+36\tau_2^2\bq_{2n,-n}\bq_{n,-2n}+72\tau_4(\bq_{3n,-2n}\bq_{-n}+cc)\\\nonumber
&&\hspace*{-0.2cm}-\;9\tau_1\bq_{n,-n}\bq_n^2\bq_{-n}^2+18\tau_2(\bq_{2n,-n}\bq_{-n}+cc)\bq_n\bq_{-n}-9\tau_1^3\bq_{2n}\bq_{-2n}\bq_{n,-n}+18\tau_3\tau_1(\bq_{3n,-n}\bq_{-2n}+cc)\\\label{eq:j9}
&&\hspace*{-0.2cm}+\;54\tau_3\tau_1\bq_{2n,-2n}\bq_{n,-n}-6\tau_1^3\bq_{n,-n}^3-120\tau_5\bq_{3n,-3n}-6\tau_2\tau_1(\bq_{3n}\bq_{-2n}\bq_{-n}+cc)+4\tau_2^2\bq_{3n}\bq_{-3n}
\end{eqnarray}
The expression for $\lim_{n\rightarrow0}\widehat{2k}_{k*n,-k*n}$ can be obtained by setting ${\bm q}_{pn,-mn}=1$, for example, 
\begin{eqnarray}
\nonumber
&&\lim_{n\rightarrow0}\hat{4}_{2*n,-2*n}=1-6\tau_1+8\tau_2+3\tau_1^2-6\tau_3\\\label{eq:j10}
&&\lim_{n\rightarrow0}\hat{6}_{3*n,-3*n}=1-15\tau_1+40\tau_2+45\tau_1^2 -90\tau_3-120\tau_2\tau_1-15\tau_1^3+144\tau_4+90\tau_3\tau_1+40\tau_2^2-120\tau_5
\end{eqnarray}
From Eqs~.\ref{eq:j9} and \ref{eq:j10}, we obtain the formula for $\lr{4}$ and $\lr{6}$ shown in Eqs.~\ref{eq:af20} and \ref{eq:af21}, respectively.

Finally the event-by-event weights are:
\begin{eqnarray}
W_{\lr{2k}_{ka|kb}} = (\frac{(\sum_iw_i)^k}{k!}\lim_{n\rightarrow0}\hat{k}_{k*n})_a(\frac{(\sum_iw_i)^k}{k!}\lim_{n\rightarrow0}\hat{k}_{k*n})_b\;,W_{\lr{2k}_{k*n,-k*n}}=\frac{(\sum_iw_i)^{2k}}{(2k)!}\lim_{n\rightarrow0}\hat{2k}_{2*n,-2*n}
\end{eqnarray}

\clearpage
\bibliography{cumu2_v6}{}

\begin{thebibliography}{55}%
\makeatletter
\providecommand \@ifxundefined [1]{%
 \@ifx{#1\undefined}
}%
\providecommand \@ifnum [1]{%
 \ifnum #1\expandafter \@firstoftwo
 \else \expandafter \@secondoftwo
 \fi
}%
\providecommand \@ifx [1]{%
 \ifx #1\expandafter \@firstoftwo
 \else \expandafter \@secondoftwo
 \fi
}%
\providecommand \natexlab [1]{#1}%
\providecommand \enquote  [1]{``#1''}%
\providecommand \bibnamefont  [1]{#1}%
\providecommand \bibfnamefont [1]{#1}%
\providecommand \citenamefont [1]{#1}%
\providecommand \href@noop [0]{\@secondoftwo}%
\providecommand \href [0]{\begingroup \@sanitize@url \@href}%
\providecommand \@href[1]{\@@startlink{#1}\@@href}%
\providecommand \@@href[1]{\endgroup#1\@@endlink}%
\providecommand \@sanitize@url [0]{\catcode `\\12\catcode `\$12\catcode
  `\&12\catcode `\#12\catcode `\^12\catcode `\_12\catcode `\%12\relax}%
\providecommand \@@startlink[1]{}%
\providecommand \@@endlink[0]{}%
\providecommand \url  [0]{\begingroup\@sanitize@url \@url }%
\providecommand \@url [1]{\endgroup\@href {#1}{\urlprefix }}%
\providecommand \urlprefix  [0]{URL }%
\providecommand \Eprint [0]{\href }%
\providecommand \doibase [0]{http://dx.doi.org/}%
\providecommand \selectlanguage [0]{\@gobble}%
\providecommand \bibinfo  [0]{\@secondoftwo}%
\providecommand \bibfield  [0]{\@secondoftwo}%
\providecommand \translation [1]{[#1]}%
\providecommand \BibitemOpen [0]{}%
\providecommand \bibitemStop [0]{}%
\providecommand \bibitemNoStop [0]{.\EOS\space}%
\providecommand \EOS [0]{\spacefactor3000\relax}%
\providecommand \BibitemShut  [1]{\csname bibitem#1\endcsname}%
\let\auto@bib@innerbib\@empty
\bibitem [{\citenamefont {Adare}\ \emph {et~al.}(2008)\citenamefont {Adare}
  \emph {et~al.}}]{Adare:2008ae}%
  \BibitemOpen
  \bibfield  {author} {\bibinfo {author} {\bibfnamefont {A.}~\bibnamefont
  {Adare}} \emph {et~al.} (\bibinfo {collaboration} {PHENIX Collaboration}),\
  }\href {\doibase 10.1103/PhysRevC.78.014901} {\bibfield  {journal} {\bibinfo
  {journal} {Phys. Rev. C}\ }\textbf {\bibinfo {volume} {78}},\ \bibinfo
  {pages} {014901} (\bibinfo {year} {2008})}\BibitemShut {NoStop}%
\bibitem [{\citenamefont {Abelev}\ \emph {et~al.}(2009)\citenamefont {Abelev}
  \emph {et~al.}}]{Abelev:2009af}%
  \BibitemOpen
  \bibfield  {author} {\bibinfo {author} {\bibfnamefont {B.~I.}\ \bibnamefont
  {Abelev}} \emph {et~al.} (\bibinfo {collaboration} {STAR Collaboration}),\
  }\href {\doibase 10.1103/PhysRevC.80.064912} {\bibfield  {journal} {\bibinfo
  {journal} {Phys. Rev. C}\ }\textbf {\bibinfo {volume} {80}},\ \bibinfo
  {pages} {064912} (\bibinfo {year} {2009})}\BibitemShut {NoStop}%
\bibitem [{\citenamefont {Alver}\ \emph {et~al.}(2010)\citenamefont {Alver}
  \emph {et~al.}}]{Alver:2009id}%
  \BibitemOpen
  \bibfield  {author} {\bibinfo {author} {\bibfnamefont {B.}~\bibnamefont
  {Alver}} \emph {et~al.} (\bibinfo {collaboration} {PHOBOS Collaboration}),\
  }\href {\doibase 10.1103/PhysRevLett.104.062301} {\bibfield  {journal}
  {\bibinfo  {journal} {Phys. Rev. Lett.}\ }\textbf {\bibinfo {volume} {104}},\
  \bibinfo {pages} {062301} (\bibinfo {year} {2010})}\BibitemShut {NoStop}%
\bibitem [{\citenamefont {Aamodt}\ \emph {et~al.}(2011)\citenamefont {Aamodt}
  \emph {et~al.}}]{ALICE:2011ab}%
  \BibitemOpen
  \bibfield  {author} {\bibinfo {author} {\bibfnamefont {K.}~\bibnamefont
  {Aamodt}} \emph {et~al.} (\bibinfo {collaboration} {ALICE Collaboration}),\
  }\href {\doibase 10.1103/PhysRevLett.107.032301} {\bibfield  {journal}
  {\bibinfo  {journal} {Phys.~Rev.~Lett.}\ }\textbf {\bibinfo {volume} {107}},\
  \bibinfo {pages} {032301} (\bibinfo {year} {2011})}\BibitemShut {NoStop}%
\bibitem [{\citenamefont {Aad}\ \emph {et~al.}(2012)\citenamefont {Aad} \emph
  {et~al.}}]{Aad:2012bu}%
  \BibitemOpen
  \bibfield  {author} {\bibinfo {author} {\bibfnamefont {G.}~\bibnamefont
  {Aad}} \emph {et~al.} (\bibinfo {collaboration} {ATLAS Collaboration}),\
  }\href {\doibase 10.1103/PhysRevC.86.014907} {\bibfield  {journal} {\bibinfo
  {journal} {Phys.~Rev.~C}\ }\textbf {\bibinfo {volume} {86}},\ \bibinfo
  {pages} {014907} (\bibinfo {year} {2012})}\BibitemShut {NoStop}%
\bibitem [{\citenamefont {Chatrchyan}\ \emph {et~al.}(2014)\citenamefont
  {Chatrchyan} \emph {et~al.}}]{Chatrchyan:2013kba}%
  \BibitemOpen
  \bibfield  {author} {\bibinfo {author} {\bibfnamefont {S.}~\bibnamefont
  {Chatrchyan}} \emph {et~al.} (\bibinfo {collaboration} {CMS Collaboration}),\
  }\href {\doibase 10.1103/PhysRevC.89.044906} {\bibfield  {journal} {\bibinfo
  {journal} {Phys.~Rev.~C}\ }\textbf {\bibinfo {volume} {89}},\ \bibinfo
  {pages} {044906} (\bibinfo {year} {2014})}\BibitemShut {NoStop}%
\bibitem [{\citenamefont {Khachatryan}\ \emph {et~al.}(2010)\citenamefont
  {Khachatryan} \emph {et~al.}}]{Khachatryan:2010gv}%
  \BibitemOpen
  \bibfield  {author} {\bibinfo {author} {\bibfnamefont {V.}~\bibnamefont
  {Khachatryan}} \emph {et~al.} (\bibinfo {collaboration} {CMS
  Collaboration}),\ }\href {\doibase 10.1007/JHEP09(2010)091} {\bibfield
  {journal} {\bibinfo  {journal} {JHEP}\ }\textbf {\bibinfo {volume} {09}},\
  \bibinfo {pages} {091} (\bibinfo {year} {2010})}\BibitemShut {NoStop}%
\bibitem [{\citenamefont {Aad}\ \emph {et~al.}(2016)\citenamefont {Aad} \emph
  {et~al.}}]{Aad:2015gqa}%
  \BibitemOpen
  \bibfield  {author} {\bibinfo {author} {\bibfnamefont {G.}~\bibnamefont
  {Aad}} \emph {et~al.} (\bibinfo {collaboration} {ATLAS Collaboration}),\
  }\href {\doibase 10.1103/PhysRevLett.116.172301} {\bibfield  {journal}
  {\bibinfo  {journal} {Phys. Rev. Lett.}\ }\textbf {\bibinfo {volume} {116}},\
  \bibinfo {pages} {172301} (\bibinfo {year} {2016})}\BibitemShut {NoStop}%
\bibitem [{\citenamefont {Aaboud}\ \emph
  {et~al.}(2017{\natexlab{a}})\citenamefont {Aaboud} \emph
  {et~al.}}]{Aaboud:2016yar}%
  \BibitemOpen
  \bibfield  {author} {\bibinfo {author} {\bibfnamefont {M.}~\bibnamefont
  {Aaboud}} \emph {et~al.} (\bibinfo {collaboration} {ATLAS Collaboration}),\
  }\href {\doibase 10.1103/PhysRevC.96.024908} {\bibfield  {journal} {\bibinfo
  {journal} {Phys. Rev. C}\ }\textbf {\bibinfo {volume} {96}},\ \bibinfo
  {pages} {024908} (\bibinfo {year} {2017}{\natexlab{a}})}\BibitemShut
  {NoStop}%
\bibitem [{\citenamefont {Chatrchyan}\ \emph
  {et~al.}(2013{\natexlab{a}})\citenamefont {Chatrchyan} \emph
  {et~al.}}]{CMS:2012qk}%
  \BibitemOpen
  \bibfield  {author} {\bibinfo {author} {\bibfnamefont {S.}~\bibnamefont
  {Chatrchyan}} \emph {et~al.} (\bibinfo {collaboration} {CMS Collaboration}),\
  }\href {\doibase 10.1016/j.physletb.2012.11.025} {\bibfield  {journal}
  {\bibinfo  {journal} {Phys.~Lett.~B}\ }\textbf {\bibinfo {volume} {718}},\
  \bibinfo {pages} {795} (\bibinfo {year} {2013}{\natexlab{a}})}\BibitemShut
  {NoStop}%
\bibitem [{\citenamefont {Abelev}\ \emph {et~al.}(2013)\citenamefont {Abelev}
  \emph {et~al.}}]{Abelev:2012ola}%
  \BibitemOpen
  \bibfield  {author} {\bibinfo {author} {\bibfnamefont {B.}~\bibnamefont
  {Abelev}} \emph {et~al.} (\bibinfo {collaboration} {ALICE Collaboration}),\
  }\href {\doibase 10.1016/j.physletb.2013.01.012} {\bibfield  {journal}
  {\bibinfo  {journal} {Phys.~Lett.~B}\ }\textbf {\bibinfo {volume} {719}},\
  \bibinfo {pages} {29} (\bibinfo {year} {2013})}\BibitemShut {NoStop}%
\bibitem [{\citenamefont {Aad}\ \emph {et~al.}(2013{\natexlab{a}})\citenamefont
  {Aad} \emph {et~al.}}]{Aad:2012gla}%
  \BibitemOpen
  \bibfield  {author} {\bibinfo {author} {\bibfnamefont {G.}~\bibnamefont
  {Aad}} \emph {et~al.} (\bibinfo {collaboration} {ATLAS Collaboration}),\
  }\href {\doibase 10.1103/PhysRevLett.110.182302} {\bibfield  {journal}
  {\bibinfo  {journal} {Phys.~Rev.~Lett.}\ }\textbf {\bibinfo {volume} {110}},\
  \bibinfo {pages} {182302} (\bibinfo {year} {2013}{\natexlab{a}})}\BibitemShut
  {NoStop}%
\bibitem [{\citenamefont {Adare}\ \emph {et~al.}(2013)\citenamefont {Adare}
  \emph {et~al.}}]{Adare:2013piz}%
  \BibitemOpen
  \bibfield  {author} {\bibinfo {author} {\bibfnamefont {A.}~\bibnamefont
  {Adare}} \emph {et~al.} (\bibinfo {collaboration} {PHENIX Collaboration}),\
  }\href {\doibase 10.1103/PhysRevLett.111.212301} {\bibfield  {journal}
  {\bibinfo  {journal} {Phys.~Rev.~Lett.}\ }\textbf {\bibinfo {volume} {111}},\
  \bibinfo {pages} {212301} (\bibinfo {year} {2013})}\BibitemShut {NoStop}%
\bibitem [{\citenamefont {Aad}\ \emph {et~al.}(2014{\natexlab{a}})\citenamefont
  {Aad} \emph {et~al.}}]{Aad:2014lta}%
  \BibitemOpen
  \bibfield  {author} {\bibinfo {author} {\bibfnamefont {G.}~\bibnamefont
  {Aad}} \emph {et~al.} (\bibinfo {collaboration} {ATLAS Collaboration}),\
  }\href {\doibase 10.1103/PhysRevC.90.044906} {\bibfield  {journal} {\bibinfo
  {journal} {Phys.~Rev.~C}\ }\textbf {\bibinfo {volume} {90}},\ \bibinfo
  {pages} {044906} (\bibinfo {year} {2014}{\natexlab{a}})}\BibitemShut
  {NoStop}%
\bibitem [{\citenamefont {Khachatryan}\ \emph
  {et~al.}(2015{\natexlab{a}})\citenamefont {Khachatryan} \emph
  {et~al.}}]{Khachatryan:2015waa}%
  \BibitemOpen
  \bibfield  {author} {\bibinfo {author} {\bibfnamefont {V.}~\bibnamefont
  {Khachatryan}} \emph {et~al.} (\bibinfo {collaboration} {CMS
  Collaboration}),\ }\href {\doibase 10.1103/PhysRevLett.115.012301} {\bibfield
   {journal} {\bibinfo  {journal} {Phys. Rev. Lett.}\ }\textbf {\bibinfo
  {volume} {115}},\ \bibinfo {pages} {012301} (\bibinfo {year}
  {2015}{\natexlab{a}})}\BibitemShut {NoStop}%
\bibitem [{\citenamefont {Heinz}\ and\ \citenamefont
  {Snellings}(2013)}]{Heinz:2013th}%
  \BibitemOpen
  \bibfield  {author} {\bibinfo {author} {\bibfnamefont {U.}~\bibnamefont
  {Heinz}}\ and\ \bibinfo {author} {\bibfnamefont {R.}~\bibnamefont
  {Snellings}},\ }\href {\doibase 10.1146/annurev-nucl-102212-170540}
  {\bibfield  {journal} {\bibinfo  {journal} {Ann.~Rev.~Nucl.~Part.~Sci.}\
  }\textbf {\bibinfo {volume} {63}},\ \bibinfo {pages} {123} (\bibinfo {year}
  {2013})}\BibitemShut {NoStop}%
\bibitem [{\citenamefont {Gale}\ \emph {et~al.}(2013)\citenamefont {Gale},
  \citenamefont {Jeon},\ and\ \citenamefont {Schenke}}]{Gale:2013da}%
  \BibitemOpen
  \bibfield  {author} {\bibinfo {author} {\bibfnamefont {C.}~\bibnamefont
  {Gale}}, \bibinfo {author} {\bibfnamefont {S.}~\bibnamefont {Jeon}}, \ and\
  \bibinfo {author} {\bibfnamefont {B.}~\bibnamefont {Schenke}},\ }\href
  {\doibase 10.1142/S0217751X13400113} {\bibfield  {journal} {\bibinfo
  {journal} {Int.~J.~Mod.~Phys.~A}\ }\textbf {\bibinfo {volume} {28}},\
  \bibinfo {pages} {1340011} (\bibinfo {year} {2013})}\BibitemShut {NoStop}%
\bibitem [{\citenamefont {Luzum}\ and\ \citenamefont
  {Petersen}(2014)}]{Luzum:2013yya}%
  \BibitemOpen
  \bibfield  {author} {\bibinfo {author} {\bibfnamefont {M.}~\bibnamefont
  {Luzum}}\ and\ \bibinfo {author} {\bibfnamefont {H.}~\bibnamefont
  {Petersen}},\ }\href {\doibase 10.1088/0954-3899/41/6/063102} {\bibfield
  {journal} {\bibinfo  {journal} {J.~Phys.~G}\ }\textbf {\bibinfo {volume}
  {41}},\ \bibinfo {pages} {063102} (\bibinfo {year} {2014})}\BibitemShut
  {NoStop}%
\bibitem [{\citenamefont {Jia}(2014)}]{Jia:2014jca}%
  \BibitemOpen
  \bibfield  {author} {\bibinfo {author} {\bibfnamefont {J.}~\bibnamefont
  {Jia}},\ }\href {\doibase 10.1088/0954-3899/41/12/124003} {\bibfield
  {journal} {\bibinfo  {journal} {J.~Phys.~G}\ }\textbf {\bibinfo {volume}
  {41}},\ \bibinfo {pages} {124003} (\bibinfo {year} {2014})}\BibitemShut
  {NoStop}%
\bibitem [{\citenamefont {Dusling}\ \emph {et~al.}(2016)\citenamefont
  {Dusling}, \citenamefont {Li},\ and\ \citenamefont
  {Schenke}}]{Dusling:2015gta}%
  \BibitemOpen
  \bibfield  {author} {\bibinfo {author} {\bibfnamefont {K.}~\bibnamefont
  {Dusling}}, \bibinfo {author} {\bibfnamefont {W.}~\bibnamefont {Li}}, \ and\
  \bibinfo {author} {\bibfnamefont {B.}~\bibnamefont {Schenke}},\ }\href
  {\doibase 10.1142/S0218301316300022} {\bibfield  {journal} {\bibinfo
  {journal} {Int. J. Mod. Phys. E}\ }\textbf {\bibinfo {volume} {25}},\
  \bibinfo {pages} {1630002} (\bibinfo {year} {2016})}\BibitemShut {NoStop}%
\bibitem [{\citenamefont {Loizides}(2016)}]{Loizides:2016tew}%
  \BibitemOpen
  \bibfield  {author} {\bibinfo {author} {\bibfnamefont {C.}~\bibnamefont
  {Loizides}},\ }\href {\doibase 10.1016/j.nuclphysa.2016.04.022} {\bibfield
  {journal} {\bibinfo  {journal} {Nucl. Phys. A}\ }\textbf {\bibinfo {volume}
  {956}},\ \bibinfo {pages} {200} (\bibinfo {year} {2016})}\BibitemShut
  {NoStop}%
\bibitem [{\citenamefont {Bozek}\ and\ \citenamefont
  {Broniowski}(2013)}]{Bozek:2013uha}%
  \BibitemOpen
  \bibfield  {author} {\bibinfo {author} {\bibfnamefont {P.}~\bibnamefont
  {Bozek}}\ and\ \bibinfo {author} {\bibfnamefont {W.}~\bibnamefont
  {Broniowski}},\ }\href {\doibase 10.1103/PhysRevC.88.014903} {\bibfield
  {journal} {\bibinfo  {journal} {Phys. Rev. C}\ }\textbf {\bibinfo {volume}
  {88}},\ \bibinfo {pages} {014903} (\bibinfo {year} {2013})}\BibitemShut
  {NoStop}%
\bibitem [{\citenamefont {Dusling}\ and\ \citenamefont
  {Venugopalan}(2013)}]{Dusling:2013qoz}%
  \BibitemOpen
  \bibfield  {author} {\bibinfo {author} {\bibfnamefont {K.}~\bibnamefont
  {Dusling}}\ and\ \bibinfo {author} {\bibfnamefont {R.}~\bibnamefont
  {Venugopalan}},\ }\href {\doibase 10.1103/PhysRevD.87.094034} {\bibfield
  {journal} {\bibinfo  {journal} {Phys. Rev. D}\ }\textbf {\bibinfo {volume}
  {87}},\ \bibinfo {pages} {094034} (\bibinfo {year} {2013})}\BibitemShut
  {NoStop}%
\bibitem [{\citenamefont {Khachatryan}\ \emph {et~al.}(2017)\citenamefont
  {Khachatryan} \emph {et~al.}}]{Khachatryan:2016txc}%
  \BibitemOpen
  \bibfield  {author} {\bibinfo {author} {\bibfnamefont {V.}~\bibnamefont
  {Khachatryan}} \emph {et~al.} (\bibinfo {collaboration} {CMS
  Collaboration}),\ }\href {\doibase 10.1016/j.physletb.2016.12.009} {\bibfield
   {journal} {\bibinfo  {journal} {Phys. Lett. B}\ }\textbf {\bibinfo {volume}
  {765}},\ \bibinfo {pages} {193} (\bibinfo {year} {2017})}\BibitemShut
  {NoStop}%
\bibitem [{\citenamefont {Borghini}\ \emph
  {et~al.}(2001{\natexlab{a}})\citenamefont {Borghini}, \citenamefont {Dinh},\
  and\ \citenamefont {Ollitrault}}]{Borghini:2000sa}%
  \BibitemOpen
  \bibfield  {author} {\bibinfo {author} {\bibfnamefont {N.}~\bibnamefont
  {Borghini}}, \bibinfo {author} {\bibfnamefont {P.~M.}\ \bibnamefont {Dinh}},
  \ and\ \bibinfo {author} {\bibfnamefont {J.-Y.}\ \bibnamefont {Ollitrault}},\
  }\href {\doibase 10.1103/PhysRevC.63.054906} {\bibfield  {journal} {\bibinfo
  {journal} {Phys.~Rev.~C}\ }\textbf {\bibinfo {volume} {63}},\ \bibinfo
  {pages} {054906} (\bibinfo {year} {2001}{\natexlab{a}})}\BibitemShut
  {NoStop}%
\bibitem [{\citenamefont {Aad}\ \emph {et~al.}(2013{\natexlab{b}})\citenamefont
  {Aad} \emph {et~al.}}]{Aad:2013fja}%
  \BibitemOpen
  \bibfield  {author} {\bibinfo {author} {\bibfnamefont {G.}~\bibnamefont
  {Aad}} \emph {et~al.} (\bibinfo {collaboration} {ATLAS Collaboration}),\
  }\href {\doibase 10.1016/j.physletb.2013.06.057} {\bibfield  {journal}
  {\bibinfo  {journal} {Phys. Lett. B}\ }\textbf {\bibinfo {volume} {725}},\
  \bibinfo {pages} {60} (\bibinfo {year} {2013}{\natexlab{b}})}\BibitemShut
  {NoStop}%
\bibitem [{\citenamefont {Jia}\ and\ \citenamefont
  {Radhakrishnan}(2015)}]{Jia:2014pza}%
  \BibitemOpen
  \bibfield  {author} {\bibinfo {author} {\bibfnamefont {J.}~\bibnamefont
  {Jia}}\ and\ \bibinfo {author} {\bibfnamefont {S.}~\bibnamefont
  {Radhakrishnan}},\ }\href {\doibase 10.1103/PhysRevC.92.024911} {\bibfield
  {journal} {\bibinfo  {journal} {Phys. Rev. C}\ }\textbf {\bibinfo {volume}
  {92}},\ \bibinfo {pages} {024911} (\bibinfo {year} {2015})}\BibitemShut
  {NoStop}%
\bibitem [{\citenamefont {Aaboud}\ \emph
  {et~al.}(2017{\natexlab{b}})\citenamefont {Aaboud} \emph
  {et~al.}}]{ATLAS:2016ics}%
  \BibitemOpen
  \bibfield  {author} {\bibinfo {author} {\bibfnamefont {M.}~\bibnamefont
  {Aaboud}} \emph {et~al.} (\bibinfo {collaboration} {ATLAS Collaboration}),\
  }\href {\doibase 10.1140/epjc/s10052-017-4988-1} {\bibfield  {journal}
  {\bibinfo  {journal} {Eur. Phys. J. C}\ }\textbf {\bibinfo {volume} {77}},\
  \bibinfo {pages} {428} (\bibinfo {year} {2017}{\natexlab{b}})}\BibitemShut
  {NoStop}%
\bibitem [{\citenamefont {Adler}\ \emph {et~al.}(2002)\citenamefont {Adler}
  \emph {et~al.}}]{Adler:2002pu}%
  \BibitemOpen
  \bibfield  {author} {\bibinfo {author} {\bibfnamefont {C.}~\bibnamefont
  {Adler}} \emph {et~al.} (\bibinfo {collaboration} {STAR Collaboration}),\
  }\href {\doibase 10.1103/PhysRevC.66.034904} {\bibfield  {journal} {\bibinfo
  {journal} {Phys. Rev. C}\ }\textbf {\bibinfo {volume} {66}},\ \bibinfo
  {pages} {034904} (\bibinfo {year} {2002})}\BibitemShut {NoStop}%
\bibitem [{\citenamefont {Voloshin}\ and\ \citenamefont
  {Zhang}(1996)}]{Voloshin:1994mz}%
  \BibitemOpen
  \bibfield  {author} {\bibinfo {author} {\bibfnamefont {S.}~\bibnamefont
  {Voloshin}}\ and\ \bibinfo {author} {\bibfnamefont {Y.}~\bibnamefont
  {Zhang}},\ }\href {\doibase 10.1007/s002880050141} {\bibfield  {journal}
  {\bibinfo  {journal} {Z.~Phys.~C}\ }\textbf {\bibinfo {volume} {70}},\
  \bibinfo {pages} {665} (\bibinfo {year} {1996})}\BibitemShut {NoStop}%
\bibitem [{\citenamefont {Borghini}\ \emph
  {et~al.}(2001{\natexlab{b}})\citenamefont {Borghini}, \citenamefont {Dinh},\
  and\ \citenamefont {Ollitrault}}]{Borghini:2001vi}%
  \BibitemOpen
  \bibfield  {author} {\bibinfo {author} {\bibfnamefont {N.}~\bibnamefont
  {Borghini}}, \bibinfo {author} {\bibfnamefont {P.~M.}\ \bibnamefont {Dinh}},
  \ and\ \bibinfo {author} {\bibfnamefont {J.-Y.}\ \bibnamefont {Ollitrault}},\
  }\href {\doibase 10.1103/PhysRevC.64.054901} {\bibfield  {journal} {\bibinfo
  {journal} {Phys.~Rev.~C}\ }\textbf {\bibinfo {volume} {64}},\ \bibinfo
  {pages} {054901} (\bibinfo {year} {2001}{\natexlab{b}})}\BibitemShut
  {NoStop}%
\bibitem [{\citenamefont {Bilandzic}\ \emph {et~al.}(2011)\citenamefont
  {Bilandzic}, \citenamefont {Snellings},\ and\ \citenamefont
  {Voloshin}}]{Bilandzic:2010jr}%
  \BibitemOpen
  \bibfield  {author} {\bibinfo {author} {\bibfnamefont {A.}~\bibnamefont
  {Bilandzic}}, \bibinfo {author} {\bibfnamefont {R.}~\bibnamefont
  {Snellings}}, \ and\ \bibinfo {author} {\bibfnamefont {S.}~\bibnamefont
  {Voloshin}},\ }\href {\doibase 10.1103/PhysRevC.83.044913} {\bibfield
  {journal} {\bibinfo  {journal} {Phys.~Rev.~C}\ }\textbf {\bibinfo {volume}
  {83}},\ \bibinfo {pages} {044913} (\bibinfo {year} {2011})}\BibitemShut
  {NoStop}%
\bibitem [{\citenamefont {Bilandzic}(2012)}]{Bilandzic:2012wva}%
  \BibitemOpen
  \bibfield  {author} {\bibinfo {author} {\bibfnamefont {A.}~\bibnamefont
  {Bilandzic}},\ }\href
  {https://inspirehep.net/record/1186272/files/CERN-THESIS-2012-018.pdf} {Ph.D.
  thesis},\ \bibinfo  {school} {Utrecht U.} (\bibinfo {year}
  {2012})\BibitemShut {NoStop}%
\bibitem [{\citenamefont {Voloshin}\ \emph {et~al.}(2008)\citenamefont
  {Voloshin}, \citenamefont {Poskanzer}, \citenamefont {Tang},\ and\
  \citenamefont {Wang}}]{Voloshin:2007pc}%
  \BibitemOpen
  \bibfield  {author} {\bibinfo {author} {\bibfnamefont {S.~A.}\ \bibnamefont
  {Voloshin}}, \bibinfo {author} {\bibfnamefont {A.~M.}\ \bibnamefont
  {Poskanzer}}, \bibinfo {author} {\bibfnamefont {A.}~\bibnamefont {Tang}}, \
  and\ \bibinfo {author} {\bibfnamefont {G.}~\bibnamefont {Wang}},\ }\href
  {\doibase 10.1016/j.physletb.2007.11.043} {\bibfield  {journal} {\bibinfo
  {journal} {Phys. Lett. B}\ }\textbf {\bibinfo {volume} {659}},\ \bibinfo
  {pages} {537} (\bibinfo {year} {2008})}\BibitemShut {NoStop}%
\bibitem [{\citenamefont {Bzdak}\ \emph {et~al.}(2014)\citenamefont {Bzdak},
  \citenamefont {Bozek},\ and\ \citenamefont {McLerran}}]{Bzdak:2013rya}%
  \BibitemOpen
  \bibfield  {author} {\bibinfo {author} {\bibfnamefont {A.}~\bibnamefont
  {Bzdak}}, \bibinfo {author} {\bibfnamefont {P.}~\bibnamefont {Bozek}}, \ and\
  \bibinfo {author} {\bibfnamefont {L.}~\bibnamefont {McLerran}},\ }\href
  {\doibase 10.1016/j.nuclphysa.2014.03.007} {\bibfield  {journal} {\bibinfo
  {journal} {Nucl.~Phys.~A}\ }\textbf {\bibinfo {volume} {927}},\ \bibinfo
  {pages} {15} (\bibinfo {year} {2014})}\BibitemShut {NoStop}%
\bibitem [{\citenamefont {Yan}\ and\ \citenamefont
  {Ollitrault}(2014)}]{Yan:2013laa}%
  \BibitemOpen
  \bibfield  {author} {\bibinfo {author} {\bibfnamefont {L.}~\bibnamefont
  {Yan}}\ and\ \bibinfo {author} {\bibfnamefont {J.-Y.}\ \bibnamefont
  {Ollitrault}},\ }\href {\doibase 10.1103/PhysRevLett.112.082301} {\bibfield
  {journal} {\bibinfo  {journal} {Phys.~Rev.~Lett.}\ }\textbf {\bibinfo
  {volume} {112}},\ \bibinfo {pages} {082301} (\bibinfo {year}
  {2014})}\BibitemShut {NoStop}%
\bibitem [{\citenamefont {Jia}\ and\ \citenamefont
  {Mohapatra}(2013)}]{Jia:2013tja}%
  \BibitemOpen
  \bibfield  {author} {\bibinfo {author} {\bibfnamefont {J.}~\bibnamefont
  {Jia}}\ and\ \bibinfo {author} {\bibfnamefont {S.}~\bibnamefont
  {Mohapatra}},\ }\href {\doibase 10.1103/PhysRevC.88.014907} {\bibfield
  {journal} {\bibinfo  {journal} {Phys. Rev. C}\ }\textbf {\bibinfo {volume}
  {88}},\ \bibinfo {pages} {014907} (\bibinfo {year} {2013})}\BibitemShut
  {NoStop}%
\bibitem [{\citenamefont {Bhalerao}\ \emph {et~al.}(2003)\citenamefont
  {Bhalerao}, \citenamefont {Borghini},\ and\ \citenamefont
  {Ollitrault}}]{Bhalerao:2003xf}%
  \BibitemOpen
  \bibfield  {author} {\bibinfo {author} {\bibfnamefont {R.~S.}\ \bibnamefont
  {Bhalerao}}, \bibinfo {author} {\bibfnamefont {N.}~\bibnamefont {Borghini}},
  \ and\ \bibinfo {author} {\bibfnamefont {J.~Y.}\ \bibnamefont {Ollitrault}},\
  }\href {\doibase 10.1016/j.nuclphysa.2003.08.007} {\bibfield  {journal}
  {\bibinfo  {journal} {Nucl. Phys. A}\ }\textbf {\bibinfo {volume} {727}},\
  \bibinfo {pages} {373} (\bibinfo {year} {2003})}\BibitemShut {NoStop}%
\bibitem [{\citenamefont {Bilandzic}\ \emph {et~al.}(2014)\citenamefont
  {Bilandzic}, \citenamefont {Christensen}, \citenamefont {Gulbrandsen},
  \citenamefont {Hansen},\ and\ \citenamefont {Zhou}}]{Bilandzic:2013kga}%
  \BibitemOpen
  \bibfield  {author} {\bibinfo {author} {\bibfnamefont {A.}~\bibnamefont
  {Bilandzic}}, \bibinfo {author} {\bibfnamefont {C.~H.}\ \bibnamefont
  {Christensen}}, \bibinfo {author} {\bibfnamefont {K.}~\bibnamefont
  {Gulbrandsen}}, \bibinfo {author} {\bibfnamefont {A.}~\bibnamefont {Hansen}},
  \ and\ \bibinfo {author} {\bibfnamefont {Y.}~\bibnamefont {Zhou}},\ }\href
  {\doibase 10.1103/PhysRevC.89.064904} {\bibfield  {journal} {\bibinfo
  {journal} {Phys. Rev. C}\ }\textbf {\bibinfo {volume} {89}},\ \bibinfo
  {pages} {064904} (\bibinfo {year} {2014})}\BibitemShut {NoStop}%
\bibitem [{\citenamefont {{Sj\"ostrand, Torbj\"orn and Mrenna, Stephen and
  Skands, Peter Z.}}(2008)}]{Sjostrand:2007gs}%
  \BibitemOpen
  \bibfield  {author} {\bibinfo {author} {\bibnamefont {{Sj\"ostrand,
  Torbj\"orn and Mrenna, Stephen and Skands, Peter Z.}}},\ }\href {\doibase
  10.1016/j.cpc.2008.01.036} {\bibfield  {journal} {\bibinfo  {journal}
  {Comput. Phys. Commun.}\ }\textbf {\bibinfo {volume} {178}},\ \bibinfo
  {pages} {852} (\bibinfo {year} {2008})}\BibitemShut {NoStop}%
\bibitem [{\citenamefont {Masera}\ \emph {et~al.}(2009)\citenamefont {Masera},
  \citenamefont {Ortona}, \citenamefont {Poghosyan},\ and\ \citenamefont
  {Prino}}]{Masera}%
  \BibitemOpen
  \bibfield  {author} {\bibinfo {author} {\bibfnamefont {M.}~\bibnamefont
  {Masera}}, \bibinfo {author} {\bibfnamefont {G.}~\bibnamefont {Ortona}},
  \bibinfo {author} {\bibfnamefont {M.~G.}\ \bibnamefont {Poghosyan}}, \ and\
  \bibinfo {author} {\bibfnamefont {F.}~\bibnamefont {Prino}},\ }\href
  {\doibase 10.1103/PhysRevC.79.064909} {\bibfield  {journal} {\bibinfo
  {journal} {Phys. Rev. C}\ }\textbf {\bibinfo {volume} {79}},\ \bibinfo
  {pages} {064909} (\bibinfo {year} {2009})}\BibitemShut {NoStop}%
\bibitem [{\citenamefont {Chatrchyan}\ \emph
  {et~al.}(2013{\natexlab{b}})\citenamefont {Chatrchyan} \emph
  {et~al.}}]{Chatrchyan:2013nka}%
  \BibitemOpen
  \bibfield  {author} {\bibinfo {author} {\bibfnamefont {S.}~\bibnamefont
  {Chatrchyan}} \emph {et~al.} (\bibinfo {collaboration} {CMS Collaboration}),\
  }\href {\doibase 10.1016/j.physletb.2013.06.028} {\bibfield  {journal}
  {\bibinfo  {journal} {Phys.~Lett.~B}\ }\textbf {\bibinfo {volume} {724}},\
  \bibinfo {pages} {213} (\bibinfo {year} {2013}{\natexlab{b}})}\BibitemShut
  {NoStop}%
\bibitem [{\citenamefont {Schlichting}\ and\ \citenamefont
  {Tribedy}(2016)}]{Schlichting:2016sqo}%
  \BibitemOpen
  \bibfield  {author} {\bibinfo {author} {\bibfnamefont {S.}~\bibnamefont
  {Schlichting}}\ and\ \bibinfo {author} {\bibfnamefont {P.}~\bibnamefont
  {Tribedy}},\ }\href {\doibase 10.1155/2016/8460349} {\bibfield  {journal}
  {\bibinfo  {journal} {Adv. High Energy Phys.}\ }\textbf {\bibinfo {volume}
  {2016}},\ \bibinfo {pages} {8460349} (\bibinfo {year} {2016})}\BibitemShut
  {NoStop}%
\bibitem [{\citenamefont {Dumitru}\ \emph {et~al.}(2015)\citenamefont
  {Dumitru}, \citenamefont {McLerran},\ and\ \citenamefont
  {Skokov}}]{Dumitru:2014yza}%
  \BibitemOpen
  \bibfield  {author} {\bibinfo {author} {\bibfnamefont {A.}~\bibnamefont
  {Dumitru}}, \bibinfo {author} {\bibfnamefont {L.}~\bibnamefont {McLerran}}, \
  and\ \bibinfo {author} {\bibfnamefont {V.}~\bibnamefont {Skokov}},\ }\href
  {\doibase 10.1016/j.physletb.2015.02.046} {\bibfield  {journal} {\bibinfo
  {journal} {Phys. Lett. B}\ }\textbf {\bibinfo {volume} {743}},\ \bibinfo
  {pages} {134} (\bibinfo {year} {2015})}\BibitemShut {NoStop}%
\bibitem [{\citenamefont {Lappi}\ \emph {et~al.}(2016)\citenamefont {Lappi},
  \citenamefont {Schenke}, \citenamefont {Schlichting},\ and\ \citenamefont
  {Venugopalan}}]{Lappi:2015vta}%
  \BibitemOpen
  \bibfield  {author} {\bibinfo {author} {\bibfnamefont {T.}~\bibnamefont
  {Lappi}}, \bibinfo {author} {\bibfnamefont {B.}~\bibnamefont {Schenke}},
  \bibinfo {author} {\bibfnamefont {S.}~\bibnamefont {Schlichting}}, \ and\
  \bibinfo {author} {\bibfnamefont {R.}~\bibnamefont {Venugopalan}},\ }\href
  {\doibase 10.1007/JHEP01(2016)061} {\bibfield  {journal} {\bibinfo  {journal}
  {JHEP}\ }\textbf {\bibinfo {volume} {01}},\ \bibinfo {pages} {061} (\bibinfo
  {year} {2016})}\BibitemShut {NoStop}%
\bibitem [{\citenamefont {Bhalerao}\ \emph {et~al.}(2015)\citenamefont
  {Bhalerao}, \citenamefont {Ollitrault},\ and\ \citenamefont
  {Pal}}]{Bhalerao:2014xra}%
  \BibitemOpen
  \bibfield  {author} {\bibinfo {author} {\bibfnamefont {R.~S.}\ \bibnamefont
  {Bhalerao}}, \bibinfo {author} {\bibfnamefont {J.-Y.}\ \bibnamefont
  {Ollitrault}}, \ and\ \bibinfo {author} {\bibfnamefont {S.}~\bibnamefont
  {Pal}},\ }\href {\doibase 10.1016/j.physletb.2015.01.019} {\bibfield
  {journal} {\bibinfo  {journal} {Phys. Lett. B}\ }\textbf {\bibinfo {volume}
  {742}},\ \bibinfo {pages} {94} (\bibinfo {year} {2015})}\BibitemShut
  {NoStop}%
\bibitem [{\citenamefont {Khachatryan}\ \emph
  {et~al.}(2015{\natexlab{b}})\citenamefont {Khachatryan} \emph
  {et~al.}}]{Khachatryan:2015oea}%
  \BibitemOpen
  \bibfield  {author} {\bibinfo {author} {\bibfnamefont {V.}~\bibnamefont
  {Khachatryan}} \emph {et~al.} (\bibinfo {collaboration} {CMS
  Collaboration}),\ }\href {\doibase 10.1103/PhysRevC.92.034911} {\bibfield
  {journal} {\bibinfo  {journal} {Phys. Rev. C}\ }\textbf {\bibinfo {volume}
  {92}},\ \bibinfo {pages} {034911} (\bibinfo {year}
  {2015}{\natexlab{b}})}\BibitemShut {NoStop}%
\bibitem [{\citenamefont {Bhalerao}\ \emph {et~al.}(2013)\citenamefont
  {Bhalerao}, \citenamefont {Ollitrault},\ and\ \citenamefont
  {Pal}}]{Bhalerao:2013ina}%
  \BibitemOpen
  \bibfield  {author} {\bibinfo {author} {\bibfnamefont {R.~S.}\ \bibnamefont
  {Bhalerao}}, \bibinfo {author} {\bibfnamefont {J.~Y.}\ \bibnamefont
  {Ollitrault}}, \ and\ \bibinfo {author} {\bibfnamefont {S.}~\bibnamefont
  {Pal}},\ }\href {\doibase 10.1103/PhysRevC.88.024909} {\bibfield  {journal}
  {\bibinfo  {journal} {Phys.~Rev.~C}\ }\textbf {\bibinfo {volume} {88}},\
  \bibinfo {pages} {024909} (\bibinfo {year} {2013})}\BibitemShut {NoStop}%
\bibitem [{\citenamefont {Gardim}\ \emph {et~al.}(2012)\citenamefont {Gardim},
  \citenamefont {Grassi}, \citenamefont {Luzum},\ and\ \citenamefont
  {Ollitrault}}]{Gardim:2011xv}%
  \BibitemOpen
  \bibfield  {author} {\bibinfo {author} {\bibfnamefont {F.~G.}\ \bibnamefont
  {Gardim}}, \bibinfo {author} {\bibfnamefont {F.}~\bibnamefont {Grassi}},
  \bibinfo {author} {\bibfnamefont {M.}~\bibnamefont {Luzum}}, \ and\ \bibinfo
  {author} {\bibfnamefont {J.-Y.}\ \bibnamefont {Ollitrault}},\ }\href
  {\doibase 10.1103/PhysRevC.85.024908} {\bibfield  {journal} {\bibinfo
  {journal} {Phys. Rev. C}\ }\textbf {\bibinfo {volume} {85}},\ \bibinfo
  {pages} {024908} (\bibinfo {year} {2012})}\BibitemShut {NoStop}%
\bibitem [{\citenamefont {Teaney}\ and\ \citenamefont
  {Yan}(2012)}]{Teaney:2012ke}%
  \BibitemOpen
  \bibfield  {author} {\bibinfo {author} {\bibfnamefont {D.}~\bibnamefont
  {Teaney}}\ and\ \bibinfo {author} {\bibfnamefont {L.}~\bibnamefont {Yan}},\
  }\href {\doibase 10.1103/PhysRevC.86.044908} {\bibfield  {journal} {\bibinfo
  {journal} {Phys. Rev. C}\ }\textbf {\bibinfo {volume} {86}},\ \bibinfo
  {pages} {044908} (\bibinfo {year} {2012})}\BibitemShut {NoStop}%
\bibitem [{\citenamefont {Aad}\ \emph {et~al.}(2014{\natexlab{b}})\citenamefont
  {Aad} \emph {et~al.}}]{Aad:2014fla}%
  \BibitemOpen
  \bibfield  {author} {\bibinfo {author} {\bibfnamefont {G.}~\bibnamefont
  {Aad}} \emph {et~al.} (\bibinfo {collaboration} {ATLAS Collaboration}),\
  }\href {\doibase 10.1103/PhysRevC.90.024905} {\bibfield  {journal} {\bibinfo
  {journal} {Phys. Rev. C}\ }\textbf {\bibinfo {volume} {90}},\ \bibinfo
  {pages} {024905} (\bibinfo {year} {2014}{\natexlab{b}})}\BibitemShut
  {NoStop}%
\bibitem [{\citenamefont {Aad}\ \emph {et~al.}(2015)\citenamefont {Aad} \emph
  {et~al.}}]{Aad:2015lwa}%
  \BibitemOpen
  \bibfield  {author} {\bibinfo {author} {\bibfnamefont {G.}~\bibnamefont
  {Aad}} \emph {et~al.} (\bibinfo {collaboration} {ATLAS Collaboration}),\
  }\href {\doibase 10.1103/PhysRevC.92.034903} {\bibfield  {journal} {\bibinfo
  {journal} {Phys. Rev. C}\ }\textbf {\bibinfo {volume} {92}},\ \bibinfo
  {pages} {034903} (\bibinfo {year} {2015})}\BibitemShut {NoStop}%
\bibitem [{\citenamefont {Qiu}\ and\ \citenamefont {Heinz}(2012)}]{Qiu:2012uy}%
  \BibitemOpen
  \bibfield  {author} {\bibinfo {author} {\bibfnamefont {Z.}~\bibnamefont
  {Qiu}}\ and\ \bibinfo {author} {\bibfnamefont {U.}~\bibnamefont {Heinz}},\
  }\href {\doibase 10.1016/j.physletb.2012.09.030} {\bibfield  {journal}
  {\bibinfo  {journal} {Phys. Lett. B}\ }\textbf {\bibinfo {volume} {717}},\
  \bibinfo {pages} {261} (\bibinfo {year} {2012})}\BibitemShut {NoStop}%
\bibitem [{\citenamefont {Di~Francesco}\ \emph {et~al.}(2017)\citenamefont
  {Di~Francesco}, \citenamefont {Guilbaud}, \citenamefont {Luzum},\ and\
  \citenamefont {Ollitrault}}]{DiFrancesco:2016srj}%
  \BibitemOpen
  \bibfield  {author} {\bibinfo {author} {\bibfnamefont {P.}~\bibnamefont
  {Di~Francesco}}, \bibinfo {author} {\bibfnamefont {M.}~\bibnamefont
  {Guilbaud}}, \bibinfo {author} {\bibfnamefont {M.}~\bibnamefont {Luzum}}, \
  and\ \bibinfo {author} {\bibfnamefont {J.-Y.}\ \bibnamefont {Ollitrault}},\
  }\href {\doibase 10.1103/PhysRevC.95.044911} {\bibfield  {journal} {\bibinfo
  {journal} {Phys. Rev. C}\ }\textbf {\bibinfo {volume} {95}},\ \bibinfo
  {pages} {044911} (\bibinfo {year} {2017})}\BibitemShut {NoStop}%
\bibitem [{\citenamefont {Bilandzic}\ \emph {et~al.}()\citenamefont
  {Bilandzic}, \citenamefont {Christensen},\ and\ \citenamefont
  {Gulbrandsen}}]{code}%
  \BibitemOpen
  \bibfield  {author} {\bibinfo {author} {\bibfnamefont {A.}~\bibnamefont
  {Bilandzic}}, \bibinfo {author} {\bibfnamefont {C.~H.}\ \bibnamefont
  {Christensen}}, \ and\ \bibinfo {author} {\bibfnamefont {K.}~\bibnamefont
  {Gulbrandsen}},\ }\href@noop {} {\enquote {\bibinfo {title}
  {\url{http://www.nbi.dk/~cholm/mcorrelations/}},}\ }\BibitemShut {NoStop}%
\end{thebibliography}%
\bibliographystyle{apsrev4-1}
\end{document}